\documentclass{aa}  

\usepackage{graphicx,nicefrac}
\usepackage[varg]{txfonts}
\usepackage[colorlinks=true,citecolor=blue]{hyperref}

\usepackage{siunitx}
\DeclareSIUnit\Lsun{\ensuremath{\mathit{L}_\odot}}
\DeclareSIUnit\Msun{\ensuremath{\mathit{M}_\odot}}
\DeclareSIUnit\Rstar{\ensuremath{\mathit{R}_\star}}
\DeclareSIUnit\parsec{pc}
\DeclareSIUnit\year{yr}
\DeclareSIUnit{\gauss}{G}
\DeclareSIUnit\magnitude{mag}
\DeclareSIUnit\milliarcsecond{mas}

\newcommand{\PA}{\mathit{P\!A}}
\newcommand{\AIC}{\mathit{A\mkern -1 mu I\!C}}

\newcommand{\hexcolor}[2]{\textcolor[HTML]{#1}{#2}}

\newcommand{\bulletnewline}{\vspace{.5em}\newline}

\usepackage{chemformula}
\usepackage{booktabs}
\usepackage{subcaption}
\usepackage{multirow}

\begin{document}

\title{The MATISSE view of the inner region of the RY Tau protoplanetary disk\thanks{Based on observations collected at the European Organisation for Astronomical Research in the Southern Hemisphere with the VLTI-MATISSE instrument under ESO programmes 106.21Q8.006 and 106.21Q8.002.}}

\author{
  J.~S.~Martin\inst{1}
  \and
  J.~Kobus\inst{1}
  \and
  J.~Varga \inst{2}
  \and
  A.~Matter \inst{3}
  \and
  S.~Wolf\inst{1}
  \and
  M.~Abello \inst{3}
  \and
  F.~Allouche \inst{3}
  \and
  J.-C.~Augereau \inst{4}
  \and
  P.~Berio \inst{3}
  \and
  F.~Bettonvil \inst{5}
  \and
  R. van Boekel \inst{6}
  \and
  P.~A.~Boley \inst{6}
  \and
  P.~Cruzalèbes \inst{3}
  \and
  W.~C.~Danchi \inst{7}
  \and
  J.~Drevon \inst{3}
  \and
  C.~Dominik \inst{8}
  \and
  V.~Fleury \inst{3}
  \and
  V.~Gámez Rosas \inst{9}
  \and
  A.~Glindemann \inst{9}
  \and
  L.~N.~A.~van Haastere \inst{11}
  \and
  M.~Heininger \inst{10}
  \and
  Th.~Henning \inst{6}
  \and
  K.-H.~Hofmann \inst{10}
  \and
  M.~Hogerheijde \inst{8,11}
  \and
  M.~Houllé \inst{3}
  \and
  J.~W.~Isbell \inst{12}
  \and
  W.~Jaffe \inst{11}
  \and
  L.~Labadie \inst{13}
  \and
  S.~Lagarde \inst{3}
  \and
  J.~H.~Leftley \inst{14}
  \and
  M.~Lehmitz \inst{6}
  \and
  M.~Letessier \inst{4}
  \and
  B.~Lopez \inst{3}
  \and
  F.~Lykou \inst{2}
  \and
  J.~Ma \inst{4}
  \and
  A.~Meilland \inst{3}
  \and
  F.~Millour \inst{3}
  \and
  C.~Paladini \inst{15}
  \and
  E.~Pantin \inst{16}
  \and
  R.~G.~Petrov \inst{3}
  \and
  P.~Priolet \inst{4}
  \and
  S.~Robbe-Dubois \inst{3}
  \and
  D.~Schertl \inst{9}
  \and
  M.~Scheuck \inst{6}
  \and
  J.~Scigliuto \inst{3}
  \and
  G.~Weigelt \inst{10}
  \and
  J.~Woillez \inst{9}
  \and
  MATISSE~Collaboration
}

\institute{Institut für Theoretische Physik und Astrophysik, Christian-Albrechts-Universität zu Kiel, Leibnizstraße 15, 24118 Kiel, Germany \\
  \email{jmartin@astrophysik.uni-kiel.de}
  \and
  Konkoly Observatory, Research Centre for Astronomy and Earth Sciences, HUN-REN, Konkoly Thege Miklós út 15-17., H-1121 Budapest, Hungary
  \and
  Laboratoire Lagrange, Université Côte d’Azur, Observatoire de la Côte d’Azur, CNRS, Boulevard de l’Observatoire, CS 34229, 06304 Nice Cedex 4, France
  \and
  Univ. Grenoble Alpes, CNRS, IPAG, 38000 Grenoble, France
  \and
  NOVA Optical IR Instrumentation Group at ASTRON, The Netherlands
  \and
  Max-Planck-Institut für Astronomie, Königstuhl 17, 69117 Heidelberg, Germany
  \and
  NASA Goddard Space Flight Center, Astrophysics Division, Greenbelt, MD, 20771, USA
  \and
  Anton Pannekoek Institute for Astronomy, University of Amsterdam, Science Park 904, 1090 GE Amsterdam, The Netherlands
  \and
  European Southern Observatory, Karl-Schwarzschild-Str. 2, 85748 Garching, Germany    
  \and
  Max-Planck-Institut für Radioastronomie, Auf dem Hügel 69, 53121, Bonn, Germany
  \and
  Leiden Observatory, Leiden University, P.O. Box 9513, 2300 RA Leiden, The Netherlands
  \and
  Steward Observatory, Department of Astronomy, University of Arizona, Tucson, AZ 85721, USA
  \and
  Physikalisches Institut, Universität zu Köln, Zülpicher Str. 77, Cologne, 50937, Germany
  \and
  School of Physics \& Astronomy, University of Southampton, University Road, Southampton SO17 1BJ, UK
  \and
  European Southern Observatory, Alonso de Córdova, 3107 Vitacura, Santiago, Chile
  \and
  AIM, CEA, CNRS, Université Paris-Saclay, Université Paris Diderot, Sorbonne Paris Cité, F-91191 Gif-sur-Yvette, France
}

\date{}

\abstract
    { The physical conditions and processes taking place in the innermost regions of protoplanetary disks are essential for planet formation and general disk evolution. In this context, we study the T-Tauri type young stellar object RY Tau, which exhibits a dust-depleted inner cavity characteristic of a transition disk.}
    { The goal of this study is to analyze spectrally resolved interferometric observations in the $L$, $M$, and $N$ bands of the RY Tau protoplanetary disk obtained with MATISSE. We aim to provide constraints on the spatial distribution and mineralogy of dust in the inner few astronomical units by producing synthetic observations fitting the interferometric observables.}
    { We employed a 2D temperature gradient disk model to estimate the orientation of the inner disk. Successively, we analyzed the chemical composition of silicates depending on spatial region in the disk. Finally, we sampled the parameter space of a viscous accretion disk model via Monte Carlo radiative transfer simulations to investigate the actual 3D dust density distribution of RY Tau.}
    { We constrained the orientation of the inner disk of RY Tau, finding no evidence of significant misalignment with respect to its outer disk. We identified several silicate species commonly found in protoplanetary disks and observed a depletion of amorphous dust grains toward the central protostar. By simultaneously considering the observed visibilities and the spectral energy distribution (SED), we found that an accretion disk and an optically thin envelope enshrouding the protostar fits the observations best. Radiative transfer simulations show that hot dust close to the protostar and in the line of sight (LOS) to the observer, either in the uppermost disk layers of a strongly flared disk or in a dusty envelope, is necessary to model the observations. The shadow cast by a dense innermost disk midplane on the dust further out explains the observed closure phases in the $L$ band and (to some extent) in the $M$ band. However, the closure phases in the $N$ band are underestimated by our model, hinting at an additional asymmetry in the flux density distribution that is not visible at shorter wavelengths.
    }
    {}

    \keywords{accretion -- astrochemistry --
      protoplanetary disks -- stars: individual: RY Tau --
      stars: variables: T Tauri, Herbig Ae/Be 
    }

    \maketitle

    \section{Introduction}\label{sec:introduction}
    Observing young stellar objects (YSOs) in active star forming regions is crucial to testing theoretical models of the star and planet formation processes. Stars form through the gravitational collapse of molecular clouds with angular momentum conservation, leading to the formation of an accretion disk \citep[e.g.,][]{WilliamsCieza2011}. These circumstellar disks are made up of gas and dust, extending across several hundred astronomical units (au) and they are the progenitors of planetary systems with planet formation taking place within them \citep[e.g.,][]{Keppler+2018}. YSOs in different evolutionary stages classified by their spectral energy distribution \citep[SED;][]{Lada1987} have been observed by direct imaging and spectroscopic measurements on large scales \citep[e.g.,][]{Smith+2005,Espaillat+2011,Garufi+2024}, revealing a diverse population of different sizes, shapes, and dust compositions. In recent years, infrared (IR) long-baseline interferometric observations have provided additional insights into the evolution of protoplanetary disks \citep[e.g.,][]{vanBoekel+2004,Monnier+2006,Weigelt+2011,Menu+2015,Brunngraeber+2016,Lazareff+2017,Varga+2018,Kobus+2020,Varga+2021,Hofmann+2022,Varga+2024}, probing their innermost regions comparable in size to the inner Solar System harboring the terrestrial planets.\newline
    The physical processes driving planet formation and the transition from the dispersal of the natal envelope to the various evolutionary stages of protoplanetary disks are not yet fully understood, but they can be extensively described with various theoretical models. A depletion of dust grains in the innermost disk regions is observed in a small population of YSOs \citep{Calvet+2002,Brown+2007,Andrews+2011,FrancisVanDerMarel2020}, indicating a short-lived transition disk state. Selected explanations of this phenomenon include extensive particle growth \citep{DullemondDominik2005}, photoevaporation driven disk winds \citep{Alexander+2006}, the gravitational impact of companions of the embedded YSO \citep{LinPapaloizou1979,Penzlin+2024}, or dead zones induced by embedded planets in the outer disks \citep{Regaly+2012,KleyNelson2012}.\newline
    An exemplary, extensively studied transition disk is that of RY Tau, which is an intermediate mass classical T-Tauri star (CTTS) \citep{StOngeBastien2008,Petrov+2019,Valegard+2022} in the Taurus-Auriga Molecular Cloud \citep{Kenyon+1994} with an estimated age of $\SI{4}{\mega\year}$ \citep{Valegard+2021}. The SED is characteristic for a Class II object \citep{Lada1987,Garufi+2018,Garufi+2019,Garufi+2024} and an inclined circumstellar disk is detected in the \mbox{(sub-)millimeter} wavelength range \citep{Isella+2010,Long+2019,Harrison+2024}, which is typical for a CTTS. These observations reveal an inner dust cavity with a radius of $\SI{27}{\astronomicalunit}$ and ring-like structures on larger scales,  possibly caused by ongoing planet formation \citep{FrancisVanDerMarel2020}.\newline 
    Observations in the optical and near-infrared (NIR) show that the protostar of RY Tau is enshrouded by dust \citep{Nakajima+1995,Takami+2013,Davies+2020,Takami+2023}, which can be attributed to the remnants of the natal envelope \citep{Garufi+2019} or a disk wind \citep{Babina+2016,Valegard+2022,Sauty+2022,Meskini+2024}. RY Tau exhibits UX Ori-type behavior, that is, strong variability in the visible wavelength range \citep{Petrov+1999}. The variability is accounted for by obstruction of the host star by a dusty environment \citep{Babina+2016}, which is a common phenomenon among YSOs \citep{Kobus+2020}. \citet{Petrov+2021} reported anticyclic variations of gas infall and outflow, which could be caused by magnetohydrodynamic processes or a close-in planet interacting with the accretion disk. A bipolar jet (HH938) is observed in X-ray \citep{Skinner+2016}, far-ultraviolet \citep[FUV;][]{Skinner+2018}, visible, and NIR \citep{StOngeBastien2008,Garufi+2019,Takami+2023} wavelengths.\newline
    Mid-infrared (MIR) observations obtained with the \emph{Spitzer Space Telescope} \citep{Espaillat+2011} show a strong silicate emission feature at a wavelength of approximately \SI{10}{\micro\meter}. The FIR to submillimeter (submm) photometry is provided by \emph{Herschel Space Observatory} observations \citep{Ribas+2017}. Interferometric observations using the MID-infrared Interferometric instrument \citep[MIDI;][]{Leinert+2003} in the $N$ band have provided the first insights into the circumstellar dust composition in the innermost disk regions, revealing an increase in crystalline silicates compared to amorphous dust toward the central protostar. \citet{Schegerer+2008} additionally found that a radiative transfer model of an accretion disk in combination with an envelope fits the SED and visibilities of two baselines in the $N$ band well.\newline
    Interferometric observations of RY Tau in the $K$ band (\SIrange{2.0}{2.4}{\micro\meter}) were collected using the Center for High Angular Resolution Astronomy (CHARA) array and the GRAVITY instrument at the VLTI \citep{GRAVITY+2021}. These observations provide evidence of an inner disk rim shaped by sublimation, and occultation of the central protostar of RY Tau by close-in dust in the LOS to the observer \citep{Davies+2020}. The characteristic sizes (i.e., the half-flux radii) of RY Tau in the NIR and MIR are \SI{2.54}{\milliarcsecond} in the $K$ band \citep{GRAVITY+2021} and \SI{8.0}{\milliarcsecond} in the $N$ band \citep{Varga+2018}.\newline
    In this work, we present the first interferometric observations of RY Tau covering the $L$ (\SIrange{3.2}{3.9}{\micro\meter}), $M$ (\SIrange{4.5}{5.0}{\micro\meter}) and $N$ (\SIrange{8}{13}{\micro\meter}) bands simultaneously. The data were collected using the Multi AperTure mid-Infrared SpectroScopic Experiment \citep[MATISSE;][]{Lopez+2022} at the Very Large Telescope Interferometer (VLTI). The goal of this study is to present constraints on the spatial distribution and mineralogy of the dust in the inner few astronomical units around the central protostar of RY Tau. To this end, we employ a model-driven approach to produce synthetic observations that fit the spectrally resolved measurements by MATISSE presented in Sect.~\ref{sec:observations}. We introduce the basic methods used in this work in Sect.~\ref{sec:methods}. We present a 2D semi-physical, radially symmetric disk model with a simple temperature prescription  to constrain the inner disk orientation in
Sect.~\ref{sec:temperature_gradient}. On the basis of the estimated orientation of the inner disk, the radial dependence of the silicate composition of the circumstellar dust is investigated in Sect.~\ref{sec:compositional_analysis}. Finally, a large set of radiative transfer simulation results is presented and analyzed in Sect.~\ref{sec:radiative_transfer}  to provide constraints on the spatial dust distribution in the circumstellar vicinity of RY Tau, which is the main goal of this work.  

    \section{MATISSE observations and data reduction}\label{sec:observations}
\begin{figure*}
      \centering
      \includegraphics[width=\textwidth]{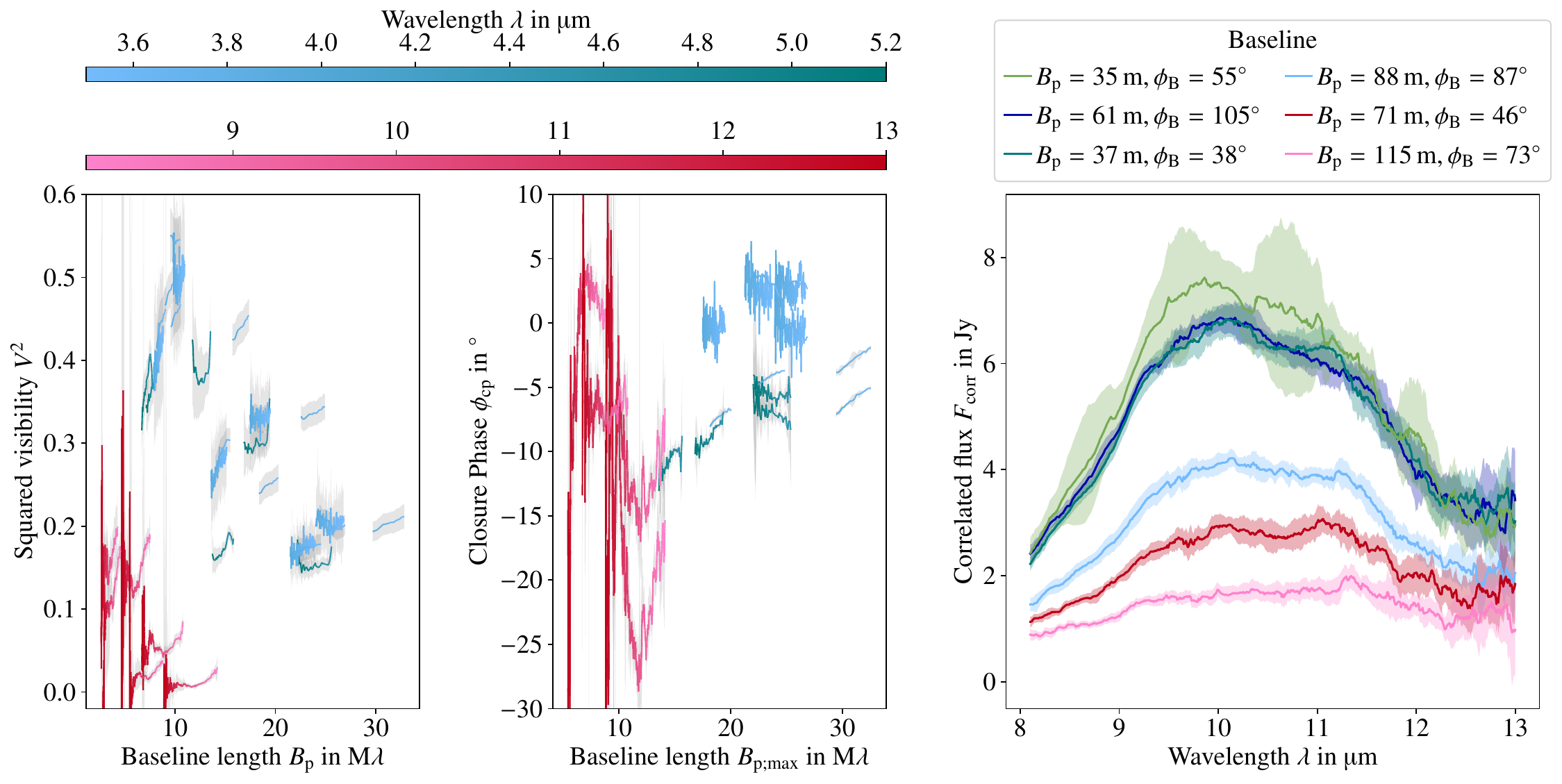}
      \caption{Overview of the MATISSE observations of RY Tau listed in Table \ref{tab:observations}. The corresponding $u$-$v$ coordinates are shown in Fig.~\ref{fig:u-v-coverage}. \emph{Left panel:} Squared visibilities, $V^2$, over projected baseline length $B_\text{p}$ (pink/red: \SIrange{8}{13}{\micro\meter} wavelengths; blue/green: \SIrange{3.5}{5.2}{\micro\meter} wavelengths; see color bars). \emph{Middle panel:} Measured closure phases, $\phi_\text{cp}$, over the longest projected baseline length in the telescope triplet, $B_\text{p;max}$. \emph{Right panel:} Correlated fluxes, $F_\text{corr}$, and corresponding projected baseline length, $B_\text{p}$, and orientation, $\phi_\text{B}$, in the $N$ band.}
      \label{fig:data_overview}
    \end{figure*}

    RY Tau was observed with MATISSE \citep{Lopez+2022} in January 2021 in the frame of a Guaranteed Time Observing (GTO) program for surveying YSOs, providing the first simultaneous \mbox{$L$-,} $M$-, and $N$-band interferometric observations. To increase the $u$-$v$ coverage in the $L$ band, additional observations were performed after 16 days. These short times between observations, compared to periods of strong changes in activity on a timescale of $\sim \si{\year}$ \citep[e.g.,][]{Petrov+2019}, allowed us to model a consistent YSO state without having to take temporal variability into account. We note, however, that \citet{Petrov+2021} reported variation of radial velocities accompanied by visible light magnitude changes $\sim \SI{0.5}{\magnitude}$ with a period of \SI{22}{\day}. Nevertheless, we take the additional $L$-band observations on January 23 into account in our modeling, assuming that the dusty environment around the protostar does not change enough to have a significant impact on the observed visibilities. An overview of the observations analyzed in this work is given in Table \ref{tab:observations}. The spectral resolutions are \num{31.5} for LOW-LM, \num{499} for MED-LM, and \num{218} for HIGH-N spectral modes, respectively \citep{Lopez+2022}. The auxiliary telescopes (ATs) observations were collected in the ``medium'' configuration with the ATs placed at the stations given in Table \ref{tab:observations}. In total, we collected data on six baselines in the $L$, $M$ and $N$ band using the unit telescopes (UTs) and, additionally, on twelve baselines in the $L$ band using the ATs. The $u$-$v$ coordinates of the baselines are shown in Fig.~\ref{fig:u-v-coverage}.\newline 
    \begin{table*}
      \small
      \centering
      \caption[]{\label{tab:observations}MATISSE observations of RY Tau analyzed in this paper.}
      \begin{tabular}{ccccccccccc}
        \toprule
        \multicolumn{5}{c}{Target} & \multicolumn{3}{c}{Calibrator} & Spectral & Band(s) & DRS \\
        & & & & & & & & mode & & version \\
        Date and time & Seeing & $\tau_0$ & Stations & Array & Name & LDD & Time & & & \\
        (UTC) & (\si{\arcsecond}) & (\si{\milli\second}) & & & & (\si{\milliarcsecond}) & (UTC) & & & \\ 
        \midrule
        2021-01-07T01:36 & \num{0.3} & \num{18.4} & U1-U2-U3-U4 & UTs & HD17361 & \num{1.8} & 00:52 & LOW-LM & $L$, $M$ & 1.5.2 \\
        2021-01-07T01:36 & \num{0.3} & \num{18.4} & U1-U2-U3-U4 & UTs & HD17361 & \num{2.0} & 00:52 & HIGH-N & $N$ & 1.7.5 \\
        2021-01-23T01:07 & \num{0.5} & \num{6.3} & K0-G2-D0-J3 & Medium & HD281813 & \num{2.5} & 00:20 & MED-LM & $L$ & 1.5.8 \\
        2021-01-23T01:20 & \num{0.7} & \num{3.9} & K0-G2-D0-J3 & Medium & HD281813 & \num{2.5} & 00:31 & LOW-LM & $L$ & 1.5.8 \\
        \bottomrule
      \end{tabular}
      \tablefoot{The atmospheric coherence time is denoted by $\tau_0$ and the limb darkened diameter (LDD) is the estimated angular diameter of the calibrator. The resolution of the spectral mode is given by \citet{Lopez+2022}.}
    \end{table*}
    The data reduction was performed as described in \citet{Millour+2016} with the standard MATISSE data reduction pipeline DRS version given in Table \ref{tab:observations}. We applied post-processing, including flux calibration and the averaging of exposures with the MATISSE Python tools\footnote{\url{https://github.com/Matisse-Consortium/tools/tree/master/mat_tools}}, and with our own Python scripts as well. More details on the data processing workflow can be found in Sect.~3.1 of \citet{Varga+2021}. To fit the 2D and 3D disk models (see Sects.~\ref{sec:temperature_gradient} and \ref{sec:radiative_transfer}) to the data, we reduced the spectral resolution of the observations by weighted averaging of the measured values and simple averaging of the errors of the observables $V^2$ and $\phi_\text{cp}$ over wavelength bins. Thanks to this simple averaging of the errors, we ensured that we could remain  conservative and avoid underestimating any possible systematic uncertainties. The wavelength bin width, $\Delta_\lambda$, was chosen depending on the band; $\Delta_\lambda(L,M) = \SI{0.05}{\micro\meter}$, $\Delta_\lambda(N) = \SI{0.1}{\micro\meter}$. The spectral averaging of the data yields a common wavelength grid for all baselines per spectral band on which the modeled observables were computed. This approach reduces the computational cost of the modeling significantly, compared to modeling the observables at the full spectral resolution.\newline
    As the quality of the measured data degrades toward the short wavelengths of the $L$ band, we used the data in the \SIrange{3.5}{3.9}{\micro\meter} range only. For the same reason, we considered measurements at \SIrange{4.5}{5.2}{\micro\meter} in the $M$ band and \SIrange{8.1}{13.0}{\micro\meter} in the $N$ band.\newline
    The measured squared visibilities, $V^2$, closure phases, $\phi_\text{cp}$, and correlated fluxes, $F_\text{corr}$, in the $N$ band are shown in Fig.~\ref{fig:data_overview}. The slope of the measured squared visibilities indicate that the apparent size of RY Tau increases with wavelength, which is typical for a protoplanetary disk in the NIR and MIR. The closure phases are nonzero for most baselines in the $L$ and $M$ band and large closure phases are observed in the $N$ band, which indicates an asymmetric intensity distribution, with a stronger asymmetry at longer wavelengths. The correlated fluxes in the $N$ band show a prominent silicate emission feature, which changes its shape with baseline length.

    \section{Methods}\label{sec:methods}
    The observations shown in Fig.~\ref{fig:data_overview} are used to constrain the spatial distribution and chemical composition of the dust in the circumstellar environment by fitting different models to the measured data. As a quality criterion of how well a given model reproduces the observations the reduced, $\chi^2$, expressed as
    \begin{equation}
      \chi^2_\text{red}(q) = \frac{1}{N_\text{data} - N_\text{free}} \sum_{i}^{N_\text{data}} \frac{\Big( q_i - m_{q;i} \Big)^2}{\sigma_{q_i}^2}
    ,\end{equation}
    is computed from the measured quantity, $q_i \pm \sigma_{q_i}$, and the modeled value, $m_{q;i}$, for the number of data points, $N_\text{data}$, and the number of free parameters, $N_\text{free}$, of the model \citep[e.g.,][]{Hogg+2010}.
    A statistically solid approach is to sample the parameter space of a given model via the Markov Chain Monte Carlo  (MCMC) algorithm \citep{GoodmanWeare2010}, which was employed to fit a 2D temperature gradient disk model (see Sect.~\ref{sec:temperature_gradient}) and a 1D single temperature model (see Sect.~\ref{sec:compositional_analysis}) to the observations. The fits are performed using the Python implementation \texttt{emcee} \citep{ForemanMackey+2013}. The resulting best-fit value is the median of the posterior probability density distribution of the fit parameter. The corresponding negative and positive errors on the best-fit value are taken as the \SI{16}{\percent} and \SI{84}{\percent} percentiles (corresponding to $\mp 1 \sigma$ error bars of a normal distribution) of the distribution, respectively.\newline
    In Sect.~\ref{sec:radiative_transfer}, we describe how we modeled the observations by simulating the appearance of a 3D dust density distribution in the circumstellar environment illuminated by the central protostar of RY Tau. This is achieved with Monte Carlo Radiative Transfer (MCRT) simulations performed with \texttt{POLARIS}\footnote{\url{https://github.com/polaris-MCRT/POLARIS}} \citep{Reissl+2016,Reissl+2018}. Flux density maps are simulated by computing the dust temperature distribution self-consistently \citep{BjorkmanWood2001} and successively deriving the thermal emission of the dust arriving at the detector via numerical integration along rays. Scattering of stellar light is taken into account employing the Monte Carlo technique \citep{Wolf+1999} optimized by utilizing peel-off photon packages \citep{YusefZadeh1984} and enforcing first scattering \citep{Mattila1970}. Self-scattering of photons emitted by the dust off the dust is neglected as experimental simulations showed that it would have little influence on synthetic MATISSE observations of RY Tau. The computational cost of a single MCRT simulation is high, which renders MCMC sampling unfeasible. We resorted to a brute force exploration of the parameter space on a coarse grid of free parameter values to find a combination of parameters that would fit the observables best with respect to chosen parameter value resolution and range. If necessary, the coarse grid was successively refined in interesting regions of the parameter space. The comparison of the synthetic observables produced by the 2D and 3D disk models to the interferometric measurements of visibilities, $V^2$, and closure phases, $\phi_\text{cp}$, requires us to compute them from the flux density maps, which is achieved using the Python package \texttt{galario} \citep{Tazzari+2018}.
    
    \section{Estimation of disk orientation}\label{sec:temperature_gradient}
    Constraining the inner disk orientation defined by the inclination, $\iota$, and position angle, $\PA$, is crucial for the further analysis of the circumstellar dust environment of RY Tau. The position angle, $\PA$, is defined as the angle of the disk major axis in the image plane measured from north to east. Atacama Large Millimeter/submillimeter Array (ALMA) observations provide a reliable value for the disk orientation on large scales, resulting in $\iota_\text{ALMA} = (\num{65.0} \pm 0.1)\si{\degree}$ and $\PA_\text{ALMA} = (\num{23.1} \pm \num{0.1})\si{\degree}$ \citep{Long+2019}; these values were adopted by \citet{Davies+2020} for the inner disk orientation. In this study, we chose to estimate the inner disk orientation independently, as \citet{Bohn+2022} and \citet{Villenave+2024} reported that a tilted inner disk might be a common phenomenon among YSOs.

    \subsection{2D star and disk model}\label{sec:2d_star_disk}
\begin{figure}
      \centering
      \includegraphics[width=\columnwidth]{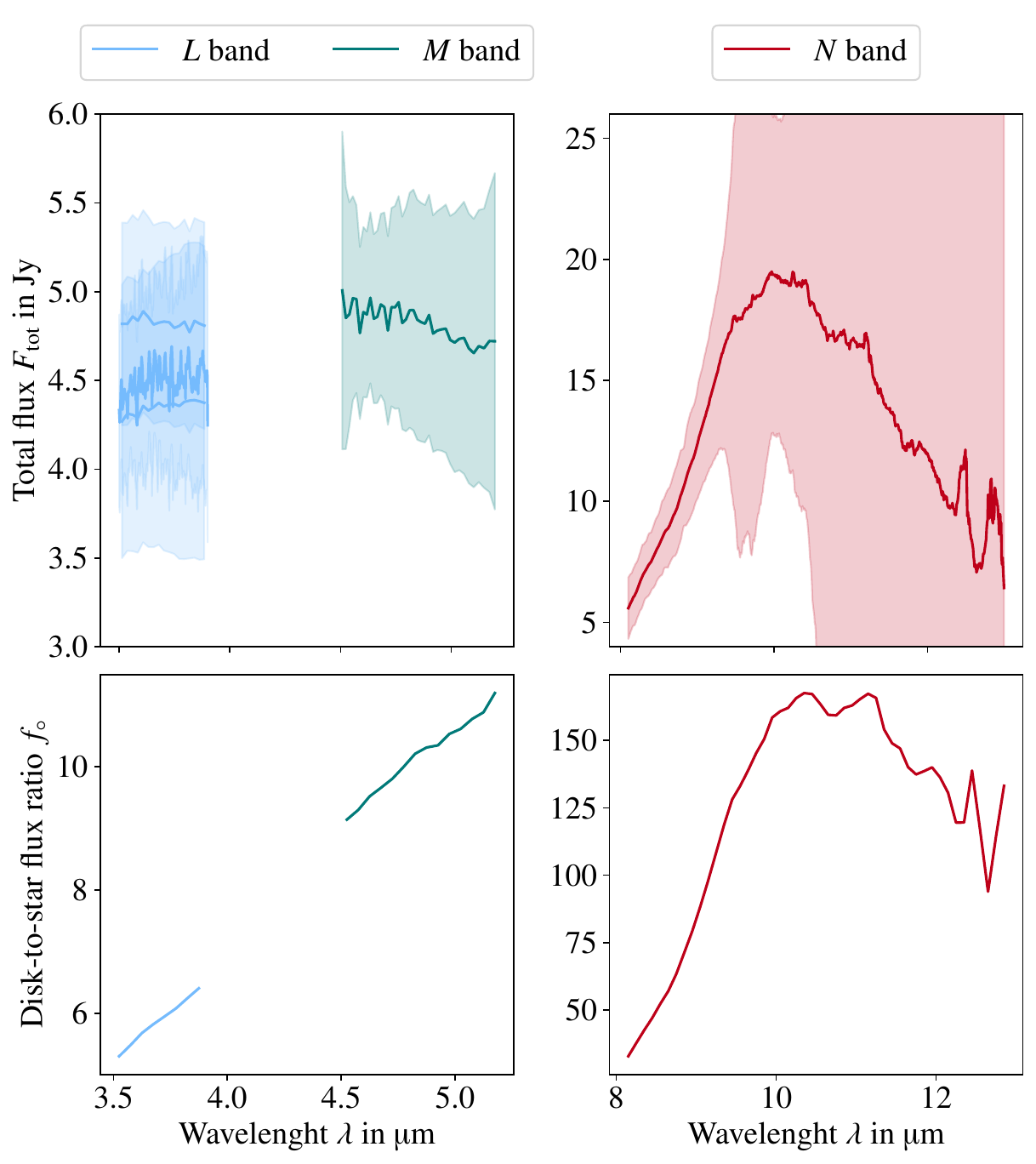}
      \caption{\emph{Top row:} Measured total flux, $F_\text{tot}$. \emph{Bottom row:} Derived disk-to-star flux ratio, $f_\circ$, assuming stellar properties given in Table \ref{tab:stellar_properties} and wavelength binning as described in Sect.~\ref{sec:observations}.}
      \label{fig:disk-to-star}
    \end{figure}

    To constrain the inner disk orientation, we adopted the semi-physical model proposed by \citet{Menu+2015}, assuming that the dominating radiation in the MIR originates from an optically thin surface layer of an accretion disk with constant optical depth, $\tau_\nu$. Neglecting scattered light and optical depth effects, the disk intensity, $I_{\text{disk};\nu}$, as a function of the radial distance, $r$, from the central object is given by
    \begin{equation}
      \label{eq:disk_intensity}
      I_{\text{disk};\nu} = \tau_\nu B_\nu\left( T(r) \right) \propto B_\nu\left( T(r) \right),
    \end{equation}
    with $B_\nu$ denoting the Planck function and assuming a radial temperature gradient following
    \begin{equation}
      \label{eq:temperature_gradient}
      T(r) = T_\text{in}\left(  \frac{r}{R_\text{in}} \right)^{-q}
    ,\end{equation}
    as used by \citet{Menu+2015} and \citet{Varga+2017,Varga+2018}. The inner temperature $T_\text{in}$ can be related to the inner radius $R_\text{in}$ of the disk via
    \begin{equation}
      \label{eq:inner_radius}
      R_\text{in} = \sqrt{\frac{L_\star}{4 \pi \sigma_\text{SB} T_\text{in}^4}}
    ,\end{equation}
    assuming an optically thick inner rim emitting as a blackbody \citep[e.g.,][]{DullemondMonnier2010} and neglecting the backwarming effect due to thermal emission of the dust \citep[e.g.,][]{BensbergWolf2022}. We note that this model is a strong simplification of the expected flared accretion disk with an optically thick disk midplane and an optically thin surface layer. However, it serves the purpose of estimating the inner disk orientation, which is then used to investigate the appearance of an illuminated 3D dust density distribution via MCRT simulations in Sect.~\ref{sec:radiative_transfer}.\newline
    The simple thin disk model needs to be completed by a stellar model in order to derive synthetic visibilities. In our study, we adopted the stellar properties used by \citet{Davies+2020} given in Table \ref{tab:stellar_properties}\footnote{The true distance to RY Tau is currently unknown and different values are used in recent literature. While \citet{Harrison+2024} used a value of $d=\SI{128}{\parsec}$, VizieR \citep{Ochsenbein+2000} lists a value of $d=\SI{161}{\parsec}$ derived from \citet{GaiaDR3} parallax measurements. As the distance impacts the luminosity estimation, we chose to follow the approach by \citet{Davies+2020} selecting a consistent set of stellar parameters. The adopted distance is very close to the value of $d=\SI{139}{\parsec}$ estimated by \citet{Bailer-Jones+2021}.}. The stellar radius, $R_\star$, was derived applying the Stefan-Boltzmann law. Assuming the central protostar as a blackbody emitter, the disk-to-star flux ratio, $f_\circ(\lambda),$ was computed from total flux measurements. As the visibilities are a measure of contrast and do not include information on the absolute flux density values, it is expedient to norm the flux density of the central object to one and scale the disk intensity (see Eq.~\ref{eq:disk_intensity}) via the disk-to-star flux ratio. Interstellar extinction is taken into account adopting the extinction law proposed by \citet{Cardelli+1989} using the Python package \texttt{extinction}\footnote{\url{https://github.com/sncosmo/extinction}} with the visual magnitude, $A_V$, given in Table \ref{tab:stellar_properties}. The resulting disk-to-star flux ratio is shown in Fig.~\ref{fig:disk-to-star}. We note that the protostar contributes $\gtrsim \SI{10}{\percent}$ to the total flux in $L$ and $M$ bands, while it contributes very slightly in the $N$ band.
    \begin{table}
      \centering
      \caption{\label{tab:stellar_properties}Stellar parameter used in this work.}
      \begin{tabular}{ccc}
        \toprule
        Quantity & Value & Reference(s) \\
        \midrule
        Luminosity $L_\star$ & \SI{11.6}{\Lsun} & 1 \\
        Effective temperature $T_{\star;\text{eff}}$ & \SI{5945}{\kelvin} & 2 \\
        Stellar mass $M_\star$ & \SI{2.0}{\Msun} & 2 \\
        Visual extinction $A_V$ & \num{1.6} & 3 \\
        Distance $d$ & \SI{140}{\parsec} & 4,5,6 \\
        \bottomrule
      \end{tabular}
      \tablebib{(1)~\citet{Davies+2020};
        (2) \citet{Calvet+2004}; (3) \citet{Petrov+2019}; 
        (4) \citet{Kenyon+1994}; (5) \citet{Galli+2018};
        (6) \citet{Bailer-Jones+2021}.
      }
    \end{table}

    \subsection{Results}\label{sec:tg_results}
\begin{figure*}
      \centering
      \includegraphics[width=\textwidth]{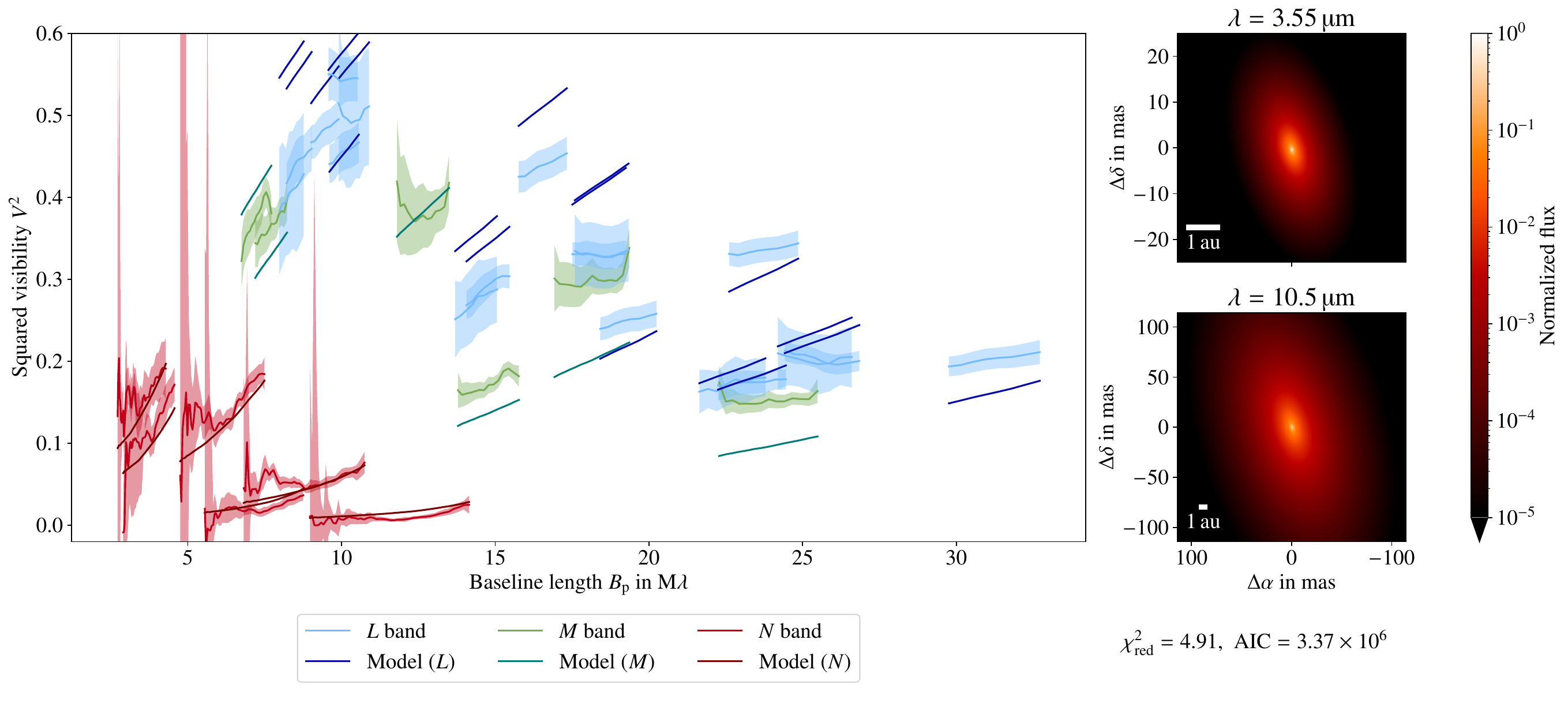}
      \caption{Best fit of the temperature gradient model with the dust sublimation temperature, $T_\text{in}$, as a free parameter.}
      \label{fig:tg_best-fit}
    \end{figure*}

    The disk orientation was constrained by fitting the free parameters of the model described in Sect.~\ref{sec:2d_star_disk}; namely, the inclination, $\iota$, position angle, $\PA$, and the disk temperature gradient exponent, $q$. The sublimation temperature of the dust was set to $T_\text{in} = \SI{1500}{\kelvin}$, which is a common choice for cosmic dust consisting mainly of silicates \citep{Vaidya+2009,Menu+2015,Hofmann+2022}. The best-fit parameters are listed in Table \ref{tab:tg_best-fit} and the modeled visibilities and normalized flux density maps at selected wavelengths are shown in Fig.~\ref{fig:tg_Tin1500K}. The resulting $\chi^2_\text{red} > 10$ indicates that the adopted model is not able to reproduce the observed visibilities well, which is also apparent in Fig.~\ref{fig:tg_Tin1500K}. To improve the fit, the inner radius of the disk $R_\text{in}$, which is related to the sublimation temperature of the dust $T_\text{in}$ via Eq.~\ref{eq:inner_radius}, was treated as free parameter. This choice was justified, as the dust sublimation temperature varies for different grain materials and silicate grains and is predicted that it may reach values of up to $\SI{2100}{\kelvin}$ \citep{Duschl+1996}. Consequently, the inner rim radius is fitted indirectly via the sublimation temperature.
    The best-fit parameters are given in Table \ref{tab:tg_best-fit} and the comparison of modeled and measured visibilities is shown in Fig.~\ref{fig:tg_best-fit}. Compared to the model with a fixed sublimation temperature, $T_\text{in} = \SI{1500}{\kelvin}$, a significantly reduced $\chi_\text{red}^2$ is achieved. As this could be the effect of overfitting, the Akaike information criterion, $\AIC$ \citep{Akaike1974}, was computed, and we found that its lower value for the model with sublimation temperature as a free parameter (see Table \ref{tab:tg_best-fit}) excludes this possibility.\newline
    \begin{table}
      \centering
      \caption{\label{tab:tg_best-fit}Maximum likelihood parameters obtained via a MCMC sampling of the temperature gradient model.}
      \begingroup
      \renewcommand{\arraystretch}{1.5}  
      \begin{tabular}{ccc}
        \toprule
        Parameter & Best fit $T_\text{in} = \SI{1500}{\kelvin}$  & Best fit ($T_\text{in}$ free)  \\
        \midrule
        $\iota$ & $\left(62.55_{-0.21}^{+0.21} \right)\si{\degree}$ &  $\left(59.55_{-0.34}^{+0.37} \right)\si{\degree}$ \\
        $\PA$ & $\left(-0.37_{-0.52}^{+0.52}\right) \si{\degree}$ & $\left(14.97_{-0.57}^{+0.68}\right) \si{\degree}$ \\
        $q$ & $0.5173_{-0.0008}^{+0.0008}$ & $0.5088_{-0.0010}^{+0.0009}$ \\
        $T_\text{sub}$ & -- &  $\left(2865_{-172}^{+444}\right) \si{\kelvin}$ \\
        \midrule
        $\chi^2_\text{red}$ & \num{11.29} & \num{4.91} \\
        $\AIC$ & \num{4.24e6} & \num{3.37e6} \\
        \bottomrule
      \end{tabular}
      \endgroup
    \end{table}
    Taking into account the simplicity of the model (2D geometry, radial symmetry, and four free parameters), the synthetic visibilities match the observed visibilities well. The fit between model and observations is especially good in the $N$ band, while in the $L$ and $M$ bands, the visibilities tend to be underestimated for longer baselines and overestimated for shorter baselines. This indicates that the emission in the shorter wavelength range of the MIR lacks some component which adds flux from the innermost part of the disk. Surprisingly, our best-fit model favors unphysically high dust sublimation temperatures of $T_\text{sub} > \SI{2100}{\kelvin}$. As the inner disk rim radius is related to the sublimation temperature via Eq.~\ref{eq:inner_radius}, this results in a corresponding inner rim radius of $R_\text{in} = \SI{0.06}{\astronomicalunit}$, corresponding to an unresolved gap between the protostar and disk (see Fig.~\ref{fig:tg_best-fit}). 
    
    \subsection{Discussion}\label{sec:tg_discussion}
    The 2D geometric modeling of a thin disk around a central star results in a best fit, yielding values for the inclination, $\iota$, and position angle, $\PA$, of the inner disk of RY Tau. Compared to the literature values of $\iota_\text{ALMA} = (\num{65.0} \pm \num{0.1}) \si{\degree}$ and $\PA_\text{ALMA} = (\num{23.1} \pm \si{0.1}) \si{\degree}$ of the orientation of the outer disk, the orientation of the inner disk is not (significantly) misaligned, although our fitting method results in a synthetic image of a less inclined ($\sim - \SI{5}{\degree}$) and rotated ($\sim - \SI{8}{\degree}$) inner disk. \cite{GRAVITY+2021} reported an inner disk inclination of $\iota_\text{GRAVITY} = (\num{60.0} \pm \num{1}) \si{\degree}$ derived from $K$-band interferometric observations, which is consistent with our analysis of MATISSE observations in the $L$, $M$, and $N$ bands. \cite{GRAVITY+2021} found an inner disk position angle of $\PA_\text{GRAVITY} = (\num{8} \pm \si{1}) \si{\degree}$, which is slightly less ($\sim - \SI{5}{\degree}$) than our estimate. In the following analysis, values of $\iota = \SI{60}{\degree}$ and $\PA = \SI{15}{\degree}$ are adopted for self-consistency of the modeling.\newline
    The unphysically high dust temperature is an artifact of the simplicity of the chosen model, with Eq. \ref{eq:inner_radius} neglecting optical depth effects and backwarming and dust opacities not taken into account. Most importantly, the 2D geometry does not allow for the effect of dust in the LOS to the observer to be modeled, such as diffuse hot circumstellar dust enshrouding the protostar as indicated in previous studies \citep[e.g.,][]{Schegerer+2008,Davies+2020} or outflows feeding the disk wind or jet of RY Tau \citep[e.g.,][]{Valegard+2022,Takami+2023}. Projected onto the image plane, the thermal emission by the enshrouding dust would fill the gap between protostar and disk; as does an unresolved inner rim (which is connected to high dust sublimation temperatures assuming a 2D geometry) employing the temperature gradient model. A 3D radiative transfer model allowing for dust in the LOS to the observer is presented in Sect.~\ref{sec:radiative_transfer}. Another explanation for an unresolved inner disk rim would be an optically thick gas disk emitting continuum flux, which is explored in Sect.~\ref{sec:accretion}.\newline
    Finally, the measured closure phases $\phi_\text{cp} \neq 0$ cannot be explained with a model producing a symmetric synthetic flux density map, as the closure phases of such an image are zero. This is the case for the 2D temperature gradient model (see Sect.~\ref{sec:2d_star_disk}). The effect of a flared disk geometry, resulting in an asymmetric brightness distribution, on the closure phases is investigated in Sect.~\ref{sec:radiative_transfer}.\newline

    \section{Spatially resolved dust mineralogy}\label{sec:compositional_analysis}
\begin{figure}
      \centering
      \includegraphics[width=\columnwidth]{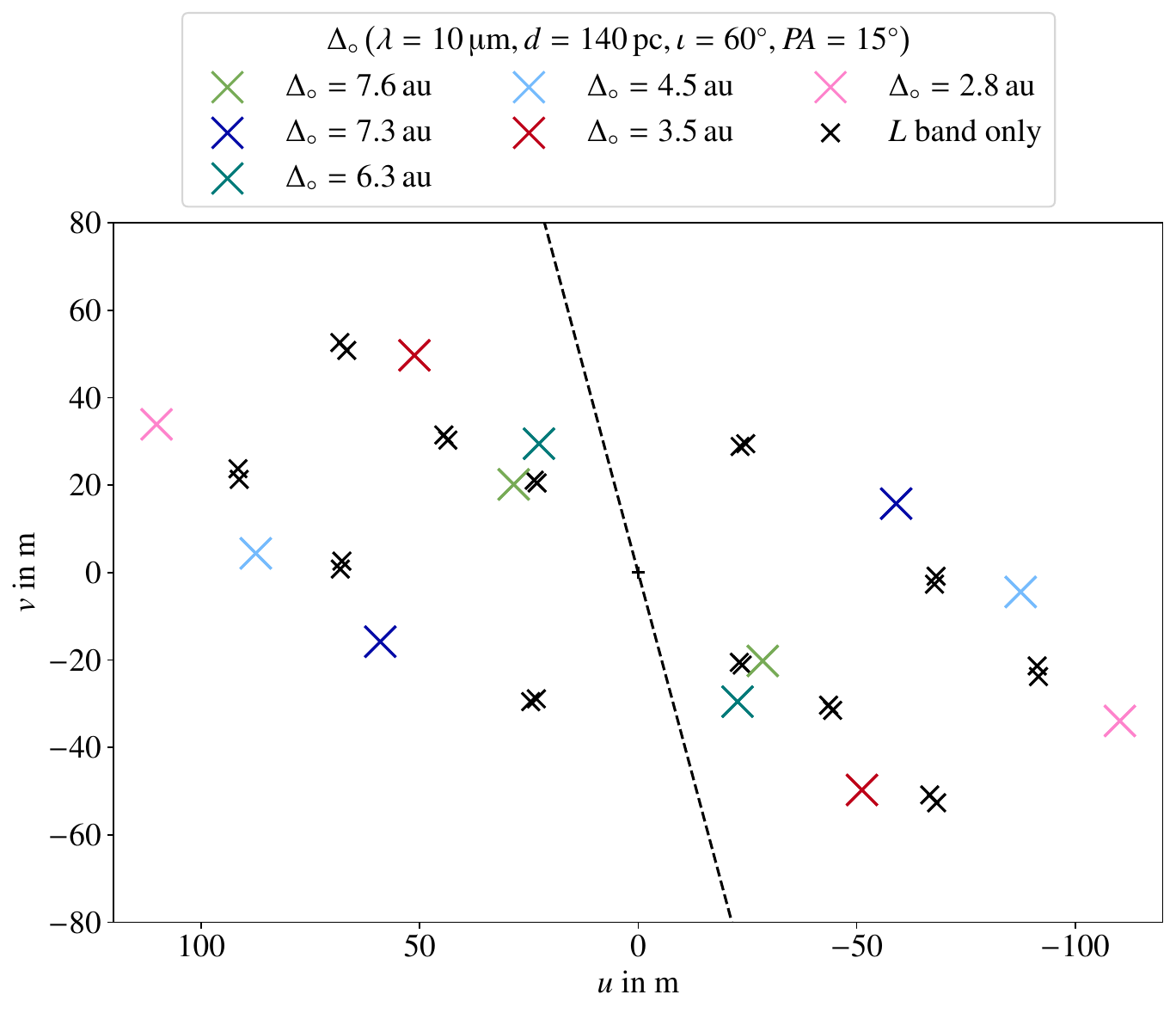}
      \caption{Baselines of the MATISSE observations described in Sect.~\ref{sec:observations}. The dashed line indicates the position angle of the disk major axis in the image plane.}
      \label{fig:u-v-coverage}
    \end{figure}

    The goal of this section is to analyze the chemical dust composition in the circumstellar environment of RY Tau. The $N$-band observations shown in Fig.~\ref{fig:data_overview} show a strong silicate emission feature in the total flux and correlated fluxes $F_\text{corr}$ at $\sim \SI{10}{\micro\meter}$, which is attributed to the Si--O stretching mode \citep[e.g.,][]{Henning2010}. The shape of the feature is a superposition of emission features of silicates and other dust materials of different chemical composition \citep[e.g.,][]{Bouwman+2001,vanBoekel+2004,vanBoekel+2005,Juhasz+2010}. Additional information contained in the correlated fluxes are the spatial scales in which the emission originates. Following \citet{Matter+2020}, we estimate the diameter of the spatial region that contributes to a correlated flux spectrum obtained with a projected baseline length $B_\text{p}$ and orientation $\phi_\text{B}$ as $ \sim \num{0.77} \nicefrac{\lambda}{B_\text{p}}$. Assuming the emission originates from a radially symmetric flat disk with an inclination, $\iota$, and orientation, $\PA$, we define the unresolved disk scale, $\Delta_\circ$, as
    \begin{equation}
      \label{eq:unresolved_disk_scale}
      \begin{aligned}
        B_\text{eff} &=  B_\text{p} \sqrt{ \cos^2 \left( \phi_\text{B} - \PA  \right) + \cos^2(\iota) \sin^2 \left( \phi_\text{B} - \PA  \right) } \, , \\
        \Delta_\circ &\approx \num{0.77} \frac{\lambda}{B_\text{eff}}, 
      \end{aligned}
    \end{equation}
    via an effective baseline $B_\text{eff}$ projected onto the disk plane following \cite{Varga+2024}. The unresolved disk scale can be interpreted as a diameter of a circular region on a flat disk from which the correlated flux of the corresponding baseline originates. The unresolved disk scale per baseline is used as a scale, which is probed by the corresponding correlated flux. The baselines of the observations described in Sect.~\ref{sec:observations} in the $u$-$v$ plane and their corresponding unresolved disk scales computed via Eq.~\ref{eq:unresolved_disk_scale} for the inclination $\iota$ and position angle, $\PA$, determined in Sect.~\ref{sec:temperature_gradient} are shown in Fig.~\ref{fig:u-v-coverage} for a wavelength of $\lambda = \SI{10}{\micro\meter}$.\newline
    The correlated fluxes are modeled per baseline $j$ with a similar approach to that in \citet{Schegerer+2006}, expressed as
    \begin{equation}\label{eq:comp_ana}
      F_{\nu;j} = B_\nu\left(T_j \right) \sum_{k=0}^N \mu_{j,k} \varkappa_{\text{abs};k}
    ,\end{equation}
    where $B_\nu\left(T_j \right)$ is the Planck function for a blackbody with the temperature, $T_j$, and $\mu_{j,k}$ are the contributions of each dust component, $k$, while $\varkappa_{\text{abs};j,k}$ is the wavelength dependent mass opacity. In this model the parameters $T_j$ and $\mu_{j,k}$ are fitting parameters and the precomputed unresolved disk scale $\Delta_\circ(B_{\text{p};j})$ is used to interpret them with respect to spacial region in the disk. The computation of mass opacities for different dust materials is described in Sect.~\ref{sec:dust_opacities}.
    
    \subsection{Dust opacity model}\label{sec:dust_opacities}
\begin{figure}
      \centering
      \includegraphics[width=\columnwidth]{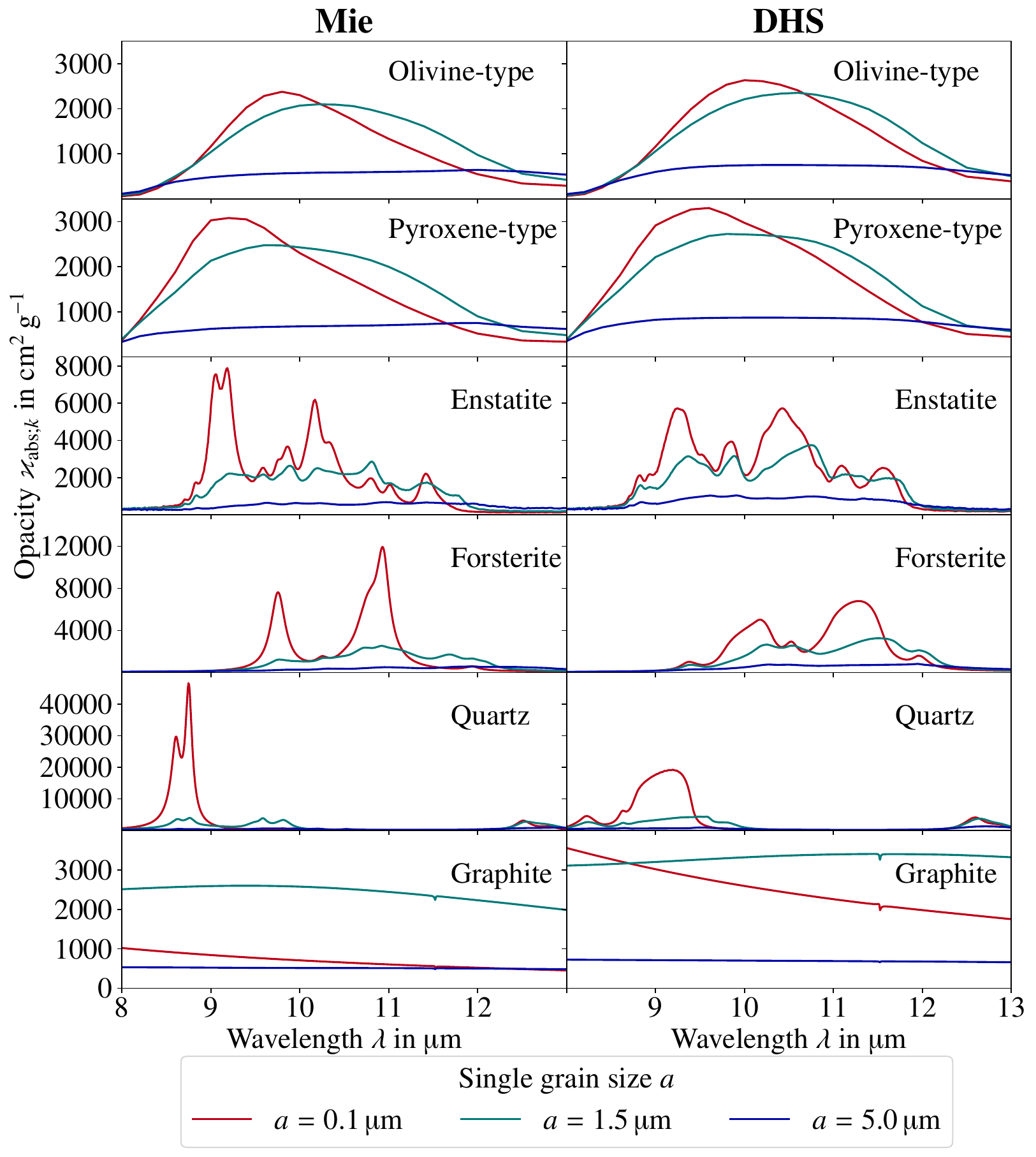}
      \caption{Wavelength-dependent mass absorption opacities derived from complex refractive index measurements for the dust species given in Table \ref{tab:optical_data}. \emph{Left column:} Opacities derived assuming spherical compact grains. \emph{Right column:} Opacities derived with the DHS approach assuming $f_\text{max} = \num{0.7}$ for the amorphous silicates and graphite and $f_\text{max} = \num{0.99}$ for the crystalline silicates.}
      \label{fig:opacities}
    \end{figure}

    The opacities of the dust species used in this work are derived from the complex refractive index measurements given in Table \ref{tab:optical_data}. The set of silicate materials was selected as by \citet{vanBoekel+2005} and \citet{Schegerer+2006} and was successfully used to fit correlated fluxes of RY Tau by \citet{Schegerer+2008}. The amorphous silicates (labeled ``amorp.'' in Table \ref{tab:optical_data}) of olivine and pyroxene stoichiometry are referred to as ``olivine-type'' and ``pyroxene-type,'' respectively.\newline
    \begin{table}
      \small
      \setlength{\tabcolsep}{4pt}
      \caption[]{\label{tab:optical_data}Optical properties of materials used in this work.}
      \begin{tabular}{lccSc}
        \toprule
        Material & Formula & Structure & {Density} & Ref. \\
        & & $s$ & {$\rho_\text{d}$ in \si{\gram\per\centi\meter\cubed}} & \\
        \midrule
        Olivine-type & \ch{Mg Fe Si O_4} & amorp. & 3.8 & \multirow{2}{*}{1, 2} \\
        Pyroxene-type & \ch{Mg Fe Si_2 O_6} & amorp. & 3.2 & \\
        \midrule
        Enstatite & \ch{Mg Si O_3} & cryst. & 3.2 & 3 \\
        Forsterite & \ch{Mg_{1.72} Fe_{0.21} Si O_4} & cryst. & 3.3 & 4 \\
        Quartz & \ch{Si O_2} & cryst. & 2.6 & 5 \\
        \midrule
        Graphite & C & conti. & 2.2 & 6 \\
        \bottomrule
      \end{tabular}
      \tablebib{(1)~\citet{Jaeger+1994};
        (2) \citet{Dorschner+1995}; (3) \citet{Jaeger+1998}; 
        (4) \citet{Zeidler+2015}; (5) \citet{Zeidler+2013};
        (6) \citet{Draine2003}.
      }
    \end{table}
    The complex refractive index of crystalline materials (labeled ``cryst.'' in Table \ref{tab:optical_data}) is measured with the symmetry axes of the crystal oriented parallel or perpendicular to the electric field vector of the wavefront separately. Depending on crystal symmetry, this results in two or three distinct sets of refractive indices per dust material, which can be combined to model different internal grain structures of dust grains consisting of the same material:\bulletnewline  
    \emph{Monocrystalline:} Each grain of the dust is assumed to be a monocrystal. The mass opacity of the material is computed from each set of complex refractive indices separately, and the resulting effective medium opacity of the dust is the averaged opacity as the grains are assumed to be arbitrarily oriented with respect to the electromagnetic field vector of the light wave \citep[e.g.,][]{Draine2016}.\bulletnewline
    \emph{Polycrystalline:} The dust grains are assumed to consist of a composite of multiple crystals of the same material, which are randomly oriented. Therefore, the complex refractive indices measured along different axes are combined into a single set of $(n, \kappa)$-values before the mass opacities are derived. To compute effective medium refractive indices, the Bruggeman rule \citep[e.g.,][]{BohrenHuffman1998} was used in this work. \bulletnewline
    Different physical processes can induce the crystallization of dust in protoplanetary disks, and the internal dust structure is not known. Furthermore the shape of the dust grains also impacts the resulting opacities \citep{Henning2010}. In this work, we chose to apply the monocrystalline approach to compute opacities for crystalline silicates, and used the polycrystalline approach for graphite. The latter is chosen because graphite contributes to the featureless continuum flux in our model and the optical properties of polycrystalline graphite better resemble a continuum opacity curve in the investigated wavelength range.\newline 
    In addition to the internal crystal structure, the dust grain shape is relevant for the resulting opacity. In this work, two approaches were employed and compared with respect to their ability to model the observed correlated fluxes of RY Tau:\bulletnewline
    \emph{Classical Mie theory (Mie):} Assuming dust as spherical compact grains \citep{Mie1908}.\bulletnewline
    \emph{Distribution of hollow spheres (DHS):} Modeling irregular shapes of grains as an effective medium integrated over a distribution of hollow spheres up to a vacuum volume fraction of $f_\text{max}$ as proposed by \citet{Min+2005}.\bulletnewline
    The opacities are computed using \texttt{optool} \citep{Dominik+2021} for different grain sizes $a \in \{ \SI{0.1}{\micro\meter}, \SI{1.5}{\micro\meter}, \SI{5}{\micro\meter} \}$. The resulting mass opacity curves for the selected set of materials, used for the fit of correlated fluxes, are shown in Fig.~\ref{fig:opacities}.
    
    \subsection{Results}
\begin{figure*}
      \centering
      \includegraphics[width=\textwidth]{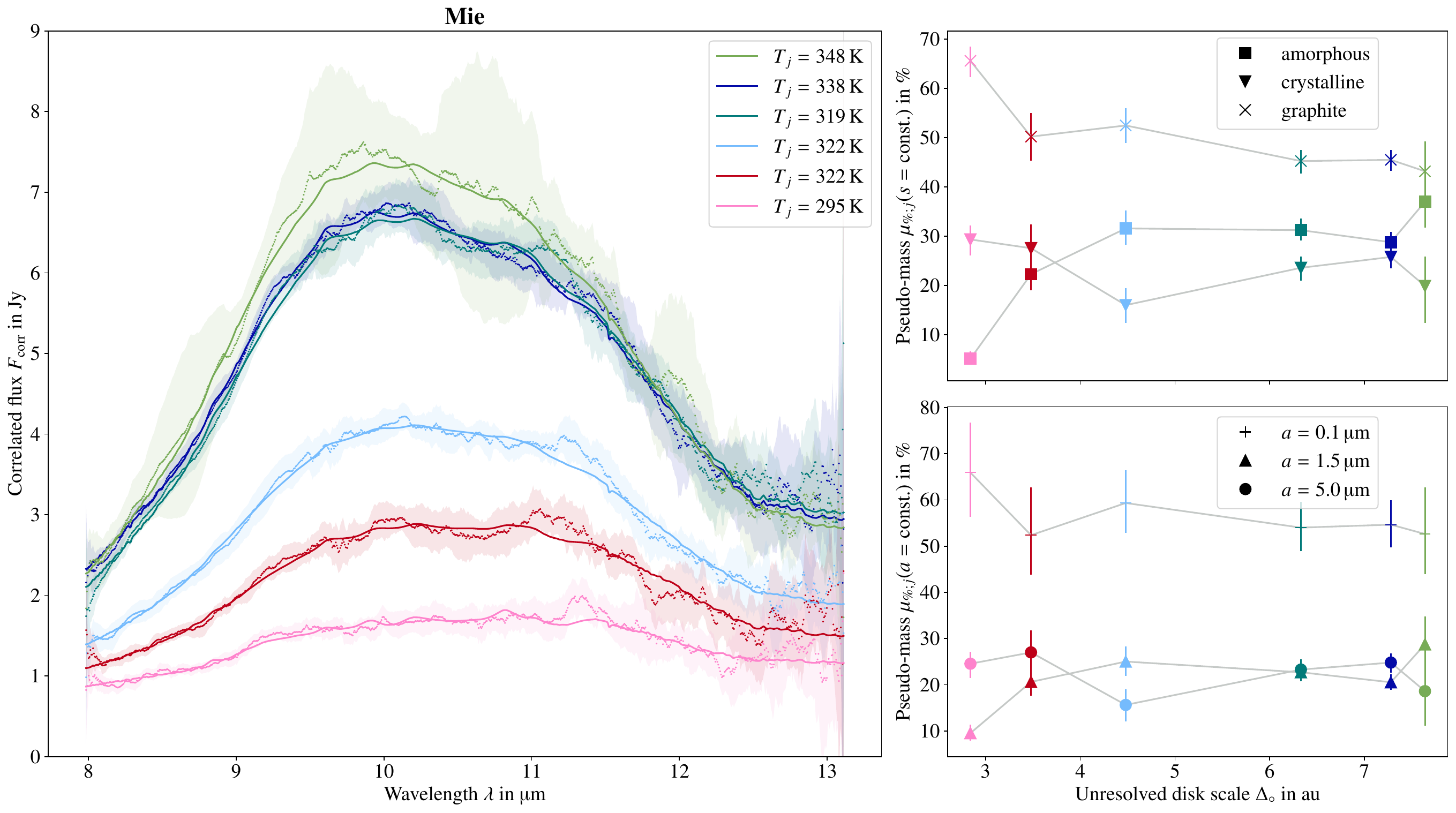}
      \caption{Best-fit model for the dust opacity model assuming spherical compact grains including only detected dust species as given in Table \ref{tab:comp_ana_mie}. \emph{Left:} Comparison of measured and fitted correlated fluxes for the $N$-band baselines. \emph{Upper right:} Trends of relative mass contribution of materials depending on their internal structure over unresolved disk scale. \emph{Lower right:} Trends of grain size over unresolved disk scale.}
      \label{fig:comp_ana_mie_only_detect}
    \end{figure*}

\begin{figure*}
      \centering
      \includegraphics[width=\textwidth]{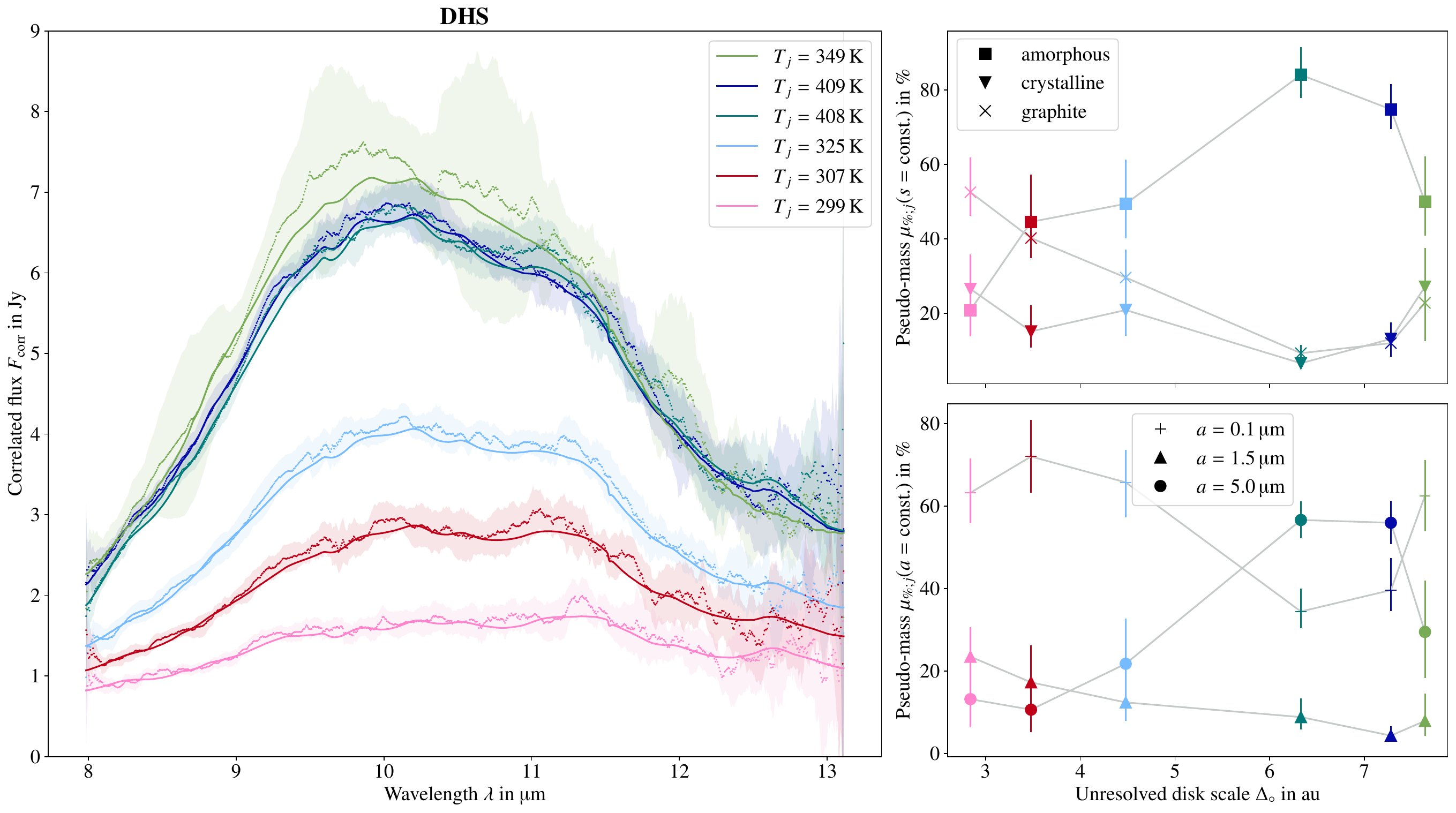}
      \caption{Best-fit model for the dust opacity model assuming a distribution of hollow spheres including only detected dust species as given in Table \ref{tab:comp_ana_dhs}. \emph{Left:} Comparison of measured and fitted correlated fluxes for the $N$-band baselines. \emph{Upper right:} Trends of relative mass contribution of materials depending on their internal structure over unresolved disk scale. \emph{Lower right:} Trends of grain size over unresolved disk scale.}
      \label{fig:comp_ana_dhs_only_detect}
    \end{figure*}

    The results of the fits on the correlated fluxes for the Mie and DHS opacity models are presented in Fig.~\ref{fig:comp_ana_mie} and Fig.~\ref{fig:comp_ana_dhs}, respectively. The fit parameters $\mu_{j,k}$ of the best-fit model are converted to pseudo-mass percentages according to
    \begin{equation}
      \mu_{\%;j,k} = \frac{\mu_{j,k}}{\sum_k \mu_{j,k}} \equiv \frac{\mu_{j,k}}{\mu_{\text{tot};j}} 
    ,\end{equation}
    which results in a (mass-)percentage of a certain dust species in the dust mixture. The resulting values are presented in Tables \ref{tab:comp_ana_mie} and \ref{tab:comp_ana_dhs}. The errors $\sigma_{\% \pm;j,k}$ are computed from the posterior distribution of probability densities of the parameters assuming a Gaussian distribution. For best-fit parameter values that are low, the estimated posterior distribution differs from a Gaussian profile, as the lower border of the allowed parameter space ($\mu_{j,k} \geq 0$) impacts the distribution. We therefore only consider a dust species detected when the following criteria are both satisfied:
    \begin{enumerate}[i)]
    \item $\mu_{\%;j,k} - \sigma_{\% -;j,k} \geq \SI{1}{\percent;}$
    \item $\mu_{\%;j,k} - \sigma_{\% +;j,k} \geq \SI{0}{\percent}$.
    \end{enumerate}
    Criterion i is motivated by a conservative approach to avoid detection of materials with low contributions as their impact on the resulting correlated flux is not relevant with respect to the measurement uncertainty. If criterion i is satisfied, criterion ii considers the posterior probability density distribution, where asymmetry is an indicator for a poorly constrained parameter close to noncontribution. In Tables \ref{tab:comp_ana_mie} and \ref{tab:comp_ana_dhs}, any nondetection of dust species or grain sizes is marked. The resulting materials in the dust mix differ significantly with respect to the employed grain shape model, described below:\ \bulletnewline
    \emph{Mie:} The silicates detected in the modeled dust mixture are small and medium-sized olivine-type, medium-sized pyroxene-type and medium-sized and large enstatite grains. The continuum contribution is limited to small grains. The $\chi^2_\text{red}$ values are $<0.5$ for all baselines, which indicates a good agreement between measured and modeled correlated fluxes. However, a systematic deviation from measured correlated flux shape is observed for unresolved disk scales of $\Delta_\circ \leq \SI{6.33}{\astronomicalunit}$ at $\lambda \sim \SI{11.2}{\micro\meter}$.\bulletnewline
    \emph{DHS:} All dust species are detected with grains of at least one size contributing to the best fit. Contrary to the Mie model best-fit dust composition, enstatite is less abundant, but small forsterite grains contribute to the correlated fluxes of all baselines. The shape of the modeled correlated fluxes at $\lambda \sim \SI{11.2}{\micro\meter}$ are closer to the measured shape, which can be attributed to the forsterite contribution. Surprisingly, the $\chi^2_\text{red}$ is higher than in the case of spherical compact grains.\bulletnewline 
    To observe trends of crystallinity and grain sizes, the quantities 
    \begin{equation}
      \begin{aligned}
        \mu_{\%;j}(s=\text{const.}) &\equiv \frac{\sum_{k \in s=\text{const.}} \mu_{j,k}}{\mu_{\text{tot};j}} \text{, and} \\
        \mu_{\%;j}(a=\text{const.}) &\equiv \frac{\sum_{k \in a=\text{const.}} \mu_{j,k}}{\mu_{\text{tot};j}} 
      \end{aligned}
    \end{equation}
    are defined and plotted over the unresolved disk scale $\Delta_\circ$ in Figs.~\ref{fig:comp_ana_mie} and \ref{fig:comp_ana_dhs}.\newline
    As even those dust species for which no significant abundance was found (i.e., which have not been detected) still contribute to the modeled correlated fluxes and the derived pseudo-masses of grain structure and size, additional MCMC fits for the Mie and DHS grain shape models were performed including only those dust species that were detected. The results are presented in Figs.~\ref{fig:comp_ana_mie_only_detect} and \ref{fig:comp_ana_dhs_only_detect} and Tables \ref{tab:comp_ana_mie_only_detect} and \ref{tab:comp_ana_dhs_only_detect}. Although the $\chi^2_\text{red}$ are similarly good for the Mie and DHS grain shape models, the latter approach results in a best fit which matches the \glqq shoulder\grqq \ at $\lambda \sim \SI{11.2}{\micro\meter}$, most prominently visible in the correlated flux for $\Delta_\circ = \SI{4.48}{\astronomicalunit}$. With respect to internal crystal structure and grain size the following general trends with \emph{increasing} unresolved disk scale can be observed:
    \begin{itemize}
    \item The amount of amorphous dust increases up to an unresolved disk scale of $\Delta_\circ \lesssim \SI{7}{\astronomicalunit}$, which is in accordance with the prediction that amorphous grains have undergone annealing, forming crystalline structures at sufficiently high temperatures \citep[e.g.,][]{Davoisne+2006}. 
    \item The contribution of graphite contributing to the featureless flux decreases (up to $\Delta_\circ \lesssim \SI{7}{\astronomicalunit}$ in the case of the DHS model).
    \item The decrease in small grain sizes is attributed to the decreasing contribution of graphite. For medium-sized and large grains, no unambiguous trends are observed.
    \end{itemize}
    These trends are observed for both Mie and DHS models, although the detailed dust species composition differs significantly.
    \begin{table*}[]
      \tiny
      \setlength{\tabcolsep}{10pt} 
      \renewcommand{\arraystretch}{1.5} 
      \centering
      \caption{Best fit of relative mass contributions of dust species.}
      \label{tab:comp_ana_mie_only_detect}
      \begin{tabular}{cccccccc}
\toprule
 & & \multicolumn{6}{c}{Unresolved disk scale $\Delta_\circ$} \\
 & Grain size in \si{\micro\meter} & \hexcolor{fe83cc}{\SI{2.84}{\astronomicalunit}} & \hexcolor{be0119}{\SI{3.48}{\astronomicalunit}} & \hexcolor{75bbfd}{\SI{4.48}{\astronomicalunit}} & \hexcolor{017a79}{\SI{6.33}{\astronomicalunit}} & \hexcolor{030aa7}{\SI{7.28}{\astronomicalunit}} & \hexcolor{77ab56}{\SI{7.64}{\astronomicalunit}} \\
\midrule
\multirow{2}{*}{Olivine} & 0.1 & {\color{lightgray}$0.3_{-0.2}^{+0.5}$} & {\color{lightgray}$2.2_{-1.3}^{+1.7}$} & $6.9_{-1.4}^{+1.6}$ & $8.8_{-1.2}^{+1.3}$ & $9.2_{-1.2}^{+1.3}$ & $9.5_{-2.1}^{+2.3}$ \\
 & 1.5 & {\color{lightgray}$0.5_{-0.3}^{+0.7}$} & $10.7_{-2.3}^{+2.4}$ & $10.0_{-1.9}^{+1.8}$ & {\color{lightgray}$0.5_{-0.4}^{+0.7}$} & {\color{lightgray}$0.5_{-0.4}^{+0.7}$} & {\color{lightgray}$3.0_{-2.1}^{+3.0}$} \\
\midrule
\multirow{1}{*}{Pyroxene} & 1.5 & $4.3_{-1.2}^{+1.2}$ & $9.4_{-1.9}^{+2.2}$ & $14.7_{-2.4}^{+2.7}$ & $22.0_{-1.7}^{+1.8}$ & $19.1_{-1.5}^{+1.5}$ & $24.5_{-4.3}^{+5.0}$ \\
\midrule
\multirow{2}{*}{Enstatite} & 1.5 & $4.8_{-1.1}^{+1.3}$ & {\color{lightgray}$0.5_{-0.4}^{+0.7}$} & {\color{lightgray}$0.3_{-0.2}^{+0.5}$} & {\color{lightgray}$0.3_{-0.2}^{+0.4}$} & {\color{lightgray}$1.0_{-0.5}^{+0.5}$} & {\color{lightgray}$1.2_{-0.9}^{+1.5}$} \\
 & 5.0 & $24.6_{-3.1}^{+2.5}$ & $27.0_{-4.8}^{+4.7}$ & $15.6_{-3.6}^{+3.4}$ & $23.3_{-2.6}^{+2.3}$ & $24.8_{-2.2}^{+2.0}$ & $18.6_{-7.4}^{+5.8}$ \\
\midrule
\multirow{1}{*}{Graphite} & 0.1 & $65.6_{-9.5}^{+10.8}$ & $50.2_{-8.5}^{+10.2}$ & $52.5_{-6.3}^{+6.9}$ & $45.2_{-5.0}^{+5.4}$ & $45.5_{-4.8}^{+5.2}$ & $43.2_{-8.4}^{+9.8}$ \\
\midrule
$\chi^2_\text{red}$ & & 0.210 & 0.192 & 0.187 & 0.238 & 0.166 & 0.082\\ \bottomrule\end{tabular}

      \tablefoot{Best-fit values of the compositional analysis fit on correlated fluxes in the $N$ band assuming spherical compact grains and only including detected dust species as given in Table \ref{tab:comp_ana_mie}. Any nondetection of a certain dust species is indicated by gray values.}
    \end{table*}

    \begin{table*}[]
      \tiny
      \setlength{\tabcolsep}{10pt} 
      \renewcommand{\arraystretch}{1.5} 
      \centering
      \caption{Best fit of relative mass contributions of dust species.}
      \label{tab:comp_ana_dhs_only_detect}
      \begin{tabular}{cccccccc}
\toprule
 & & \multicolumn{6}{c}{Unresolved disk scale $\Delta_\circ$} \\
 & Grain size in \si{\micro\meter} & \hexcolor{fe83cc}{\SI{2.84}{\astronomicalunit}} & \hexcolor{be0119}{\SI{3.48}{\astronomicalunit}} & \hexcolor{75bbfd}{\SI{4.48}{\astronomicalunit}} & \hexcolor{017a79}{\SI{6.33}{\astronomicalunit}} & \hexcolor{030aa7}{\SI{7.28}{\astronomicalunit}} & \hexcolor{77ab56}{\SI{7.64}{\astronomicalunit}} \\
\midrule
\multirow{2}{*}{Olivine} & 0.1 & {\color{lightgray}$2.2_{-1.6}^{+2.7}$} & $17.8_{-4.7}^{+4.4}$ & $20.3_{-3.1}^{+3.0}$ & $20.7_{-3.1}^{+3.4}$ & $20.3_{-2.7}^{+3.5}$ & $13.5_{-3.4}^{+3.5}$ \\
 & 1.5 & $7.7_{-4.8}^{+5.1}$ & $7.5_{-5.0}^{+6.3}$ & {\color{lightgray}$3.5_{-2.6}^{+4.3}$} & $3.5_{-2.3}^{+3.0}$ & {\color{lightgray}$1.1_{-0.8}^{+1.6}$} & {\color{lightgray}$2.8_{-2.1}^{+3.9}$} \\
\midrule
\multirow{3}{*}{Pyroxene} & 0.1 & {\color{lightgray}$1.9_{-1.3}^{+2.4}$} & $8.7_{-4.3}^{+4.5}$ & $12.3_{-5.0}^{+5.1}$ & {\color{lightgray}$2.0_{-1.4}^{+2.8}$} & $4.8_{-2.6}^{+4.3}$ & $23.8_{-6.4}^{+6.7}$ \\
 & 1.5 & {\color{lightgray}$3.8_{-2.7}^{+4.2}$} & {\color{lightgray}$5.8_{-4.1}^{+6.3}$} & $5.1_{-3.5}^{+4.8}$ & {\color{lightgray}$2.6_{-1.8}^{+3.4}$} & {\color{lightgray}$0.9_{-0.7}^{+1.4}$} & {\color{lightgray}$4.0_{-2.9}^{+5.2}$} \\
 & 5.0 & {\color{lightgray}$5.3_{-3.9}^{+7.1}$} & {\color{lightgray}$4.9_{-3.6}^{+6.5}$} & $8.3_{-5.8}^{+8.1}$ & $55.2_{-4.3}^{+4.1}$ & $47.8_{-3.7}^{+3.2}$ & {\color{lightgray}$6.0_{-4.3}^{+7.0}$} \\
\midrule
\multirow{2}{*}{Enstatite} & 0.1 & $2.7_{-0.9}^{+1.0}$ & {\color{lightgray}$0.8_{-0.5}^{+0.7}$} & {\color{lightgray}$0.3_{-0.2}^{+0.4}$} & {\color{lightgray}$0.3_{-0.2}^{+0.2}$} & {\color{lightgray}$0.8_{-0.2}^{+0.3}$} & {\color{lightgray}$0.5_{-0.4}^{+0.6}$} \\
 & 5.0 & {\color{lightgray}$7.9_{-5.6}^{+8.8}$} & {\color{lightgray}$5.8_{-4.2}^{+6.8}$} & $13.5_{-6.9}^{+7.4}$ & {\color{lightgray}$1.4_{-1.1}^{+2.0}$} & $8.2_{-3.7}^{+4.3}$ & $23.6_{-10.3}^{+10.3}$ \\
\midrule
\multirow{2}{*}{Forsterite} & 0.1 & $3.9_{-1.0}^{+1.1}$ & $4.6_{-0.8}^{+0.9}$ & $3.2_{-0.6}^{+0.6}$ & $2.2_{-0.4}^{+0.5}$ & $1.7_{-0.3}^{+0.5}$ & $1.9_{-0.6}^{+0.6}$ \\
 & 1.5 & $5.5_{-2.3}^{+2.3}$ & {\color{lightgray}$0.7_{-0.5}^{+1.0}$} & {\color{lightgray}$1.2_{-0.8}^{+1.0}$} & {\color{lightgray}$0.3_{-0.2}^{+0.4}$} & {\color{lightgray}$0.6_{-0.4}^{+0.8}$} & {\color{lightgray}$0.8_{-0.6}^{+1.0}$} \\
\midrule
\multirow{1}{*}{Quartz} & 1.5 & $6.5_{-1.1}^{+1.2}$ & $3.3_{-0.8}^{+0.9}$ & $2.6_{-0.7}^{+0.7}$ & $2.5_{-0.4}^{+0.5}$ & $1.8_{-0.4}^{+0.5}$ & {\color{lightgray}$0.4_{-0.3}^{+0.5}$} \\
\midrule
\multirow{1}{*}{Graphite} & 0.1 & $52.5_{-7.0}^{+7.4}$ & $40.2_{-6.0}^{+6.2}$ & $29.6_{-6.1}^{+5.2}$ & $9.3_{-2.2}^{+3.5}$ & $12.0_{-3.4}^{+5.3}$ & $22.8_{-4.6}^{+4.2}$ \\
\midrule
$\chi^2_\text{red}$ & & 0.267 & 0.250 & 0.272 & 0.273 & 0.233 & 0.160\\ \bottomrule\end{tabular}

      \tablefoot{Best-fit values of the compositional analysis fit on correlated fluxes in the $N$ band employing the DHS opacity model and only including detected dust species as given in Table \ref{tab:comp_ana_dhs}. Any nondetection of a certain dust species is indicated by gray values.}
    \end{table*}
    
    \subsection{Discussion}
    \citet{Schegerer+2008} reported an increase in crystallinity and a decrease in the contribution of amorphous dust species with increasing baseline length. In the present work, measurements with six baselines instead of only two have been analyzed, and the trend in crystallinity cannot be confirmed unambiguously. However, a decrease in amorphous silicates close to the central object is found, which suggests that radial mixing, transporting crystalline grains outwards, is taking place. Amorphous silicates close to the inner rim of the disk have undergone annealing, which explains the decrease in their contribution. We extended our dust model by including large grains ($a=\SI{5}{\micro\meter}$), which appear to be relevant in our fit and are not included in the analysis of the silicate composition done by \citet{Schegerer+2008}. We also included graphite as representative species for continuum emission and estimated the relative mass of nonsilicate material. A positive trend of continuum emission contribution with decreasing unresolved disk scale is observed.\newline
    Surprisingly, using the DHS approach to compute opacities did not result in a better fit than assuming spherical compact grains, although it is expected that the dust in protoplanetary disks is irregularly shaped in reality \citep[e.g.,][]{Min+2016}. It is, however, noted, that the DHS approach allows us to model correlated flux shapes with a visible ``shoulder'' at $\lambda = \SI{11.2}{\micro\meter}$, which is not possible assuming spherical compact grains. This ``shoulder'' is related to the tentative detection of forsterite. We conclude that, given the degeneracy of the model and the measurement uncertainties, which are dominated by the large errors of the total flux measurement shown in Fig.~\ref{fig:disk-to-star}, we cannot unambiguously determine which grain shape model is better suited to model the measured correlated fluxes.\newline
    It is also noted that Eq.~\ref{eq:comp_ana} is a strong simplification of the temperature structure in an accretion disk, as a single effective temperature $T_j$ per baseline is assumed. A 2D model to investigate the mineralogy of the dust taking into account the spatial structure and temperature gradient of a protoplanetary disk is presented by \citet{Varga+2024}.
    
    \section{Constraining the spatial dust distribution via MCRT simulations}\label{sec:radiative_transfer}
    In this section, we present constraints on the 3D dust distribution of the protoplanetary disk of RY Tau. As discussed in Sect.~\ref{sec:tg_discussion}, the semi-physical 2D modeling approach, assuming a flat disk temperature gradient model, fails to explain the observed closure phases and requires an unphysically high dust temperature to fit the visibilities well. To overcome these issues, we perform MCRT simulations with \texttt{POLARIS} (see Sect.~\ref{sec:methods}) adopting a realistic 3D density distribution, and compute synthetic observations from first principles. The dust distribution of the MCRT model is fitted not only to the measured visibilities, but also to the total flux and closure phases. 

    \subsection{YSO model}\label{sec:mcrt_model}
    The dust density distribution in the inner regions around the protostar of RY Tau is assumed to follow the density distribution of a viscous accretion disk \citep{ShakuraSunyaev1973,LyndenBellPringle1974}. The gas density profile is given by
    \begin{equation}
      \label{eq:disk_density}
      \begin{aligned}
        \varrho_\text{disk}(r>R_\text{in}, z) &= \varrho_{0,\text{disk}} \left(  \frac{R_\text{ref}}{r} \right)^\alpha \exp\left[ - \frac{1}{2} \left( \frac{z}{h(r)}\right)^2 \right], \\
        h(r) &= h_\text{ref} \left( \frac{r}{R_\text{ref}}\right)^\beta \, , \ R_\text{ref} = \SI{100}{\astronomicalunit}, 
      \end{aligned}
    \end{equation}
    in cylindrical coordinates $r, z$ with the following parameters to be constrained:
    \[
    \begin{array}{lp{0.8\linewidth}}
      R_\text{in}           & inner rim radius; \\
      \varrho_{0,\text{disk}} & reference density; \\
      \alpha               & midplane density exponent $\varrho_\text{disk}(r,0) \propto r^{-\alpha}$; \\
      h_\text{ref}          & reference scale height $h_\text{ref} \equiv h(R_\text{ref})$; \\
      \beta                & flaring exponent.
    \end{array}
    \]
    The disk density is discretized on a spherical grid with an extent of $R_\text{max} = \sqrt{2} R_\text{out} = \sqrt{2} \times \SI{35}{\astronomicalunit}$, with $R_\text{out}$ chosen sufficiently large to model the circumstellar environment in the field of view of the MATISSE observations \citep{Varga+2024}. The reference density $\varrho_{0,\text{disk}}$ is related to the total disk mass $M_\text{disk}$ within a cylindrical radius of $R_\text{out}$ via
    \begin{equation}
      \label{eq:mass_normalization}
      \varrho_{0,\text{disk}} = M_\text{disk} \frac{2-\alpha+\beta}{ 2 \sqrt{2} \pi^{\frac{3}{2}} h_\text{ref}  \left( R_\text{out}^2 \left( \frac{R_\text{ref}}{R_\text{out}} \right)^{\alpha - \beta} - R_\text{in}^2 \left(  \frac{R_\text{ref}}{R_\text{in}} \right)^{\alpha-\beta} \right) },
    \end{equation}
    as given by \cite{Burrows+1996}. This density distribution of the accretion disk (Eq.~\ref{eq:disk_density}) has been successfully used to model the appearance of protoplanetary disks \citep[e.g.,][]{Woitke+2016,Robitaille2017,Woitke+2019}.\newline
    As a goal of this study is to investigate the influence of diffuse dust elevated above the accretion disk on the appearance of RY Tau in the MIR, a simple-structured envelope is added \citep[e.g.,][]{Schegerer+2008}. We note that we investigate the dust density distribution of an accretion disk without an envelope in Sect.~\ref{sec:mcrt_disk_only}. The envelope component accounts for a remnant of the natal envelope or a disk wind enshrouding the protostar. A radial symmetric distribution following 
    \begin{equation}
      \label{eq:envelope}
      \varrho_\text{env}(r>R_\text{in}) = \varrho_{0,\text{env}} \left( \frac{r}{\SI{1}{\astronomicalunit}} \right)^\gamma , \gamma \leq 0
    \end{equation}
    is adopted which is combined with the disk density to a model density distribution of
    \begin{equation}
      \label{eq:model_density}
      \varrho_\text{model}(r>R_\text{in}, z) = \begin{cases}
        \varrho_\text{disk}(r,z) &\text{ if } \varrho_\text{disk}(r,z) \geq \varrho_\text{env}(r) \\
        \varrho_\text{env}(r) &\text{ else,}
      \end{cases}
    \end{equation}
    with $\varrho_{0,\text{env}}$ and $\gamma$ as additional fitting parameters. We note that we assume a spherical sublimation radius of $R_\text{in}$ with a model density of $\varrho_\text{model} = 0$ for $r < R_\text{in}$.
    
    \subsection{Dust model}\label{sec:mcrt_dust_model}
    The mineralogy of the dust in the circumstellar environment of RY Tau is analyzed in Sec.~\ref{sec:compositional_analysis}. It is, however, difficult to incorporate the specific results into our radiative transfer simulations for the following reasons:
    \begin{itemize}
    \item Incorporating a spatially varying mineralogy into the radiative transfer model would require determining regions in the dust density model in which a certain dust mixture is present. Although the baselines contain information on the spatial region, they also probe a spatial orientation (see Fig.~\ref{fig:u-v-coverage}) and assume a thin disk emitting at a single temperature. In our 3D density distribution, the temperature is computed self-consistently, resulting in a vertical temperature gradient at a given radial distance from the central object. Keeping in mind that the chemical composition of dust is assumed to be temperature-dependent \citep[e.g.,][]{Henning2010}, corresponding radiative transfer models would be required \citep[e.g.,][]{Jang+2024}.
    \item The complex refractive indices are used to compute the temperature distribution in a protoplanetary disk self-consistently. Therefore, the optical data is not only required in the IR, but also in optical and ultraviolet wavelengths in order to compute passive heating by the central protostar. These data are not available for all considered dust species.
    \end{itemize}
    Given the challenges above, we decided to only investigate the density distribution rather than chemical composition of the dust around RY Tau, and adopt a dust model similar to the interstellar medium (ISM) dust composition using astronomical silicate \citep{Draine2003} as silicate component, which is expected to model the optical properties of dust in protoplanetary disks well in the MIR \citep{Pollack+1994}. As in Sect.~\ref{sec:compositional_analysis}, we use graphite as additional dust component. The mass ratio of astronomical silicate to graphite in the ISM is assumed as \nicefrac{5}{8} to \nicefrac{3}{8} \citep{Draine2003}. The relative mass ratio for spherical compact graphite grains given in Table \ref{tab:comp_ana_mie} is consistently above a fraction of $\nicefrac{3}{8}$ with a ratio of $\SI{60}{\percent}$ at the shortest unresolved disk scale.\newline
    In Sect.~\ref{sec:compositional_analysis} single grain sizes are assumed to investigate radial trends in grain size distribution. In our radiative transfer modeling, however, we opt for a continuous grain size distribution according to the Mathis-Rumpl-Nordsieck \citep[MRN;][]{Mathis+1977} distribution with the number density $n_\text{dust}(a)$ of grain sizes $a$ following the power law $dn_\text{dust} (a) \propto a^{\num{-3.5}} \, da$ \citep{WeingartnerDraine2001}. This continuous distribution is commonly adopted for radiative transfer simulations \citep[e.g.,][]{Wolf+2003,Schegerer+2008,Sauter+2009,Brunngraeber+2016,Hofmann+2022}. In order to investigate the effect of grain growth, the maximum grain size $a_\text{max}$ is varied while the minimum grain size is fixed to $a_\text{min} = \SI{5}{\nano\meter}$ as suggested by \citet{Mathis+1977}.\newline
    The absorption and scattering properties of the selected dust mixture is computed with the \texttt{miex} algorithm \citep{WolfVoshchinnikov2004}, assuming spherical compact grains. As shown in Sect.~\ref{sec:compositional_analysis}, this simple approach is sufficient to model the observed correlated fluxes in the $N$ band, indicating that choosing a more sophisticated approach is not neccessary. Finally, assuming ideal dust-to-gas coupling, the dust density is set to $\varrho_\text{dust} = \num{0.01} \varrho_\text{model}$.
    
\subsection{Constraints on the density distribution of the disk}\label{sec:mcrt_disk_only}
In order to provide constraints on the density distribution of an accretion disk (Eq.~\ref{eq:disk_density}) given the dust model described in Sec.~\ref{sec:mcrt_dust_model}, a set of \num{1568} parameter combinations was sampled via MCRT simulations. The chosen ranges of the parameter values given in Table \ref{tab:parameter_space_disk} cover a broad range of parameters, ensuring that possible best-fit values are included in the covered range. Notably, the reference scale height $h_\text{ref}$ and midplane density exponent $\alpha$ values exceed common choices of parameter values used to model protoplanetary disks \citep[e.g.,][]{Sauter+2009}. We note that the disk mass $M_\text{disk}$ refers to the mass included in a radius of $R_\text{out} = \SI{35}{\astronomicalunit}$, and is therefore not comparable to disk masses derived from observations of the full accretion disk with a radial extent $>\SI{70}{\astronomicalunit}$ \citep[e.g.,][]{Long+2019,FrancisVanDerMarel2020,Valegard+2022}.\newline
\begin{table*}
  \centering
  \caption{Parameter combinations used to explore the parameter space of an accretion disk model.}\label{tab:parameter_space_disk}
  \begin{tabular}{cccc}\toprule
    Parameter & Value (range)  & Number of values & Notes \\
    \midrule
    Inclination $\iota$ & \SI{60}{\degree} & 1 & Sect.~\ref{sec:temperature_gradient} \\
    Position angle $PA$ & \SI{15}{\degree} & 1 & Sect.~\ref{sec:temperature_gradient} \\
    Lower grain size limit $a_\text{min}$ & \SI{5}{\nano\meter} & 1 & \citet{WeingartnerDraine2001} \\
    Upper grain size limit $a_\text{max}$ & \SIrange{0.25}{1.5}{\micro\meter} & 2 & \citet{WeingartnerDraine2001}, Sect.~\ref{sec:compositional_analysis} \\
    Inner radius $R_\text{in}$ & \SI{.3}{\astronomicalunit} & 1 & $T_\text{sub} \approx \SI{1500}{\kelvin}$ \\
    Disk mass $M_\text{disk}$ & \SIrange{1e-7}{1e-4}{\Msun} & 4 & logarithmic spacing \\
    Reference scale heigh $h_\text{ref}$ & \SIrange{5}{35}{\astronomicalunit} & 7 &  \\
    Flaring exponent $\beta$ & \numrange{1.0}{1.3} & 4 &  \\
    Midplane density exponent $\alpha$ & \numrange{0.6}{2.4} & 7 &  \\
    \bottomrule
  \end{tabular}
  \tablefoot{The values of the free parameters are chosen equidistant.}
\end{table*}
To assess the influence of single parameters and their mutual correlations, $\chi^2_\text{red}$ maps are presented for visibilities (Fig.~\ref{fig:x2map_vis_diskOnly}), the total flux (Fig.~\ref{fig:x2map_sed_diskOnly}), and closure phases (Fig.\ref{fig:x2map_cp_diskOnly}) for a maximum grain size of $a_\text{max} = \SI{250}{\nano\meter}$. The lowest $\chi^2_\text{red}$ values for a maximum grain size of $a_\text{max}=\SI{1.5}{\micro\meter}$ are higher for squared visibilities $V^2$ and total flux $F_\text{tot}$. The maximum grain size has little impact on the general structure of the $\chi^2_\text{red}(V^2)$ distribution in the parameter space. The $\chi^2_\text{red}(F_\text{tot})$ maps, however, show that the smaller maximum grain size leads to a larger region in the parameter space which yield an acceptable $\chi^2_\text{red}$. This is consistent with the depletion of large dust grains in the vicinity of the central protostar of RY Tau reported by \citet{FrancisVanDerMarel2020}. In the following we therefore only consider a maximum grain size of $a_\text{max} = \SI{250}{\nano\meter}$.\newline
The degeneracy of the parameter space is well illustrated in Figs.~\ref{fig:x2map_vis_diskOnly}, \ref{fig:x2map_sed_diskOnly}, and \ref{fig:x2map_cp_diskOnly} as there are multiple distinct low $\chi^2_\text{red}$ regions visible for all modeled quantities. Although a good fit ($\chi^2_\text{red}(V^2)=3.41$ and $\chi^2_\text{red}(F_\text{tot}) = 0.19$) is found for the individual observables $V^2$ and $F_\text{tot}$, the regions of low(est) $\chi^2_\text{red}$ values in the parameter space are disjoint. Additionally, the closure phases are not reproduced well, and the low $\chi^2_\text{red}(\phi_\text{cp})$ value regions in the parameter space are disjoint to the other observables.\newline
The disconnected regions of low $\chi^2_\text{red}$ values in the parameter space for different observables indicate that the model is not sufficiently complex to yield synthetic observations matching the measurements. However, the results of the presented parameter study are highly valuable, as they illustrate the relevant density constraints to model the observations adequately.\bulletnewline
\emph{Visibilities:} The best fit is achieved for the lowest disk mass of $M_\text{disk} = \SI{1e-7}{\Msun}$, resulting in an optically thin disk. In this regime the modeled closure phases are (almost) zero, as there is little to no asymmetry induced by optical depth effects. MIR observations of highly inclined accretion disks show a shadow cast by the high density region in the disk midplane, which commonly results in an asymmetric flux density distribution. The low $\chi^2_\text{red}(V^2)$ value regions for higher disk masses are found for a combination of rather high values of the reference scale height $h_\text{ref}$ combined with a low midplane density exponent $\alpha$, leading to an extended distribution of material, which also results in an optically thin density distribution. The best-fit models for different disk masses share the underestimation of total flux in all bands for $M_\text{disk} \leq \SI{1e-6}{\Msun}$ and in the $L$ and $M$ bands for $M_\text{disk} \geq \SI{1e-5}{\Msun}$, while the $N$-band total flux fits the observation.\bulletnewline
\emph{Total flux:} The low $\chi^2_\text{red}(F_\text{tot})$ value regions are located at large reference scale heights of $h_\text{ref} \geq \SI{25}{\astronomicalunit}$, and moderate to high values of the disk midplane density exponent $\alpha$ resulting in a density distribution, which can be heated efficiently. With increasing reference scale height and a fixed disk mass, more dust is distributed in regions with low to medium optical depth above the midplane of the disk, resulting in an increased absorption and thermal reemission efficiency. A higher value of the midplane density exponent increases the amount of dust close to the inner disk rim, where it is heated to higher temperatures compared to regions further away from the central protostar. However, such a density distribution is not in agreement with the observed visibilities.\bulletnewline
\emph{Closure phases:} More than half of the sampled parameter combinations yield closure phases close to zero. With higher disk masses $M_\text{disk}$ and midplane density exponents $\alpha$, the midplane gets optically thick and the cast shadow induces an asymmetry, which can explain the observed closure phases which significantly differ from zero.\bulletnewline
As indicated by earlier studies \citep[e.g.,][]{Schegerer+2008,Takami+2013,Davies+2020}, the protostar of RY Tau is enshrouded in dust material, which increases the total flux in the IR if the dust is optically thin. The findings in this section strongly suggest that an optically thin component is necessary to model the observed visibilities. We therefore combine an optically thick accretion disk, which can possibly explain the observed closure phases with an optically thin envelope, as given in Eq.~\ref{eq:model_density}.

    \subsection{The influence of a dusty envelope}\label{sec:envelope}
\begin{figure*}
      \centering
      \includegraphics[width=\textwidth]{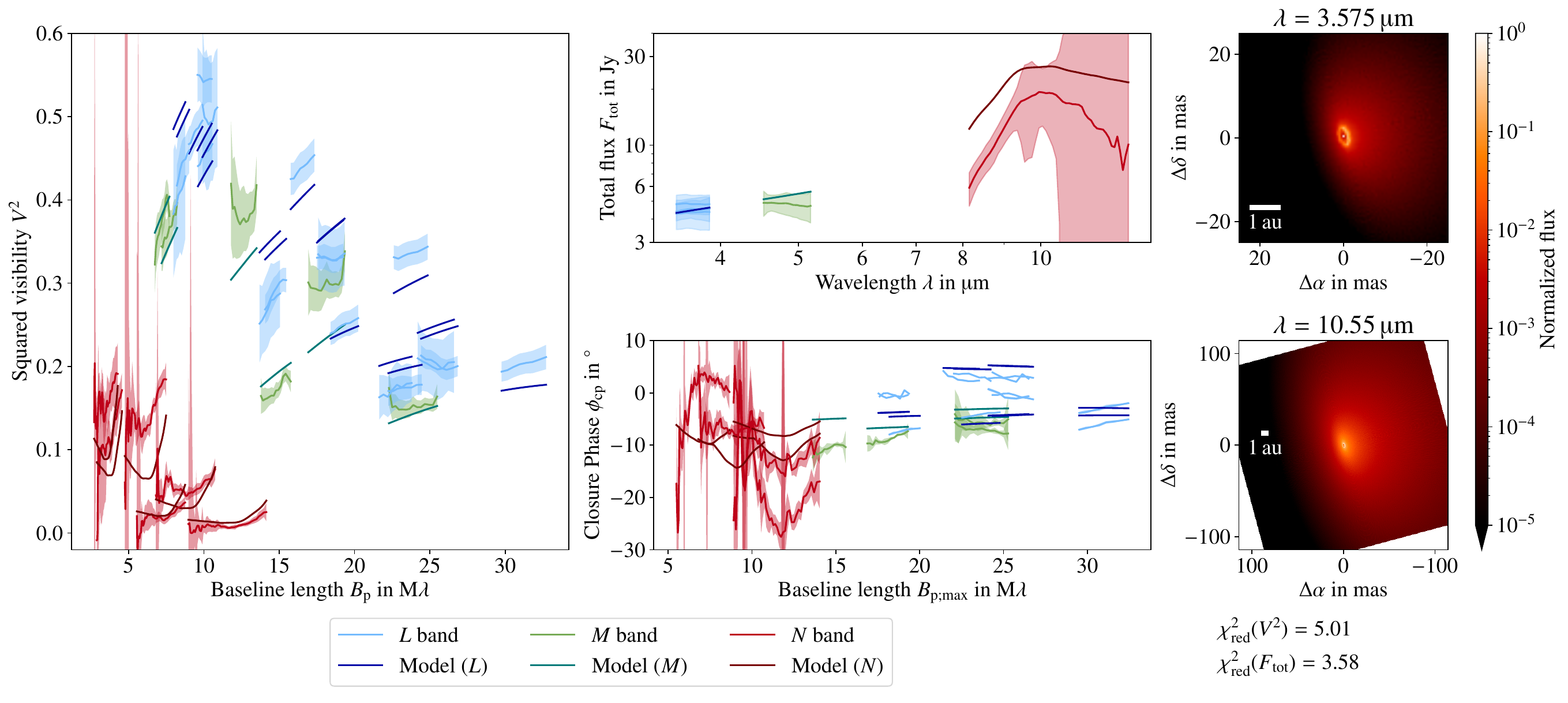}
      \caption{Best-fit model of an accretion disk with a dense midplane and an additional optically thin envelope. The model parameters are listed in Table \ref{tab:mcrt_best_fit_models}.}
      \label{fig:mcrt_best_fit_envII}
    \end{figure*}

    Adding an envelope to the model (see Eq.~\ref{eq:model_density}) results in two additional fitting parameters, namely the reference density $\varrho_{0;\text{env}}$ and density exponent $\gamma$ of the envelope. A set of simulations to sample the parameter space with the same resolution as in Sec.~\ref{sec:mcrt_disk_only} multiplied by the number of different combinations of $\varrho_{0;\text{env}}$ and $\gamma$ would result in a massive computational effort. We therefore opt for a lower resolution parameter sampling to identify regions in the parameter space which are promising with respect to modeling the visibilities, total flux and closure phases. As nonzero closure phases only result from optically thick disks, only high disk masses were taken into account for the combination with an optically thin envelope. The chosen values of $\gamma$ are motivated by the density power law for an infalling envelope of $\gamma = -0.5$ \citep{Ulrich1976}. The coarse resolution parameter space sampling value ranges are given in Table \ref{tab:parameter_space_env} and result in a total of \num{1536} simulations.\newline
    \begin{table*}
      \centering
      \caption{Parameter combinations on a coarse grid used to explore the parameter space of an accretion disk combined with a dusty envelope.}\label{tab:parameter_space_env}
      \begin{tabular}{cccc}\toprule
        Parameter & Value (range)  & Number of values & Notes \\
        \midrule
        Inclination $\iota$ & \SI{60}{\degree} & 1 & Sect.~\ref{sec:temperature_gradient} \\
        Position angle $PA$ & \SI{15}{\degree} & 1 & Sect.~\ref{sec:temperature_gradient} \\
        Lower grain size limit $a_\text{min}$ & \SI{5}{\nano\meter} & 1 & \citet{WeingartnerDraine2001} \\
        Upper grain size limit $a_\text{max}$ & \SI{0.25}{\micro\meter} & 1 & \citet{WeingartnerDraine2001}, Sect.~\ref{sec:mcrt_disk_only} \\
        Disk mass $M_\text{disk}$ & \SIrange{1e-5}{1e-4}{\Msun} & 2 & $\phi_\text{cp} \neq 0$ \\
        Reference scale height $h_\text{ref}$ & \SIrange{10}{30}{\astronomicalunit} & 3 &  \\
        Flaring exponent $\beta$ & \numrange{1.0}{1.3} & 4 &  \\
        Midplane density exponent $\alpha$ & \numrange{1.5}{2.4} & 4 & $\phi_\text{cp} \neq 0$ \\
        \midrule
        Envelope reference density $\varrho_{0,\text{env}}$ & \SIrange{1e-15}{1e-12}{\kilogram\per\meter\cubed} & 4 & logarithmic spacing \\
        Envelope density exponent $\gamma$ & \numrange{-0.9}{-0.3} & 4 & \citet{Ulrich1976} \\
        \bottomrule
      \end{tabular}
      \tablefoot{The values of the free parameters are chosen equidistant.}
    \end{table*}
    The resulting low(est) $\chi^2_\text{red}$ values do not coincide for the observables $V^2$, $F_\text{tot}$, and $\phi_\text{cp}$ as for the accretion disk model described in Sect.~\ref{sec:mcrt_disk_only}. To assess which set of parameters models the observations best, we define the best-fit model as the model which minimizes the sum of $\chi^2_\text{red}(V^2)$ and $\chi^2_\text{red}(F_\text{tot})$. As the $\chi^2_\text{red}(\phi_\text{cp})$ values are much higher than the $\chi^2_\text{red}$ values of the other two observables, it is unfeasible to take them into account as they would dominate the criterion. The results for the best-fit model with an accretion disk and envelope are presented in Fig.~\ref{fig:mcrt_best_fit_envI}. The parameters of the best-fit model are given in Table \ref{tab:mcrt_best_fit_models}.\newline
    \begin{table*}
      \centering
      \caption{Best-fit values for the different parameter spaces sampled via MCRT simulations.}\label{tab:mcrt_best_fit_models}
      \begingroup
      \renewcommand{\arraystretch}{1.2}  
      \begin{tabular}{cccc}\toprule
        Parameter & Disk only & Disk and envelope  & Dense disk and envelope \\
        \midrule
        Disk mass $M_\text{disk}$ & \SI{1e-5}{\Msun} & \SI{1e-5}{\Msun}  & \SI{1e-4}{\Msun}  \\
        Reference scale height $h_\text{ref}$ & \SI{35}{\astronomicalunit} & \SI{30}{\astronomicalunit} & \SI{7.5}{\astronomicalunit}  \\
        Flaring exponent $\beta$ & \num{1.3} & \num{1.3} & \num{1.1} \\
        Midplane density exponent $\alpha$ & \num{1.2} & \num{1.5} & \num{1.8} \\
        \midrule
        Envelope reference density $\varrho_{0,\text{env}}$ & -- & \SI{1e-14}{\kilogram\per\meter\cubed} & \SI{4.642e-14}{\kilogram\per\meter\cubed} \\
        Envelope density exponent $\gamma$ & -- & \num{-0.7} & $(\num{-0.5})$ \\
        \midrule
        Synthetic observables & \ref{fig:mcrt_best_fit_disk} & \ref{fig:mcrt_best_fit_envI} & \ref{fig:mcrt_best_fit_envII} \\
        Density distribution & \ref{fig:dens_disk_only} & \ref{fig:dens_disk_env} & \ref{fig:dens_dens_disk_env} \\
        Temperature distribution & \ref{fig:temp_disk_only} & \ref{fig:temp_disk_env} & \ref{fig:temp_dens_disk_env} \\
        \midrule
        $\chi^2_\text{red}(V^2)$ & \num{5.49} & \num{4.95} & \num{5.00} \\
        $\chi^2_\text{red}(F_\text{tot})$ & \num{1.86} & \num{1.57} & \num{3.58} \\
        $\chi^2_\text{red}(\phi_\text{cp})$ & \num{683} & \num{321} & \num{2976} \\
        \bottomrule
      \end{tabular}
      \endgroup
      \tablefoot{The best-fit values yield $\min \left( \chi^2_\text{red}(V^2) + \chi^2_\text{red}(F_\text{tot}) \right)$. Values in parentheses are fixed in the corresponding parameter space sampling.}
    \end{table*}
    As the closure phases in Fig.~\ref{fig:mcrt_best_fit_envI} indicate that the best-fit model lacks some asymmetry, an additional set of simulations was run with a refined resolution of sampled parameter values in the region of the parameter space, where shading of the outer disk regions from starlight by a dense inner disk midplane results in an asymmetric synthetic flux density map. This is achieved with disk masses $M_\text{disk} \geq \SI{1e-5}{\Msun}$, reference scale heights $h_\text{ref} < \SI{25}{\astronomicalunit}$ and a large midplane density exponent $\alpha \geq \num{1.8}$. In order to reduce the number of free parameters, which allows us to choose a higher resolution for the remaining free parameters, the envelope density exponent was fixed to the theoretical value $\gamma = \num{-0.5}$, as the coarse parameter space sampling showed that its influence on the resulting synthetic observables is small compared to the influence of the reference density of the envelope $\varrho_{0,\text{env}}$. Additionally, the range of parameter values of the latter is reduced to the relevant region in the parameter space. The resolution and range of free parameter values used to explore the parameter space in combination with a dense midplane of the disk is given in Table \ref{tab:parameter_space_thick_disk}.
    \begin{table}
      \centering
      \caption{Parameter combinations on a refined grid used to explore the parameter space of a dense midplane accretion disk combined with a dusty envelope.}\label{tab:parameter_space_thick_disk}
      \begin{tabular}{ccc}\toprule
        Parameter & Value range  & \# of values \\
        \midrule
        $M_\text{disk}$ & \SIrange{1e-5}{1e-4}{\Msun} & 2 \\
        $h_\text{ref}$ & \SIrange{5}{25}{\astronomicalunit} & 9 \\
        $\beta$ & \numrange{1.0}{1.3} & 3  \\
        $\alpha$ & \numrange{1.8}{2.4} & 3 \\
        \midrule
        $\varrho_{0,\text{env}}$ & \SIrange{1e-15}{4.462e-13}{\kilogram\per\meter\cubed} & 9 \\
        \bottomrule
      \end{tabular}
      \tablefoot{The values of fixed parameters are $\gamma = \num{0.5}$ and the ones given in Table \ref{tab:parameter_space_env}. The values are spaced equidistant with $\varrho_{0,\text{env}}$ on a logarithmic scale.}
    \end{table}
    The resulting best fit with respect to visibilities and total flux is shown in Fig.~\ref{fig:mcrt_best_fit_envII}. The corresponding $\chi^2_\text{red}$ maps for a disk mass of $M_\text{disk} = \SI{1e-4}{\Msun}$ on visibilities, total flux and closure phases are shown in Figs.~\ref{fig:x2map_vis_thickDisk}, \ref{fig:x2map_sed_thickDisk}, and \ref{fig:x2map_cp_thickDisk}, respectively. The disk mass $M_\text{disk}$ has little impact on the position of the low $\chi^2_\text{red}(V^2)$ region in the parameter space, while shifting the regions of low $\chi^2_\text{red}\left( F_\text{tot} \right)$ toward higher reference scale heights of the disk, which leads to disjoint regions of low $\chi^2_\text{red}$ on visibilities and total flux. The structure of the four-dimensional parameter space for $M_\text{disk} = \SI{1e-4}{\Msun}$ shows that the quality of the fit only weakly depends on the midplane density exponent $\alpha$ within the chosen parameter range. Consequently, the parameters $M_\text{disk}$ and $\alpha$ are not well constrained if the midplane of the innermost disk is optically thick (i.e.,~$M_\text{disk} > \SI{1e-5}{\Msun}$, $\alpha \geq \num{1.8}$ and $h_\text{ref} < \SI{30}{\astronomicalunit}$). Additionally, a strong correlation between reference scale height $h_\text{ref}$ and reference density of the envelope $\varrho_{0,\text{env}}$ with respect to the resulting $\chi^2_\text{red}$ on the visibilities is revealed in the $\chi^2_\text{red}$ maps. This further illustrates the degeneracy of the parameter space. The regions with low $\chi^2_\text{red}$ values for visibilities and total flux, although generally disjoint, slightly overlap for small disk reference scale heights and a flaring parameter of $\beta = \num{1.1}$, where the best-fit parameters of the model shown in Fig.~\ref{fig:mcrt_best_fit_envII} are located in the parameter space (see Figs.~\ref{fig:x2map_vis_thickDisk} and \ref{fig:x2map_sed_thickDisk}). Compared to the best fit found in the coarse parameter study with an envelope, the best-fit criterion $\min \left( \chi^2_\text{red}(V^2) + \chi^2_\text{red}(F_\text{tot}) \right)$ is higher, which can be mainly attributed to the total flux, which is overestimated in the $N$ band. However, the synthetic closure phases at long baselines in the $L$ and $M$ bands fit the observations well. In the $N$ band, the synthetic closure phases are significantly larger than for the model shown in Fig.~\ref{fig:mcrt_best_fit_envI}, but the model shown in Fig.~\ref{fig:mcrt_best_fit_envII} fails to fully reproduce the observed asymmetry. We note that the modeled $N$-band closure phases for the shorter two triplets with shortest maximum baseline lengths $B_\text{p;max}$ fit the observations poorly, which explains the larger value of $\chi^2_\text{red}(\phi_\text{cp})$ compared to the ``disk only'' and ``disk and envelope'' model (see Table \ref{tab:mcrt_best_fit_models}).\newline
    Conclusively, we found two different parameter combinations for the accretion disk and envelope model, which fit the observed visibilities similarly well:\bulletnewline
    \emph{Lower mass, strongly flared disk:} Best fit with respect to the numerical value of the best-fit criterion including visibilities and total flux. However, the modeled closure phases are close to zero, which contradicts the observations.\bulletnewline
    \emph{Higher mass, thin disk:} The observed closure phases can be modeled to some extent, maintaining a good fit on the visibilities at the cost of total flux overestimation in the $N$ band. The minimum value of the best-fit criterion is slightly smaller than for the best-fit disk model without an envelope shown in Fig.~\ref{fig:mcrt_best_fit_disk}.\bulletnewline
    The corresponding model density and temperature distributions are shown in Figs.~\ref{fig:density_distribution_disk_models} and \ref{fig:temperature_distribution_disk_models}, respectively. The density distributions show that hot dust is present in the LOS to the central protostar of RY Tau, confirming the results by \citet{Davies+2020}. The observed closure phases suggest that this dust is part of an envelope or disk wind, and not of the upper disk layers. We conclude that an optically thin envelope in combination with an accretion disk is the best model to produce synthetic observations that fit the MIR appearance of the inner disk of RY Tau, confirming the results by \citet{Schegerer+2008}. We finally note that an envelope model without a disk employing the density distribution given in Eq.~\ref{eq:envelope} is able to model the total flux well, while it fails to explain the observed visibilities and closure phases in a reasonable value range of $\varrho_{0,\text{env}} \in \left[ \SI{1e-15}{\kilogram\per\meter\cubed}, \SI{1e-12}{\kilogram\per\meter\cubed} \right]$ and $\gamma \in [ \num{-1.2}, \num{-0.3} ]$.\newline
    An open question remaining is the origin of the discrepancy between the observed and modeled $N$ band closure phases. Possible explanations include an induced shadow by infalling material as proposed by \citet{Krieger+2024} or an intrinsic asymmetric density distribution in the disk \citep[e.g.,][]{Varga+2021,Juhasz+2025}. Moreover, our modeling suggests that the asymmetry in the flux density in the $L$ band can be explained by an accretion disk combined with an envelope, while the asymmetry in the $N$ band is of a different origin and not visible in the $L$ band. As longer wavelengths allow us to probe regions closer to the disk midplane, an embedded accreting companion visible in the $N$ band, but obscured by the upper disk layers in the $L$ band, represents another possible origin of the observed closure phases.

    \subsection{Apparent size of RY Tau}
\begin{figure*}
      \centering
      \includegraphics[width=\textwidth]{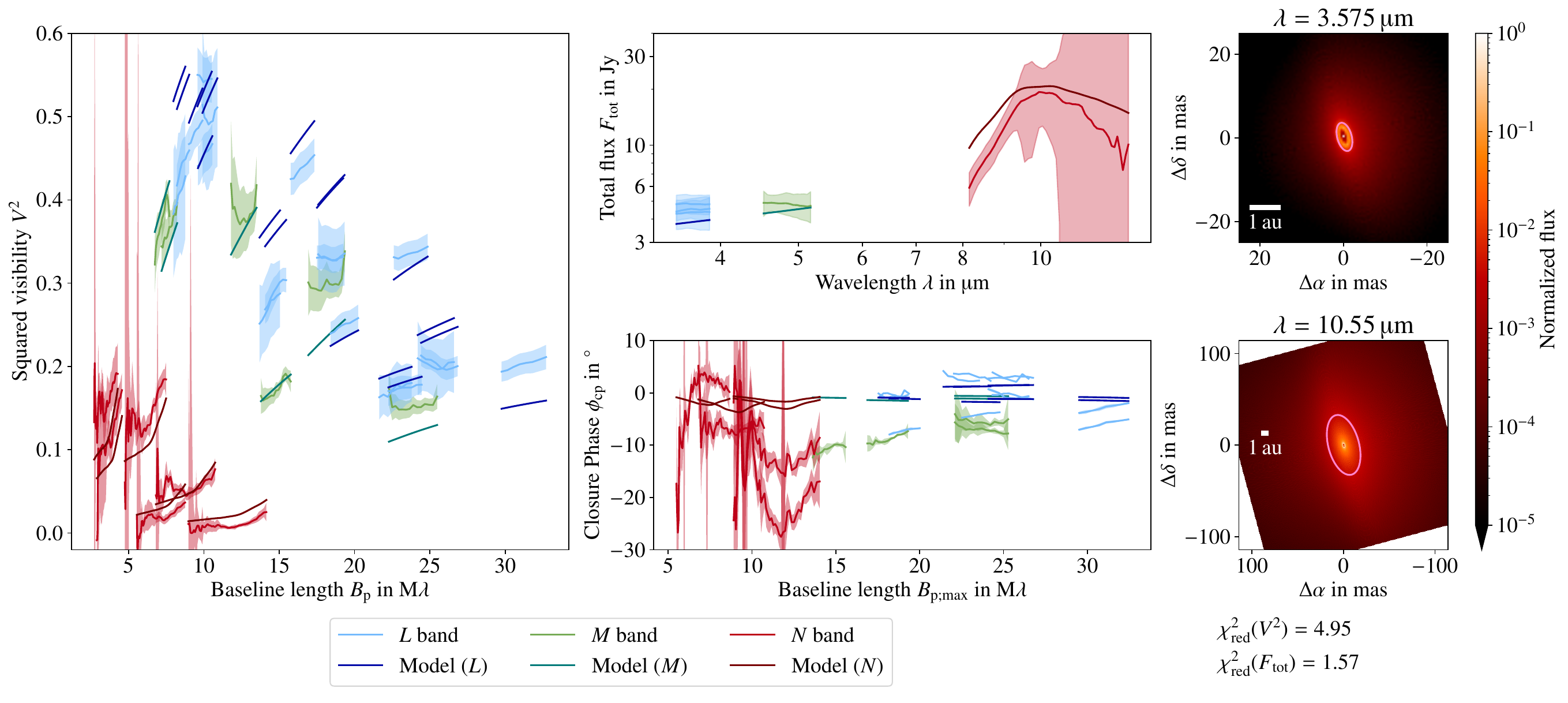}
      \caption{Best-fit model of the accretion disk model with an optically thin envelope. The model parameters are listed in Table \ref{tab:mcrt_best_fit_models}. The pink ellipses in the intensity maps indicate the half-light radii of the emitting region measured in the disk midplane as defined by \citet{Varga+2018}.}
      \label{fig:mcrt_best_fit_envI}
    \end{figure*}

    Using the best-fit disk and envelope model as basis for the size estimation in the $L$, $M$, and $N$ bands, the half-light radius $r_\text{hl}(\lambda)$ as defined in \citet{Varga+2018} was computed numerically on the intensity maps at wavelengths of \SI{3.58}{\micro\meter}, \SI{4.78}{\micro\meter}, and \SI{10.6}{\micro\meter}, respectively. The resulting $L$-band and $M$-band half-light radii are $r_\text{hl}(\SI{3.58}{\micro\meter}) = \num{3.5}^{+0.5}_{-0.5} \si{\milliarcsecond}$ and $r_\text{hl}(\SI{4.58}{\micro\meter}) = \num{4.9}^{+0.5}_{-0.5} \si{\milliarcsecond}$. For comparison, the $K$-band half-light radius found by \citet{GRAVITY+2021} is $\num{2.45}^{+0.24}_{-0.22}\si{\milliarcsecond}$.\newline
    In the $N$ band, our radiative transfer model results in a significantly larger value of $r_\text{hl}(\SI{10.6}{\micro\meter}) = \num{33.6}^{+0.5}_{-0.5} \si{\milliarcsecond}$ compared to the estimate of $\num{8.0}^{+0.7}_{-0.7} \si{\milliarcsecond}$ found by \citet{Varga+2018}. Considering the rather high $\chi^2$ of the best-fit model found by \citet{Varga+2018} used to compute the half-light radius, our estimate is better constrained by the MATISSE observations. The ellipses in the image plane, corresponding to $r_\text{hl}(\lambda)$ from within $\nicefrac{F_\text{tot}(\lambda)}{2}$ is emitted, are shown in Fig.~\ref{fig:mcrt_best_fit_envI}.
    
    \subsection{Photometry from the optical to MIR}
\begin{figure}
      \centering
      \includegraphics[width=\columnwidth]{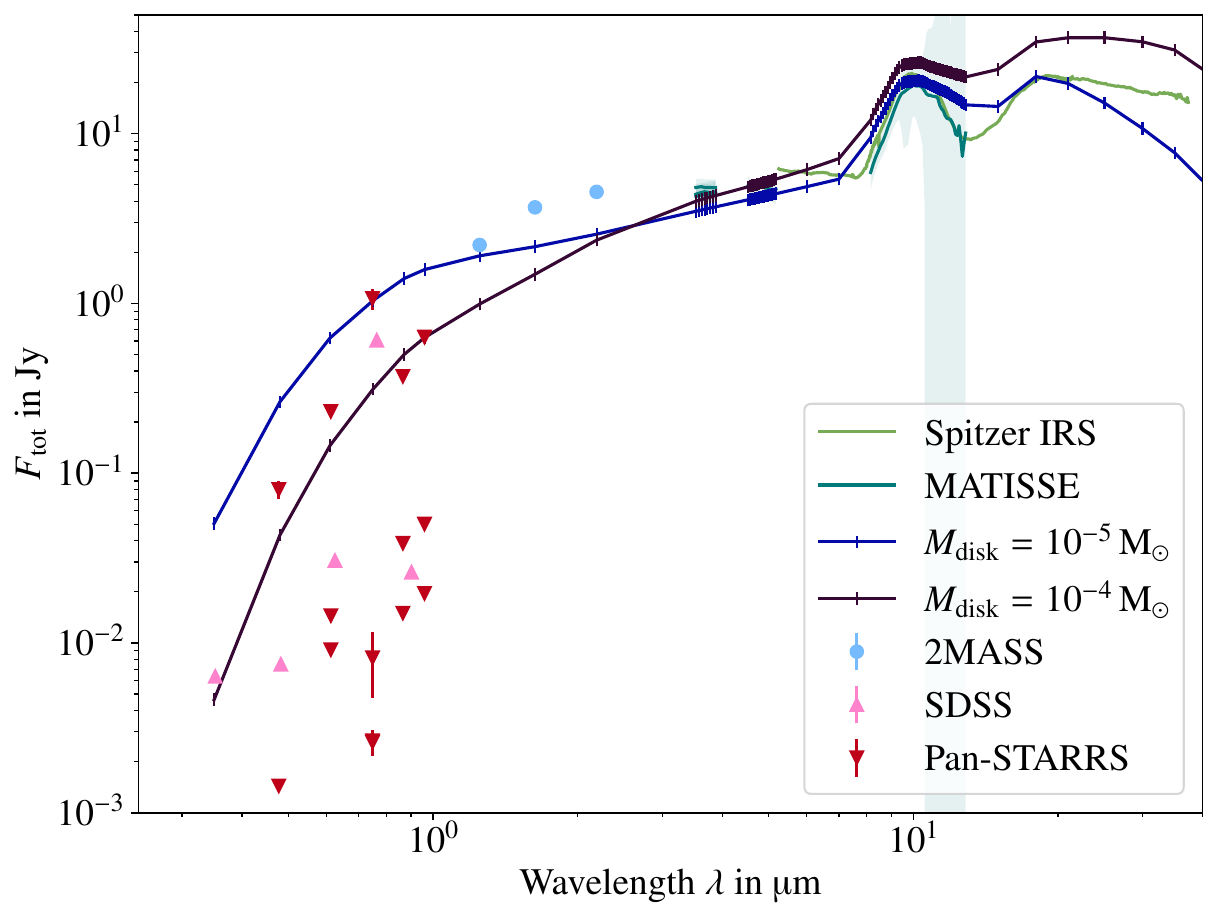}
      \caption{Comparison of the synthetic SED produced by the MCRT best-fits models presented in Sect.~\ref{sec:envelope} with photometry measurements. The models are distinguished by their disk mass $M_\text{disk}$ and the remaining parameters of the accretion disk and envelope model are given in Table \ref{tab:mcrt_best_fit_models}.}
      \label{fig:fullSED}
    \end{figure}

    As a consistency check of the best-fit models of an accretion disk and envelope given in Table \ref{tab:mcrt_best_fit_models}, the total flux measurements in $L$, $M$, and $N$ bands shown in Fig.~\ref{fig:disk-to-star} are supplemented with photometry in the optical and NIR \citep[SDSS, Pan-STARRS, 2MASS][]{Ahn+2012, Chambers+2016, Skrutskie+2006} and the Spitzer spectrum \citep{Lebouteiller+2011,Espaillat+2011} observed with the Infrared Spectrograph \citep[IRS;][]{Houck+2004} instrument in the MIR to compare the synthetic SED produced by our models to observations in a broad wavelength range. The photometry and synthetic SEDs for the lower mass and strongly flared disk ($M_\text{disk} = \SI{1e-5}{\Msun}$) and the higher mass and thin and dense midplane disk ($M_\text{disk} = \SI{1e-4}{\Msun}$) are shown in Fig.~\ref{fig:fullSED}. As expected, the photometry is highly variable in the optical, indicating that both models represent an active state of RY Tau. Comparing the MATISSE total flux measurements to the \emph{Spitzer} spectrum, no evidence for MIR variability is found. The model with the lower disk mass overestimates the total flux at short wavelengths, while the higher mass model fits the photometry in the optical well. In the NIR, both models underestimate the total flux. Toward longer wavelengths in the MIR beyond the $N$ band, the higher mass disk model overestimates the total flux. As we expect that larger grains settled to the midplane of the accretion disk, which are not included in our modeling, would contribute to the total flux in this wavelength range, the overestimation indicates that a disk mass of $M_\text{disk} = \SI{1e-4}{\Msun}$ is too high. As the total disk mass is poorly constrained if the disk midplane is optically thick, the flux overestimation toward longer wavelengths in the MIR of the ``dense disk and envelope'' (see Table \ref{tab:mcrt_best_fit_models}) model does not rule out the possibility of RY Tau hosting a dense midplane and geometrically thin accretion disk. A possibility to explain the mismatch between our models and the 2MASS photometry in the NIR is the accretion luminosity, which is reported to be relevant to model the appearance of RY Tau by \citet{Akeson+2005} and \citet{Schegerer+2008}.\newline
    We also note that the shape of the SED with respect to the broad wavelength range shown in Fig.~\ref{fig:fullSED} depends on the dust composition. While we only used polycrystalline graphite as continuum emitting material, iron and organic materials are expected to exist in protoplanetary disks \citep[e.g.,][]{Birnstiel+2018,Varga+2024}, possibly impacting the broad range shape of the SED. Additionally, the internal structure of graphite itself, while irrelevant in the $L$, $M$ and $N$ bands, has an impact on the resulting SED toward longer wavelengths.

    \section{The impact of accretion}\label{sec:accretion}
\begin{figure}
      \centering
      \includegraphics[width=\columnwidth]{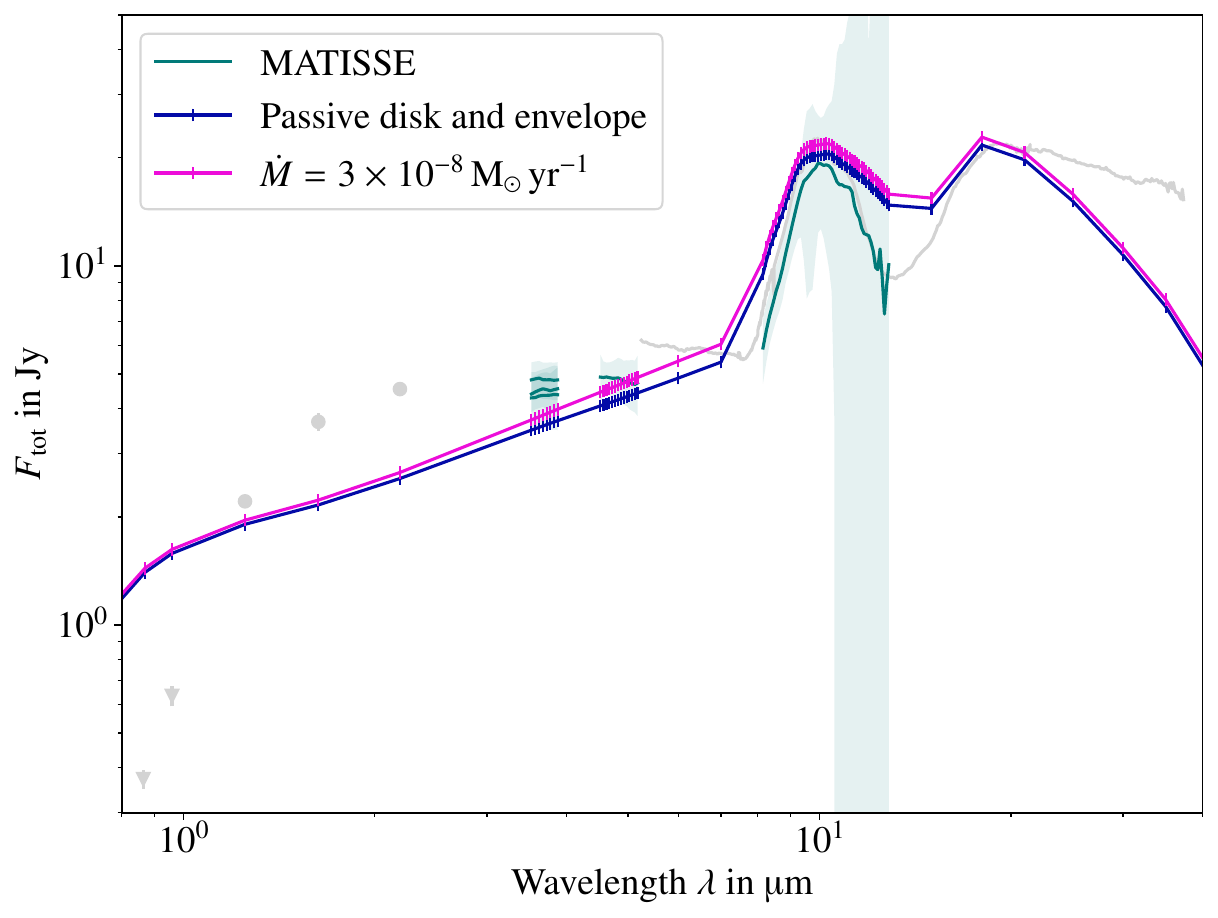}
      \caption{Contribution of accretion luminosity to the modeled SED considering i) emission by shock-heated gas accreting onto the stellar photosphere and ii) an optically thick gas disk between the magnetic boundary radius of the central protostar and the dust sublimation rim of the accretion disk.}
      \label{fig:accretion}
    \end{figure}

    To estimate the impact of accretion luminosity on the observables in the $L$, $M$ and $N$ bands, we employ a simple accretion model similar to the one implemented by \citet{Schegerer+2008}. To justify the negligence of accretion effects in the modeling presented in Sect.~\ref{sec:radiative_transfer}, we calculate the total flux released in the accretion process and assess its contribution to the observed SED. The mass accretion rate $\dot{M}$ of RY Tau is reported to be \SI{3.6e-8}{\Msun\per\year} by \citet{Petrov+2019} and \SI{2.7e-8}{\Msun\per\year} by \citet{Alcala+2021} which are both in the error interval of the value range \SIrange{6.4e-8}{9.1e-8}{\Msun\per\year} given in an earlier study by \citet{Petrov+1999}. We therefore assumed a value of \SI{3e-8}{\Msun\per\year} to estimate the flux contribution by the energy release in the accretion process. To this means, we used the best-fit accretion disk and envelope model with a disk mass of $M_\text{disk} = \SI{1e-5}{\Msun}$ (Fig.~\ref{fig:mcrt_best_fit_envI}) and included the accretion luminosity into the model to asses its influence on the resulting SED.\newline
    Following, e.g., \citet{Pringle1981}, the accretion luminosity is given by
    \begin{equation}
      \label{eq:accretion_energy}
      \begin{aligned}
        L_\text{acc}(R_\star \leq r < \infty) &= \int_{R_\star}^\infty D_\text{acc}(r) 2 \pi r \, d r = \frac{G \dot{M} M_\star}{2 R_\star}, \\
        D_\text{acc}(r) &= \frac{3GM_\star \dot{M}}{4 \pi r^3}\left( 1 - \sqrt{ \frac{R_\star}{r} } \right) . 
      \end{aligned}
    \end{equation}
    The total gravitational potential loss of accreting material is
    \begin{equation}
      \label{eq:grav_energy}
      E_\text{grav} = \frac{G M \dot{M}}{R_\star} = 2 L_\text{acc}(R_\star \leq r < \infty) ,
    \end{equation}
    with $\nicefrac{E_\text{grav}}{2}$ converted to kinetic energy, which is realeased in a boundary layer above the stellar photosphere \citep{Pringle1981}.  
    As \citet{Schegerer+2008}, we take accretion effects in different parts of our YSO model into account:
    \begin{enumerate}[i)]
    \item $R_\star \leq r < R_\text{bnd}$: The infalling gas interacts with the magnetic field forming strong shocks \citep{CalvetGullbring1998}. The shock layers above the stellar photosphere are assumed to emit blackbody radiation at a temperature of $T_\text{bnd} = \SI{8000}{\kelvin}$ \citep{Muzerolle+2003}. Assuming magnetic field strengths $\sim \si{\kilo\gauss}$, we adopted a value for the magnetic boundary radius of $R_\text{bnd} = \SI{5}{\Rstar}$ as \citet{Akeson+2005}. To test if this region has a significant impact on the observed SED in the NIR and MIR, we performed a MCRT simulation of the best-fit model including an additional energy release of $L_\text{acc}(R_\star \leq r < R_\text{bnd}) + \nicefrac{E_\text{grav}}{2} = \SI{0.17}{\Lsun} + \SI{0.29}{\Lsun}$ from the stellar photosphere assuming an accretion rate of $\dot{M} = \SI{3e-8}{\Msun\per\year}$. The simulation showed that this has a negligible impact on the SED from the optical to the MIR. We note that the impact of accretion needs to be considered if the accretion rate is in the range of the upper limit of the error interval given by \citet{Petrov+1999}, which is $\dot{M} = \SI{1.4e-7}{\Msun\per\year}$.
    \item $R_\text{bnd} \leq r < R_\text{in}$: The accretion luminosity is modeled as an infinitely thin, optically thick gas disk emitting as a blackbody with the radial temperature given by \citet{Pringle1981}
    \begin{equation}
      T_\text{gas}(r) = \left( \frac{3 G M_\star \dot{M}}{8 \pi R^3 \sigma_\text{SB}} \left( 1 - \sqrt{\frac{R_\star}{r}} \right) \right)^\frac{1}{4}.
    \end{equation}
    A similar approach has been implemented by \citet{Akeson+2005}. The flux arriving at the detector is computed by 
    \begin{equation}
      F_\text{acc} = \frac{1}{d^2} \int_{R_\text{bnd}}^{R_\text{in}} B_\nu(T_\text{gas}) \, dr \, ,
    \end{equation}
    resulting in an additional total flux in the NIR, as shown in Fig.~\ref{fig:accretion}.\newline
    We note that the accretion flux originating from this region is added to the detector without taking into account the interaction with the dust in the circumstellar environment. As the dust of the best-fit YSO model in the LOS from observer to the direct stellar vicinity is optically thin, the attenuation of $F_\text{acc}$ is neglected in order to estimate the upper limit of the impact of accretion luminosity on the NIR and MIR photometry.
  \item $R_\text{in} \leq r$: As the dissipation rate $D_\text{acc}(r)$ of infalling material given in Eq.~\ref{eq:accretion_energy} drops rapidly with increasing radial distance from the central star $r$, the energy release in this region can be neglected \citep{Akeson+2005,Schegerer+2008}.
    \end{enumerate}
    The SED taking the energy release in the accretion process in regions i) and ii) into account is compared to the SED of the best-fit model of a passively heated disk with an envelope in Fig.~\ref{fig:accretion}.
    We conclude that the impact of accretion can be neglected for the modeling of RY Tau observations in the $L$, $M$ and $N$ bands, as it has very little impact on the SED, and consequently on the visibilities. We note that the unrealistic scenario of a higher accretion rate and a smaller magnetic boundary radius would increase the NIR and MIR flux significantly.
    
    \section{Conclusions}
    We present the first MATISSE observations in the $L$, $M$, and $N$ bands of the inner disk of the YSO RY Tau. The data were analyzed with respect to mineralogy and spatial distribution of the dust in the immediate vicinity of the central protostar of the accretion disk.
    \begin{itemize}
    \item We estimated the accretion disk orientation fitting a 2D temperature gradient disk model to the observed visibilities, resulting in an inclination of $\iota = \SI{60}{\degree}$ and position angle of $\PA = \SI{15}{\degree}$, which suggests no significant misalignment of inner and outer disk. However, the model suggests an unresolved inner rim of the accretion disk, which would result in unphysically high dust sublimation temperatures. Employing 3D radiative transfer models, we demonstrated that hot dust in the LOS to the protostar of RY Tau resolves this issue.
    \item The dust composition was analyzed by applying a simple single-temperature model. Various silicate species, commonly found in protoplanetary disk, have been detected. Within the measurement accuracy and degeneracy of the modeling results, we cannot determine the dust grain shape, and found similar good fits on the $N$-band correlated fluxes with spherical compact grain and DHS models. We report a strong dependence on the continuum, crystalline, and amorphous contribution, as well as that of grain size on baseline length and orientation. We found an increase in the graphite abundance, representing the contribution of featureless flux, along with a decrease in the amount of amorphous silicate grains with decreasing unresolved disk scale, with the latter attributed to annealing.
    \item The spatial distribution of dust around the protostar of RY Tau has been constrained with MCRT simulations. We confirmed earlier results by \citet{Schegerer+2008}, that an accretion disk with an envelope fits the observations best. On the basis of our best-fit model, we estimate the half-light radii of the emitting region measured in the disk midplane as $r_{\text{hl},L} = \SI{0.5}{\astronomicalunit}$, $r_{\text{hl},M} = \SI{0.7}{\astronomicalunit}$, and $r_{\text{hl},N} = \SI{4.7}{\astronomicalunit}$ in the $L$, $M$, and $N$ bands, respectively. Based on recent accretion rate results by \citet{Petrov+2019} and \citet{Alcala+2021}, we find that accretion luminosity is negligible in the MIR. Our systematic exploration of the multidimensional parameter space revealed a strong degeneracy. We demonstrate that a strongly flared and thick disk, as well as a flat and dense disk, with scale heights of $h(\SI{5}{\astronomicalunit}) = \SI{0.61}{\astronomicalunit}$ and $h(\SI{5}{\astronomicalunit}) = \SI{0.28}{\astronomicalunit}$, respectively, in combination with an envelope fit the observed visibilities well. The latter is in agreement with the results by \citet{Davies+2020} analyzing interferometric observations of RY Tau in the $K$ band. This model also explains the observed closure phases in the $L$ band, but overestimates the total flux toward longer wavelengths. However, the specific origin of the observed asymmetry, appearing only in the $N$ band, cannot be further constrained on the basis of our observations.
 \end{itemize}
      Our analysis suggests that moderately inclined transition disks, such as the protoplanetary disk of RY Tau, cannot be necessarily modeled well assuming a flat disk. Taking into account the effect of hot dust in the LOS to the observer, possibly making up part of a disk wind, is required to interpret the MATISSE observations of RY Tau. The use of radiative transfer models, taking into account 3D effects of the dust distribution, scattering, and accretion luminosity, would plausibly be required to interpret similar observations of other protoplanetary disks. In the case of RY Tau, closure phases in the $L$ and $M$ band up to $~ \pm \SI{5}{\degree}$ can be explained by a flared disk geometry, without requiring any disk substructure. 
    
    \begin{acknowledgements}
      MATISSE was designed, funded and built in close collaboration with ESO, by a consortium composed of institutes in France (J.-L. Lagrange Laboratory -– INSU-CNRS -– Côte d’Azur Observatory -– University of Côte d’Azur), Germany (MPIA, MPIfR and University of Kiel), The Netherlands (NOVA and University of Leiden), and Austria (University of Vienna). The Konkoly Observatory and Cologne University have also provided some support in the manufacture of the instrument. This research was supported in part through high-performance computing resources available at the Kiel University Computing Centre. JV is funded from the Hungarian NKFIH OTKA projects no. K-132406, and K-147380. This work was also supported by the NKFIH NKKP grant ADVANCED 149943 and the NKFIH excellence grant TKP2021-NKTA-64. Project no.149943 has been implemented with the support provided by the Ministry of Culture and Innovation of Hungary from the National Research, Development and Innovation Fund, financed under the NKKP ADVANCED funding scheme. JV acknowledges support from the Fizeau exchange visitors programme. The research leading to these results has received funding from the European Union’s Horizon 2020 research and innovation programme under Grant Agreement 101004719 (ORP). LL gratefully acknowledges the Collaborative Research Center 1601 funded by the Deutsche Forschungsgemeinschaft (DFG, German Research Foundation)—SFB 1601 [sub-project A3]— 500700252. This research has made use of the services of the ESO Science Archive Facility. This research has made use of the OIFITS 2 data exchange standard for optical interferometry \citep{Duvert+2017} using the Python module \texttt{oifits} publicly available via \url{https://github.com/pboley/oifits}. This research has made use of the VizieR catalogue access tool, CDS, Strasbourg, France (DOI : 10.26093/cds/vizier). The original description of the VizieR service was published in \citet{Ochsenbein+2000}. Funding for SDSS-III has been provided by the Alfred P.~Sloan Foundation, the Participating Institutions, the National Science Foundation, and the U.S. Department of Energy Office of Science. The SDSS-III web site is \url{http://www.sdss3.org/}. SDSS-III is managed by the Astrophysical Research Consortium for the Participating Institutions of the SDSS-III Collaboration including the University of Arizona, the Brazilian Participation Group, Brookhaven National Laboratory, Carnegie Mellon University, University of Florida, the French Participation Group, the German Participation Group, Harvard University, the Instituto de Astrofisica de Canarias, the Michigan State/Notre Dame/JINA Participation Group, Johns Hopkins University, Lawrence Berkeley National Laboratory, Max Planck Institute for Astrophysics, Max Planck Institute for Extraterrestrial Physics, New Mexico State University, New York University, Ohio State University, Pennsylvania State University, University of Portsmouth, Princeton University, the Spanish Participation Group, University of Tokyo, University of Utah, Vanderbilt University, University of Virginia, University of Washington, and Yale University.
    \end{acknowledgements}

    \bibliographystyle{aa} 
    \bibliography{bib} 

    \begin{appendix}
    \onecolumn 
    \section{Additional material}
    \begin{figure*}[h!]
      \centering
      \includegraphics[width=\textwidth]{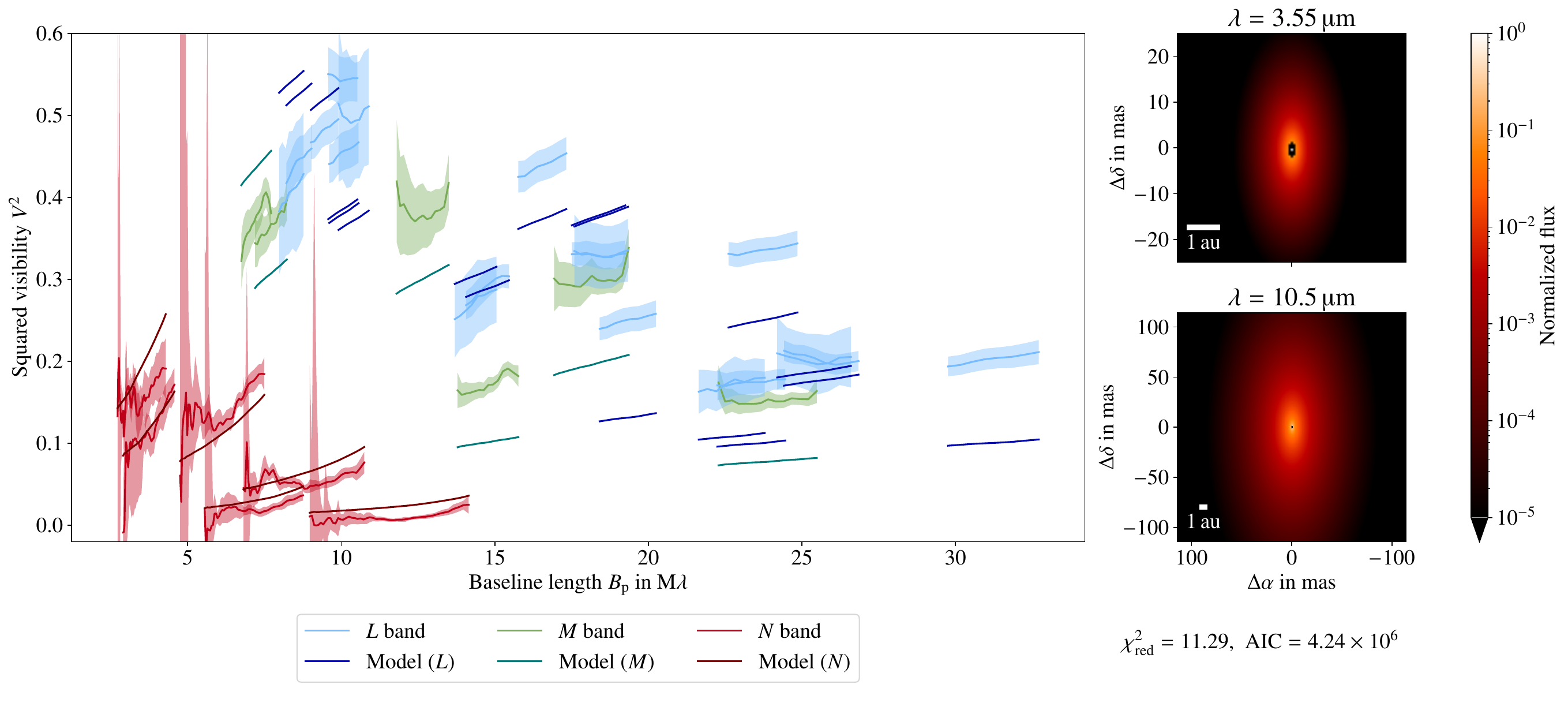}
      \caption{Best fit of the temperature gradient model with a fixed dust sublimation temperature of $T_\text{in} = \SI{1500}{\kelvin}$.}
      \label{fig:tg_Tin1500K}
    \end{figure*}
    \begin{figure*}[h!]
      \centering
      \includegraphics[width=\textwidth]{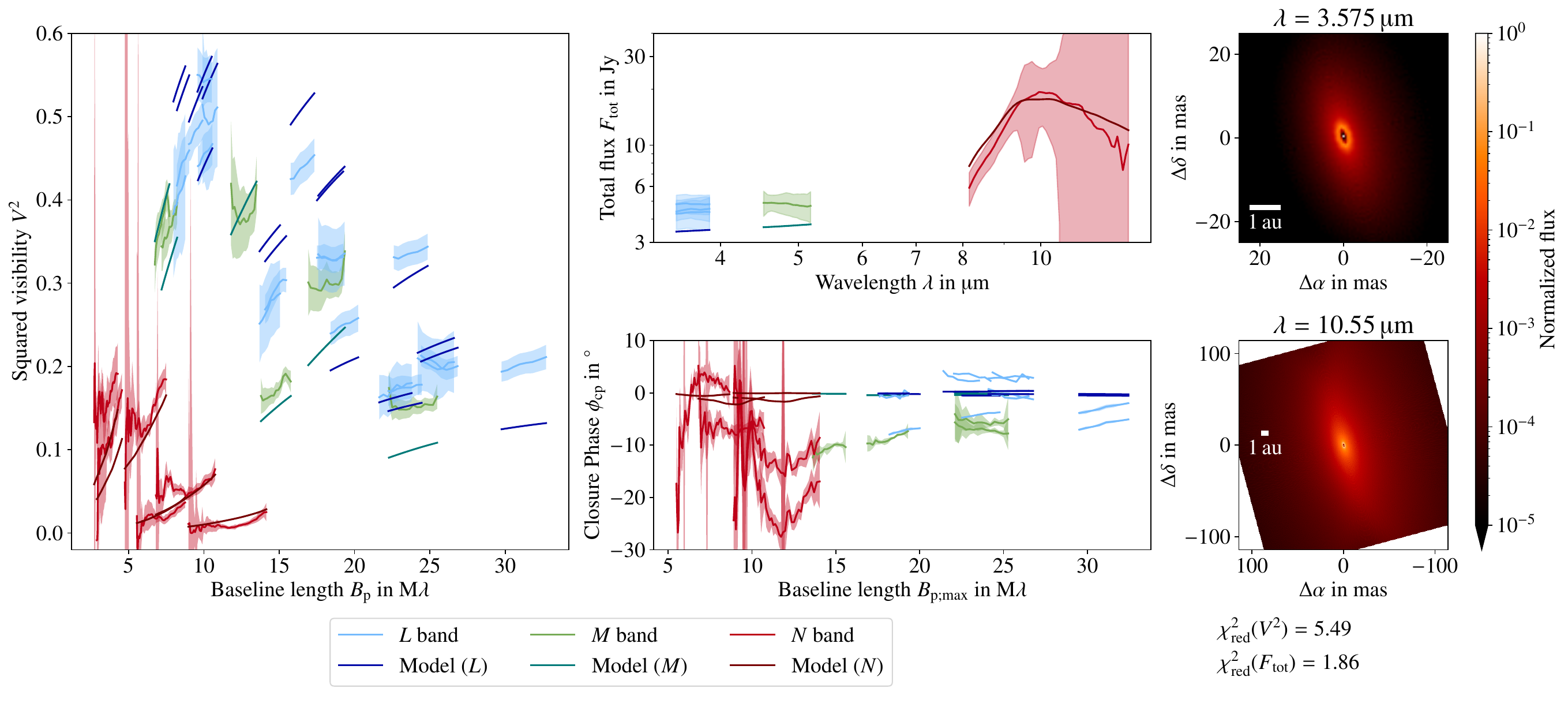}
      \caption{Best-fit model of the accretion disk model. The model parameters are listed in Table \ref{tab:mcrt_best_fit_models}.}
      \label{fig:mcrt_best_fit_disk}
    \end{figure*}

    \begin{figure*}
      \centering
      \includegraphics[width=\textwidth]{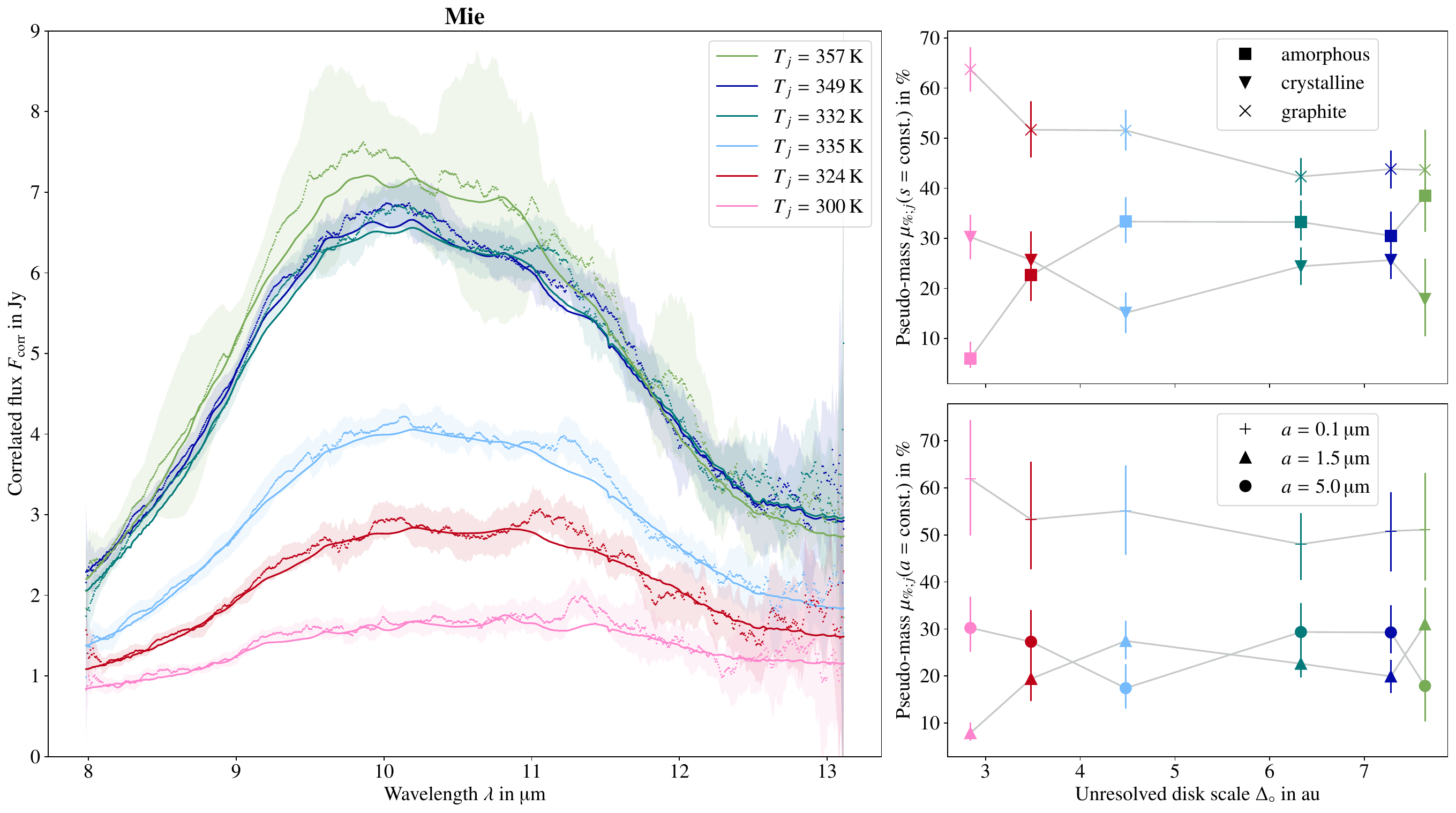}
      \caption{Best-fit model for the dust opacity model assuming spherical compact grains. \emph{Left:} Comparison of measured and fitted correlated fluxes for the $N$-band baselines. \emph{Upper right:} Trends of relative mass contribution of materials depending on their internal structure over unresolved disk scale. \emph{Lower right:} Trends of grain size over unresolved disk scale.}
      \label{fig:comp_ana_mie}
    \end{figure*}
    \begin{table*}[]
      \tiny
      \setlength{\tabcolsep}{10pt} 
      \renewcommand{\arraystretch}{1.5} 
      \centering
      \caption{Best fit of relative mass contributions of dust species in the compositional analysis fit on correlated fluxes in the $N$ band assuming spherical compact grains. Any nondetection of a certain dust species is indicated by gray values.}
      \label{tab:comp_ana_mie}
      \begin{tabular}{cccccccc}
\toprule
 & & \multicolumn{6}{c}{Unresolved disk scale $\Delta_\circ$} \\
 & Grain size in \si{\micro\meter} & \hexcolor{fe83cc}{\SI{2.84}{\astronomicalunit}} & \hexcolor{be0119}{\SI{3.48}{\astronomicalunit}} & \hexcolor{75bbfd}{\SI{4.48}{\astronomicalunit}} & \hexcolor{017a79}{\SI{6.33}{\astronomicalunit}} & \hexcolor{030aa7}{\SI{7.28}{\astronomicalunit}} & \hexcolor{77ab56}{\SI{7.64}{\astronomicalunit}} \\
\midrule
\multirow{3}{*}{Olivine} & 0.1 & {\color{lightgray}$0.4_{-0.3}^{+0.5}$} & $3.6_{-2.0}^{+2.6}$ & $5.4_{-1.9}^{+2.1}$ & $9.0_{-1.4}^{+1.4}$ & $9.3_{-1.8}^{+1.8}$ & $8.8_{-2.9}^{+3.1}$ \\
 & 1.5 & {\color{lightgray}$0.6_{-0.5}^{+0.9}$} & $10.5_{-3.9}^{+3.5}$ & $14.8_{-2.4}^{+2.5}$ & {\color{lightgray}$1.2_{-0.9}^{+1.4}$} & {\color{lightgray}$2.6_{-1.6}^{+2.0}$} & $5.3_{-3.5}^{+4.4}$ \\
 & 5.0 & {\color{lightgray}$1.2_{-0.9}^{+1.7}$} & {\color{lightgray}$0.8_{-0.6}^{+1.1}$} & {\color{lightgray}$0.6_{-0.5}^{+0.9}$} & {\color{lightgray}$0.9_{-0.7}^{+1.3}$} & {\color{lightgray}$1.1_{-0.8}^{+1.5}$} & {\color{lightgray}$0.9_{-0.7}^{+1.4}$} \\
\midrule
\multirow{3}{*}{Pyroxene} & 0.1 & {\color{lightgray}$1.2_{-0.7}^{+0.8}$} & {\color{lightgray}$0.6_{-0.4}^{+0.9}$} & {\color{lightgray}$0.4_{-0.3}^{+0.6}$} & {\color{lightgray}$0.2_{-0.2}^{+0.4}$} & {\color{lightgray}$0.5_{-0.3}^{+0.7}$} & {\color{lightgray}$1.0_{-0.8}^{+1.4}$} \\
 & 1.5 & {\color{lightgray}$1.1_{-0.8}^{+1.3}$} & $6.4_{-2.7}^{+2.9}$ & $11.3_{-3.0}^{+3.3}$ & $19.6_{-2.7}^{+2.4}$ & $15.0_{-3.0}^{+2.7}$ & $21.2_{-5.4}^{+5.9}$ \\
 & 5.0 & {\color{lightgray}$1.5_{-1.1}^{+2.2}$} & {\color{lightgray}$0.8_{-0.6}^{+1.3}$} & {\color{lightgray}$0.8_{-0.6}^{+1.1}$} & {\color{lightgray}$2.2_{-1.6}^{+2.6}$} & {\color{lightgray}$2.0_{-1.4}^{+2.4}$} & {\color{lightgray}$1.2_{-0.9}^{+1.7}$} \\
\midrule
\multirow{3}{*}{Enstatite} & 0.1 & {\color{lightgray}$0.3_{-0.2}^{+0.2}$} & {\color{lightgray}$0.3_{-0.2}^{+0.2}$} & {\color{lightgray}$0.2_{-0.1}^{+0.2}$} & {\color{lightgray}$0.2_{-0.1}^{+0.1}$} & {\color{lightgray}$0.3_{-0.1}^{+0.1}$} & {\color{lightgray}$0.1_{-0.1}^{+0.1}$} \\
 & 1.5 & $5.1_{-1.2}^{+1.3}$ & {\color{lightgray}$0.6_{-0.4}^{+0.8}$} & {\color{lightgray}$0.4_{-0.3}^{+0.5}$} & {\color{lightgray}$0.4_{-0.3}^{+0.5}$} & {\color{lightgray}$1.4_{-0.7}^{+0.7}$} & $2.7_{-1.6}^{+2.0}$ \\
 & 5.0 & $23.6_{-4.3}^{+4.1}$ & $22.3_{-5.4}^{+5.4}$ & $12.9_{-4.0}^{+3.9}$ & $22.1_{-3.7}^{+3.5}$ & $23.0_{-3.8}^{+3.5}$ & $12.2_{-7.2}^{+7.6}$ \\
\midrule
\multirow{3}{*}{Forsterite} & 0.1 & {\color{lightgray}$0.0_{-0.0}^{+0.0}$} & {\color{lightgray}$0.0_{-0.0}^{+0.1}$} & {\color{lightgray}$0.0_{-0.0}^{+0.0}$} & {\color{lightgray}$0.0_{-0.0}^{+0.0}$} & {\color{lightgray}$0.0_{-0.0}^{+0.0}$} & {\color{lightgray}$0.3_{-0.2}^{+0.3}$} \\
 & 1.5 & {\color{lightgray}$0.2_{-0.2}^{+0.3}$} & {\color{lightgray}$1.3_{-0.8}^{+1.3}$} & {\color{lightgray}$0.2_{-0.1}^{+0.3}$} & {\color{lightgray}$0.5_{-0.3}^{+0.5}$} & {\color{lightgray}$0.1_{-0.1}^{+0.2}$} & {\color{lightgray}$0.9_{-0.7}^{+1.1}$} \\
 & 5.0 & {\color{lightgray}$0.7_{-0.5}^{+1.0}$} & {\color{lightgray}$0.7_{-0.5}^{+1.0}$} & {\color{lightgray}$0.4_{-0.3}^{+0.6}$} & {\color{lightgray}$0.7_{-0.5}^{+1.1}$} & {\color{lightgray}$0.4_{-0.3}^{+0.7}$} & {\color{lightgray}$0.7_{-0.5}^{+1.1}$} \\
\midrule
\multirow{3}{*}{Quartz} & 0.1 & {\color{lightgray}$0.0_{-0.0}^{+0.0}$} & {\color{lightgray}$0.0_{-0.0}^{+0.0}$} & {\color{lightgray}$0.0_{-0.0}^{+0.0}$} & {\color{lightgray}$0.0_{-0.0}^{+0.0}$} & {\color{lightgray}$0.0_{-0.0}^{+0.0}$} & {\color{lightgray}$0.1_{-0.1}^{+0.1}$} \\
 & 1.5 & {\color{lightgray}$0.1_{-0.1}^{+0.1}$} & {\color{lightgray}$0.1_{-0.1}^{+0.1}$} & {\color{lightgray}$0.2_{-0.1}^{+0.2}$} & {\color{lightgray}$0.1_{-0.0}^{+0.1}$} & {\color{lightgray}$0.1_{-0.1}^{+0.1}$} & {\color{lightgray}$0.2_{-0.1}^{+0.2}$} \\
 & 5.0 & {\color{lightgray}$0.2_{-0.2}^{+0.4}$} & {\color{lightgray}$0.5_{-0.3}^{+0.7}$} & {\color{lightgray}$0.7_{-0.5}^{+1.0}$} & {\color{lightgray}$0.3_{-0.2}^{+0.5}$} & {\color{lightgray}$0.3_{-0.2}^{+0.5}$} & {\color{lightgray}$0.7_{-0.5}^{+1.0}$} \\
\midrule
\multirow{3}{*}{Graphite} & 0.1 & $60.1_{-12.0}^{+12.6}$ & $48.8_{-10.4}^{+12.0}$ & $49.0_{-9.1}^{+9.4}$ & $38.5_{-7.5}^{+6.4}$ & $40.7_{-8.3}^{+8.0}$ & $40.8_{-10.4}^{+11.6}$ \\
 & 1.5 & {\color{lightgray}$0.7_{-0.5}^{+1.0}$} & {\color{lightgray}$0.6_{-0.4}^{+0.9}$} & {\color{lightgray}$0.5_{-0.4}^{+0.7}$} & {\color{lightgray}$0.8_{-0.6}^{+1.0}$} & {\color{lightgray}$0.7_{-0.5}^{+1.0}$} & {\color{lightgray}$0.6_{-0.5}^{+0.9}$} \\
 & 5.0 & {\color{lightgray}$3.0_{-2.2}^{+4.3}$} & {\color{lightgray}$2.3_{-1.7}^{+3.4}$} & {\color{lightgray}$2.0_{-1.5}^{+2.9}$} & {\color{lightgray}$3.1_{-2.2}^{+4.0}$} & {\color{lightgray}$2.5_{-1.8}^{+3.5}$} & {\color{lightgray}$2.2_{-1.7}^{+3.3}$} \\
\midrule
$\chi^2_\text{red}$ & & 0.360 & 0.310 & 0.321 & 0.448 & 0.430 & 0.202\\ \bottomrule\end{tabular}
    \end{table*}

    \begin{figure*}
      \centering
      \includegraphics[width=\textwidth]{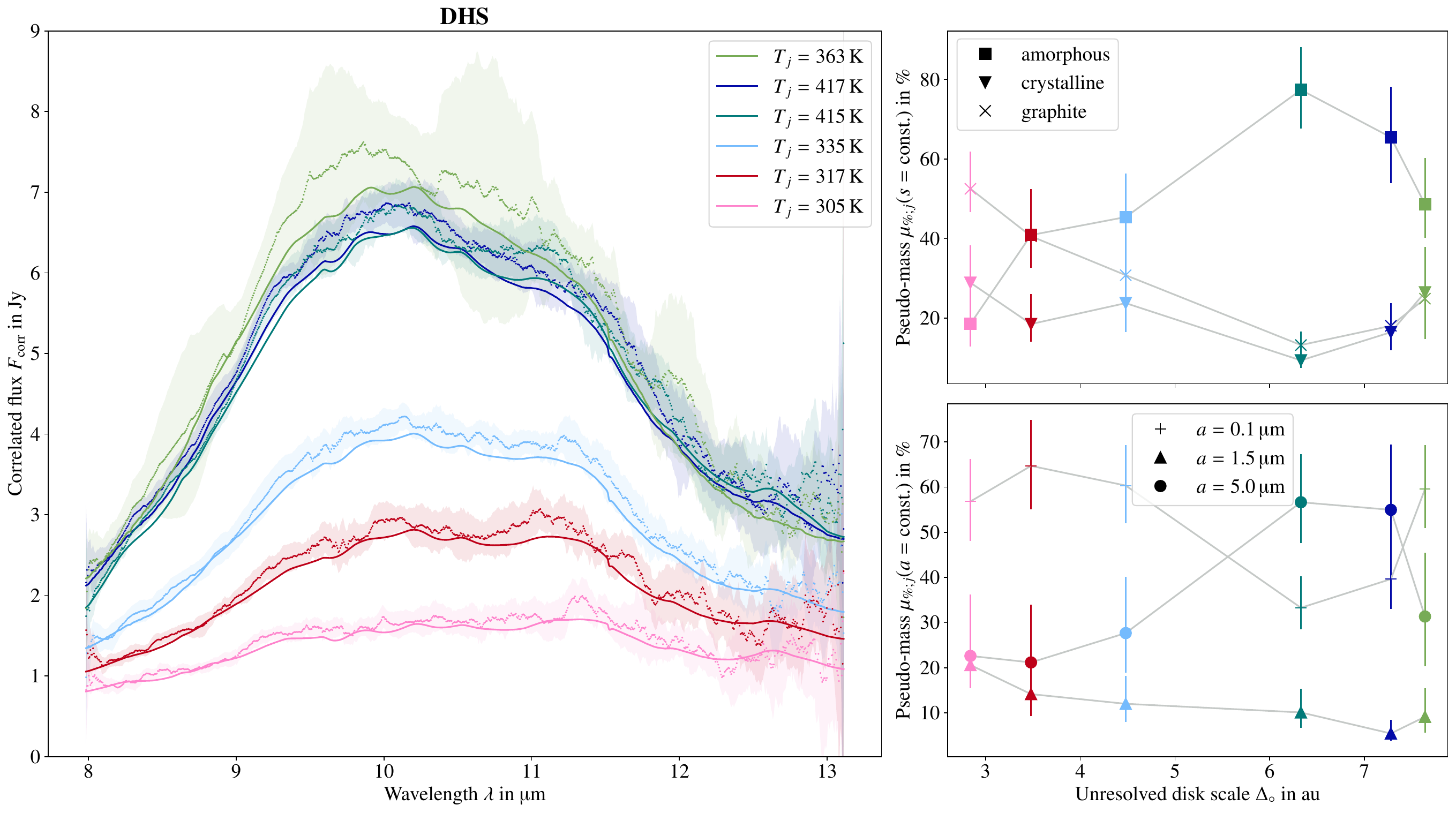}
      \caption{Best-fit model for the dust opacity model assuming a distribution of hollow spheres. \emph{Left:} Comparison of measured and fitted correlated fluxes for the $N$-band baselines. \emph{Upper right:} Trends of relative mass contribution of materials depending on their internal structure over unresolved disk scale. \emph{Lower right:} Trends of grain size over unresolved disk scale.}
      \label{fig:comp_ana_dhs}
    \end{figure*}
    \begin{table*}[]
      \tiny
      \setlength{\tabcolsep}{10pt} 
      \renewcommand{\arraystretch}{1.5} 
      \centering
      \caption{Best fit of relative mass contributions of dust species in the compositional analysis fit on correlated fluxes in the $N$ band employing the DHS opacity model. Any nondetection of a certain dust species is indicated by gray values.}
      \label{tab:comp_ana_dhs}
      \begin{tabular}{cccccccc}
\toprule
 & & \multicolumn{6}{c}{Unresolved disk scale $\Delta_\circ$} \\
 & Grain size in \si{\micro\meter} & \hexcolor{fe83cc}{\SI{2.84}{\astronomicalunit}} & \hexcolor{be0119}{\SI{3.48}{\astronomicalunit}} & \hexcolor{75bbfd}{\SI{4.48}{\astronomicalunit}} & \hexcolor{017a79}{\SI{6.33}{\astronomicalunit}} & \hexcolor{030aa7}{\SI{7.28}{\astronomicalunit}} & \hexcolor{77ab56}{\SI{7.64}{\astronomicalunit}} \\
\midrule
\multirow{3}{*}{Olivine} & 0.1 & {\color{lightgray}$2.4_{-1.7}^{+2.7}$} & $17.6_{-4.4}^{+4.4}$ & $19.1_{-3.4}^{+3.5}$ & $20.0_{-3.4}^{+4.0}$ & $19.8_{-3.5}^{+4.6}$ & $13.8_{-3.5}^{+3.7}$ \\
 & 1.5 & $4.7_{-3.3}^{+4.5}$ & {\color{lightgray}$4.6_{-3.3}^{+5.1}$} & {\color{lightgray}$2.8_{-2.0}^{+3.6}$} & {\color{lightgray}$2.8_{-2.0}^{+3.1}$} & {\color{lightgray}$1.1_{-0.9}^{+1.7}$} & {\color{lightgray}$2.3_{-1.7}^{+3.3}$} \\
 & 5.0 & {\color{lightgray}$3.0_{-2.2}^{+4.4}$} & {\color{lightgray}$3.0_{-2.3}^{+4.4}$} & {\color{lightgray}$2.6_{-1.9}^{+3.7}$} & {\color{lightgray}$2.4_{-1.8}^{+3.5}$} & {\color{lightgray}$4.6_{-3.3}^{+6.0}$} & {\color{lightgray}$2.3_{-1.7}^{+3.3}$} \\
\midrule
\multirow{3}{*}{Pyroxene} & 0.1 & {\color{lightgray}$1.9_{-1.4}^{+2.3}$} & $7.8_{-4.1}^{+4.6}$ & $12.1_{-4.7}^{+5.3}$ & {\color{lightgray}$2.4_{-1.7}^{+3.4}$} & $6.1_{-3.4}^{+5.9}$ & $23.0_{-6.2}^{+7.2}$ \\
 & 1.5 & {\color{lightgray}$3.0_{-2.2}^{+3.5}$} & {\color{lightgray}$4.8_{-3.4}^{+5.3}$} & $4.7_{-3.3}^{+4.6}$ & {\color{lightgray}$3.8_{-2.6}^{+4.1}$} & {\color{lightgray}$1.2_{-0.9}^{+1.7}$} & {\color{lightgray}$3.8_{-2.8}^{+4.9}$} \\
 & 5.0 & {\color{lightgray}$3.6_{-2.7}^{+5.2}$} & {\color{lightgray}$3.1_{-2.3}^{+4.6}$} & {\color{lightgray}$4.3_{-3.1}^{+5.7}$} & $45.8_{-8.1}^{+7.1}$ & $32.7_{-9.8}^{+8.0}$ & {\color{lightgray}$3.4_{-2.5}^{+4.9}$} \\
\midrule
\multirow{3}{*}{Enstatite} & 0.1 & $2.0_{-0.9}^{+1.0}$ & {\color{lightgray}$0.6_{-0.4}^{+0.6}$} & {\color{lightgray}$0.2_{-0.2}^{+0.3}$} & {\color{lightgray}$0.2_{-0.1}^{+0.2}$} & {\color{lightgray}$0.7_{-0.3}^{+0.4}$} & {\color{lightgray}$0.4_{-0.3}^{+0.5}$} \\
 & 1.5 & {\color{lightgray}$2.3_{-1.5}^{+2.0}$} & {\color{lightgray}$1.0_{-0.7}^{+1.2}$} & {\color{lightgray}$0.9_{-0.6}^{+1.0}$} & {\color{lightgray}$0.4_{-0.3}^{+0.5}$} & {\color{lightgray}$0.2_{-0.1}^{+0.3}$} & {\color{lightgray}$1.3_{-0.9}^{+1.6}$} \\
 & 5.0 & {\color{lightgray}$5.6_{-4.0}^{+7.0}$} & {\color{lightgray}$4.9_{-3.5}^{+5.9}$} & $12.7_{-6.9}^{+7.4}$ & {\color{lightgray}$1.3_{-1.0}^{+1.9}$} & $9.0_{-4.2}^{+5.0}$ & $18.1_{-9.9}^{+10.7}$ \\
\midrule
\multirow{3}{*}{Forsterite} & 0.1 & $4.2_{-1.0}^{+1.1}$ & $4.7_{-0.8}^{+0.9}$ & $3.5_{-0.6}^{+0.7}$ & $2.4_{-0.4}^{+0.6}$ & $1.9_{-0.4}^{+0.6}$ & $2.3_{-0.6}^{+0.7}$ \\
 & 1.5 & $4.4_{-2.2}^{+2.2}$ & {\color{lightgray}$0.6_{-0.4}^{+0.8}$} & {\color{lightgray}$1.0_{-0.7}^{+0.9}$} & {\color{lightgray}$0.3_{-0.2}^{+0.4}$} & {\color{lightgray}$0.7_{-0.5}^{+0.9}$} & {\color{lightgray}$0.8_{-0.6}^{+1.0}$} \\
 & 5.0 & {\color{lightgray}$2.4_{-1.8}^{+3.6}$} & {\color{lightgray}$1.9_{-1.4}^{+2.8}$} & {\color{lightgray}$1.3_{-1.0}^{+2.0}$} & {\color{lightgray}$1.4_{-1.0}^{+2.0}$} & {\color{lightgray}$1.0_{-0.7}^{+1.6}$} & {\color{lightgray}$2.4_{-1.8}^{+3.5}$} \\
\midrule
\multirow{3}{*}{Quartz} & 0.1 & {\color{lightgray}$0.1_{-0.1}^{+0.1}$} & {\color{lightgray}$0.1_{-0.1}^{+0.1}$} & {\color{lightgray}$0.0_{-0.0}^{+0.1}$} & {\color{lightgray}$0.0_{-0.0}^{+0.0}$} & {\color{lightgray}$0.0_{-0.0}^{+0.0}$} & {\color{lightgray}$0.1_{-0.0}^{+0.1}$} \\
 & 1.5 & $5.3_{-1.2}^{+1.3}$ & $2.3_{-0.9}^{+0.9}$ & $1.9_{-0.8}^{+0.8}$ & $2.1_{-0.5}^{+0.6}$ & {\color{lightgray}$1.3_{-0.5}^{+0.6}$} & {\color{lightgray}$0.3_{-0.2}^{+0.4}$} \\
 & 5.0 & {\color{lightgray}$2.7_{-2.0}^{+3.7}$} & {\color{lightgray}$2.5_{-1.8}^{+3.2}$} & {\color{lightgray}$2.2_{-1.6}^{+2.6}$} & {\color{lightgray}$1.3_{-0.9}^{+1.6}$} & {\color{lightgray}$1.6_{-1.2}^{+2.0}$} & {\color{lightgray}$0.9_{-0.7}^{+1.2}$} \\
\midrule
\multirow{3}{*}{Graphite} & 0.1 & $46.3_{-8.3}^{+8.6}$ & $33.9_{-7.4}^{+7.9}$ & $25.4_{-5.9}^{+6.3}$ & $8.2_{-2.8}^{+4.4}$ & $11.1_{-4.3}^{+7.3}$ & $20.0_{-4.9}^{+5.4}$ \\
 & 1.5 & {\color{lightgray}$0.9_{-0.7}^{+1.3}$} & {\color{lightgray}$0.9_{-0.7}^{+1.3}$} & {\color{lightgray}$0.8_{-0.6}^{+1.1}$} & {\color{lightgray}$0.7_{-0.5}^{+1.0}$} & {\color{lightgray}$0.9_{-0.7}^{+1.4}$} & {\color{lightgray}$0.7_{-0.5}^{+1.0}$} \\
 & 5.0 & {\color{lightgray}$5.3_{-4.0}^{+8.0}$} & {\color{lightgray}$5.7_{-4.3}^{+8.4}$} & {\color{lightgray}$4.6_{-3.4}^{+6.7}$} & {\color{lightgray}$4.4_{-3.3}^{+6.4}$} & {\color{lightgray}$6.1_{-4.6}^{+8.9}$} & {\color{lightgray}$4.2_{-3.2}^{+6.1}$} \\
\midrule
$\chi^2_\text{red}$ & & 0.461 & 0.503 & 0.647 & 0.766 & 1.109 & 0.305\\ \bottomrule\end{tabular}
    \end{table*}
    

\begin{figure*}
  \centering
  \includegraphics[width=\textwidth]{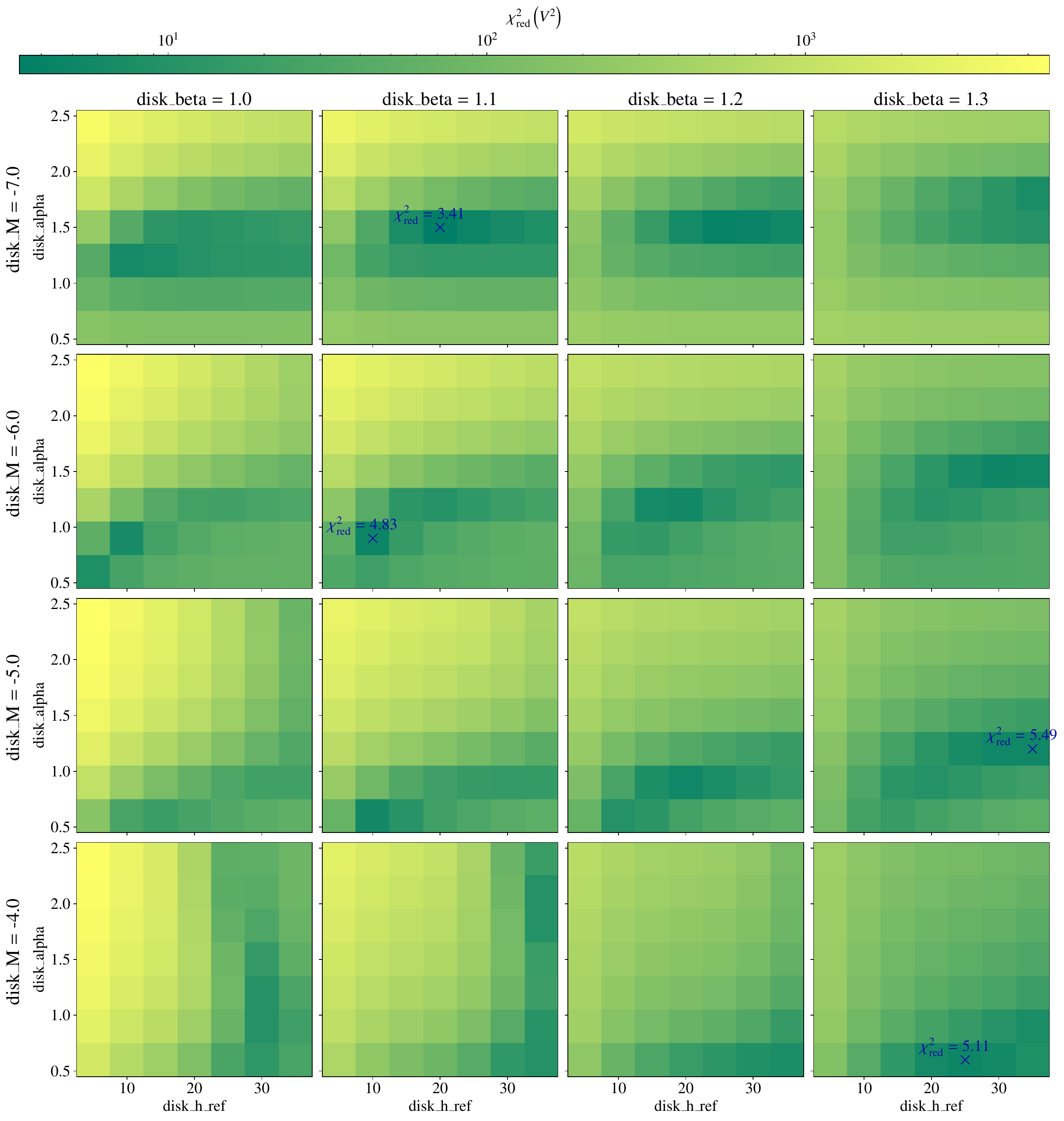}
  \caption{$\chi^2_\text{red}(V^2)$ map for $a_\text{max}=\SI{250}{\nano\meter}$ visualizing the structure of the four dimensional parameter space of $\log_{10} \left( M_\text{disk} \right)$ (``disk\_M''), $\beta$ (``disk\_beta''), $\alpha$ (``disk\_alpha'') and $h_\text{ref}$ (``disk\_h\_ref'') in \si{\astronomicalunit}. The location of the best fit per total disk mass is marked by a blue cross and the value of $\chi^2_\text{red}$ is given.}
  \label{fig:x2map_vis_diskOnly}
\end{figure*}

\begin{figure*}
  \centering
  \includegraphics[width=\textwidth]{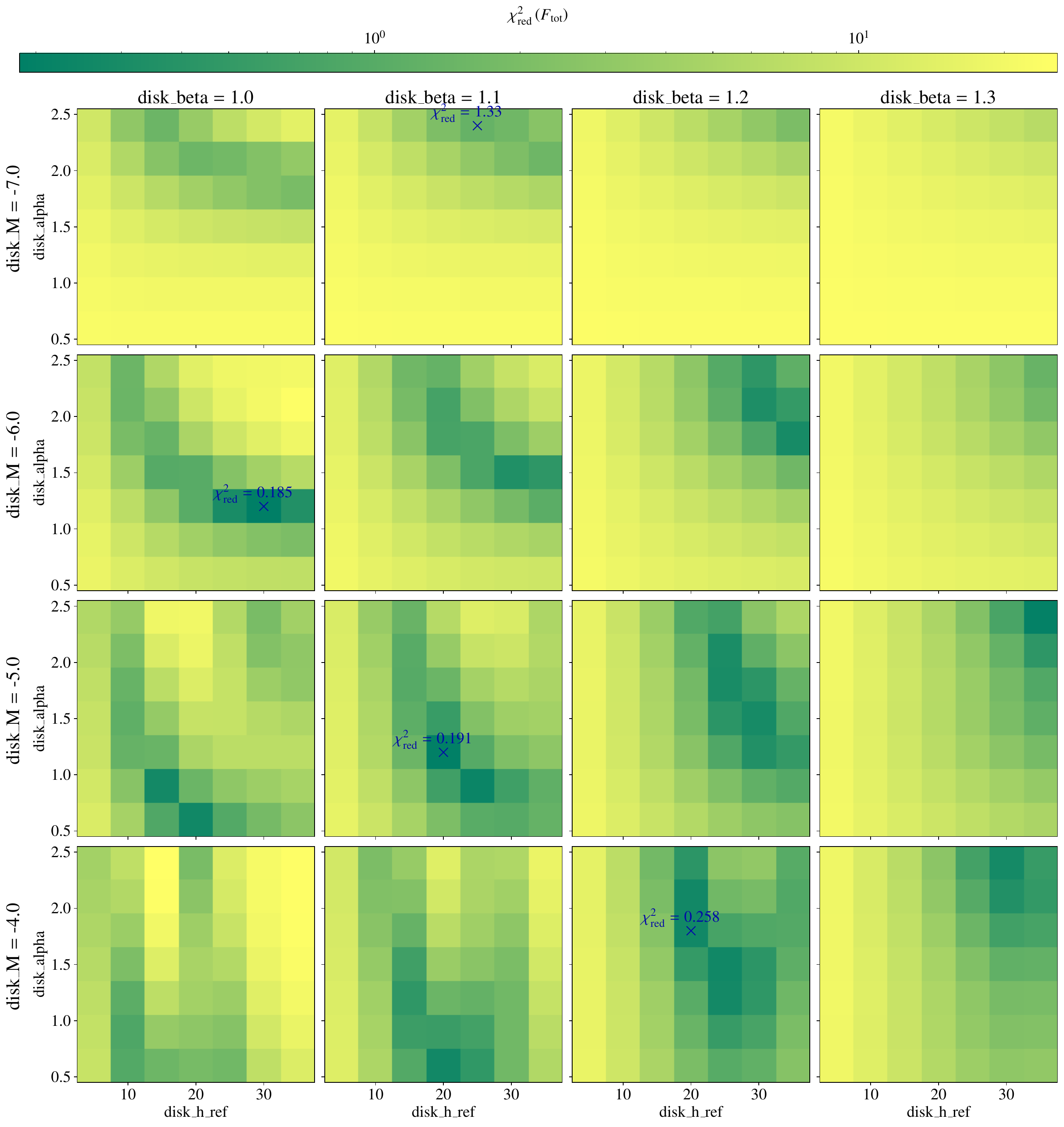}
  \caption{$\chi^2_\text{red}(F_\text{tot})$ map for $a_\text{max}=\SI{250}{\nano\meter}$ visualizing the structure of the four dimensional parameter space of $\log_{10} \left( M_\text{disk} \right)$ (``disk\_M''), $\beta$ (``disk\_beta''), $\alpha$ (``disk\_alpha'') and $h_\text{ref}$ (``disk\_h\_ref'') in \si{\astronomicalunit}. The location of the best fit per total disk mass is marked by a blue cross and the value of $\chi^2_\text{red}$ is given.}
  \label{fig:x2map_sed_diskOnly}
\end{figure*}

\begin{figure*}
  \centering
  \includegraphics[width=\textwidth]{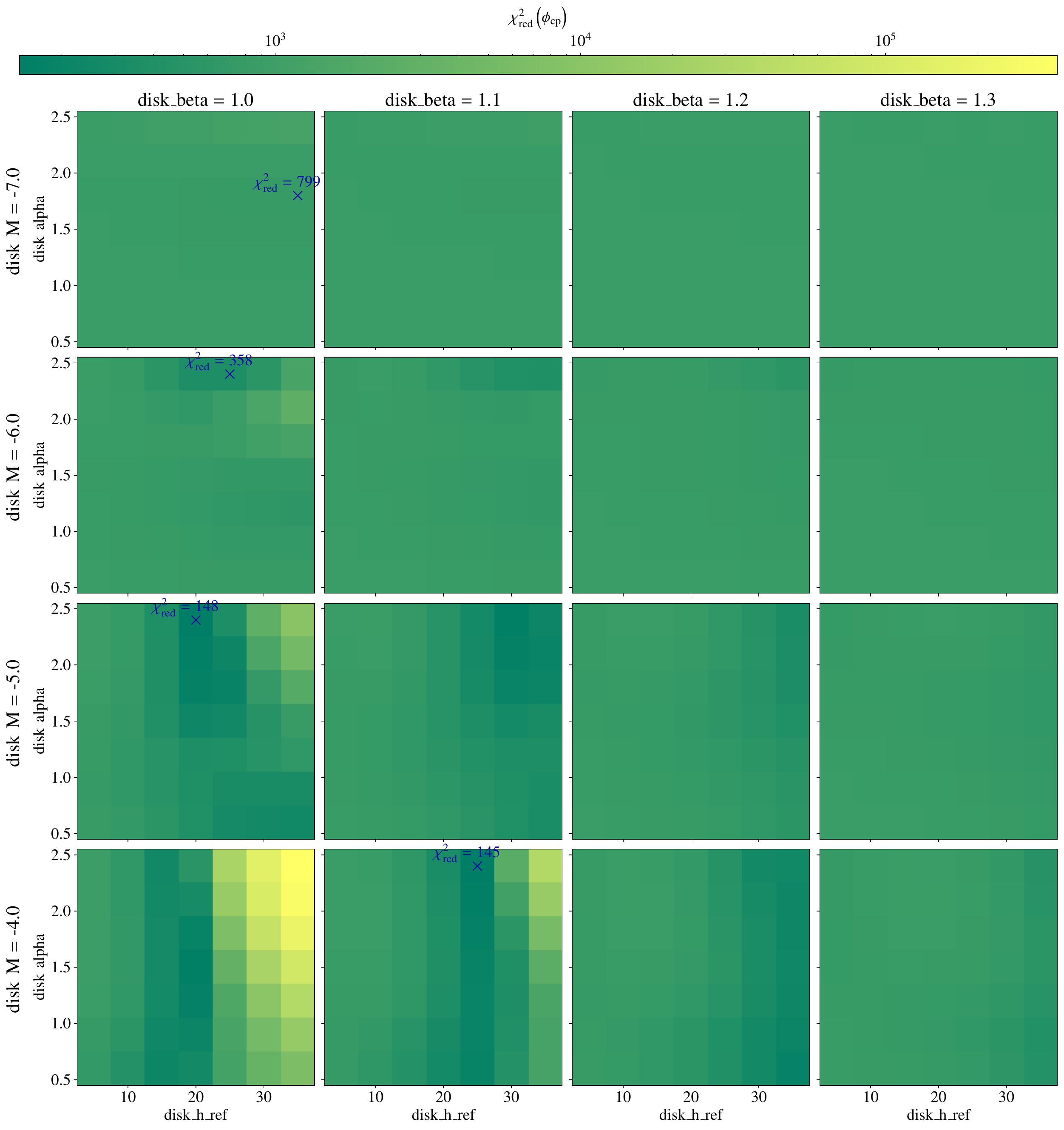}
  \caption{$\chi^2_\text{red}(\phi_\text{cp})$ map for $a_\text{max}=\SI{250}{\nano\meter}$ visualizing the structure of the four dimensional parameter space of $\log_{10} \left( M_\text{disk} \right)$ (``disk\_M''), $\beta$ (``disk\_beta''), $\alpha$ (``disk\_alpha'') and $h_\text{ref}$ (``disk\_h\_ref'') in \si{\astronomicalunit}. The location of the best fit per total disk mass is marked by a blue cross and the value of $\chi^2_\text{red}$ is given.}
  \label{fig:x2map_cp_diskOnly}
\end{figure*}

\begin{figure*}
  \centering
  \includegraphics[width=\textwidth]{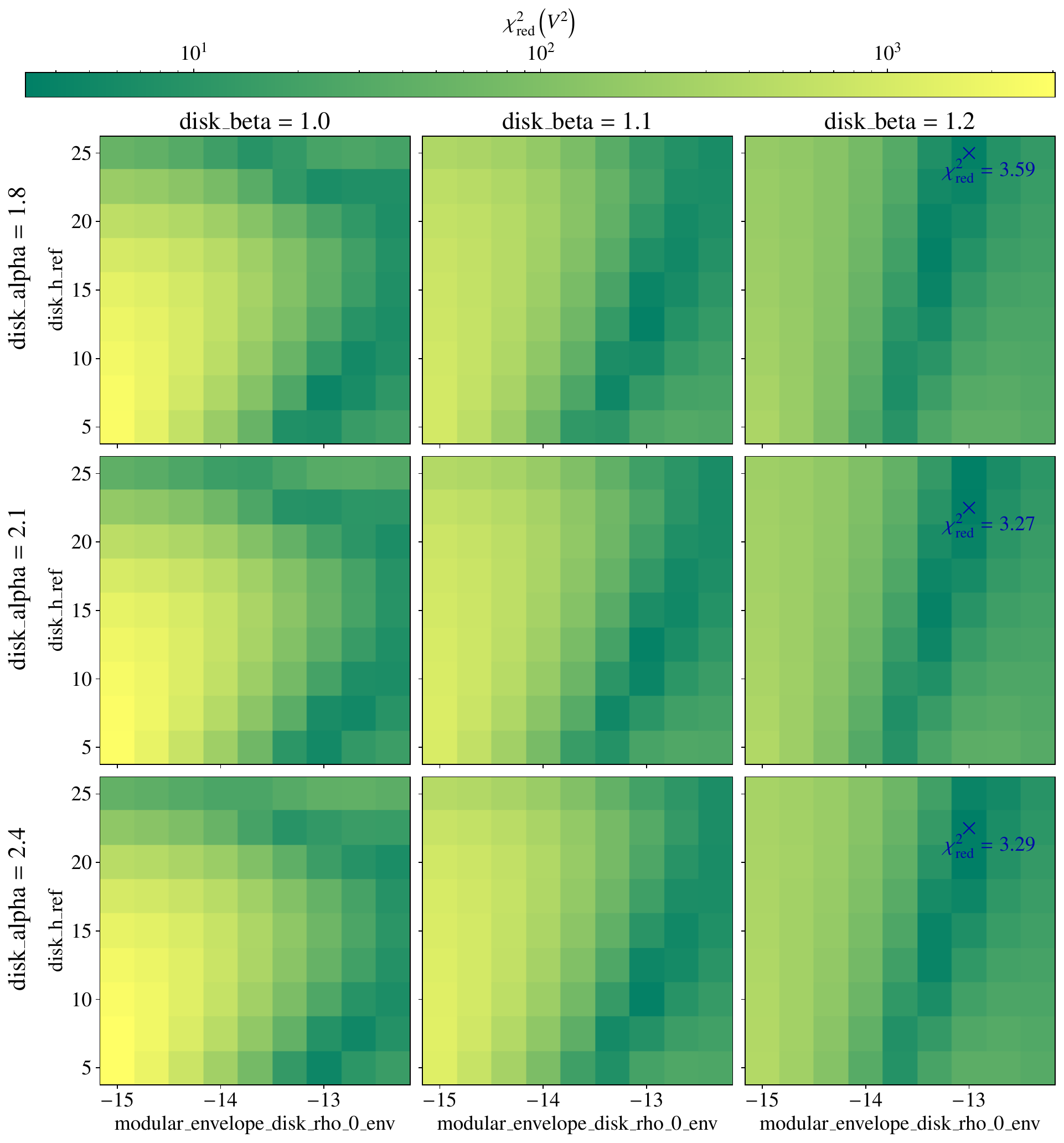}
  \caption{$\chi^2_\text{red}(V^2)$ map for $M_\text{disk} = \SI{1e-4}{\Msun}$ visualizing the structure of the four dimensional parameter space of $\alpha$ (``disk\_alpha''), $\beta$ (``disk\_beta''), $h_\text{ref}$ (``disk\_h\_ref'') in \si{\astronomicalunit} and $\log_{10} \left( \varrho_{0,\text{env}} \right)$ (``modular\_envelope\_disk\_rho\_0\_env''). The location of the best fit per disk midplane density exponent is marked by a blue cross and the value of $\chi^2_\text{red}$ is given.}
  \label{fig:x2map_vis_thickDisk}
\end{figure*}

\begin{figure*}
  \centering
  \includegraphics[width=\textwidth]{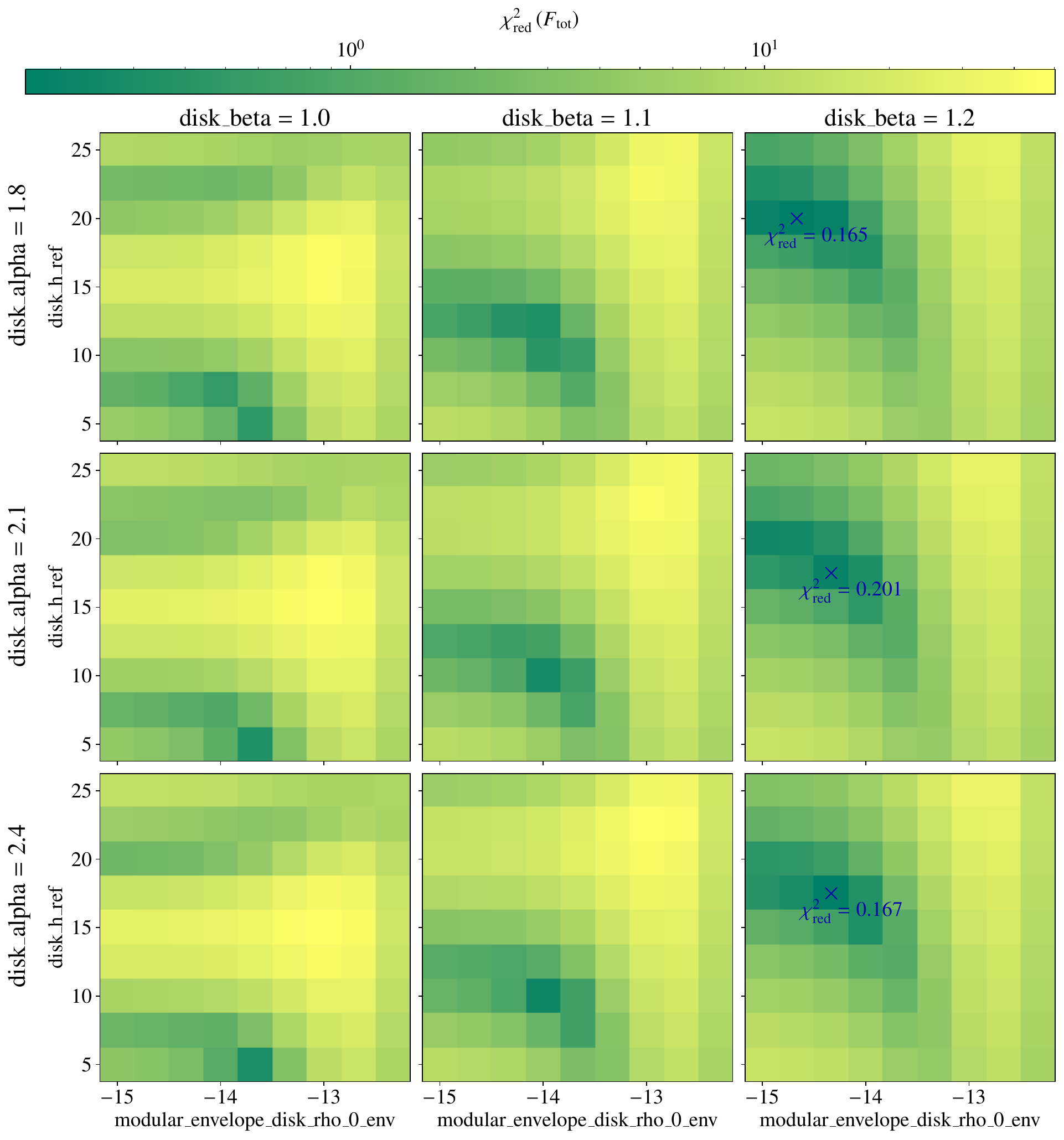}
  \caption{$\chi^2_\text{red}(F_\text{tot})$ map for $M_\text{disk} = \SI{1e-4}{\Msun}$ visualizing the structure of the four dimensional parameter space of $\alpha$ (``disk\_alpha''), $\beta$ (``disk\_beta''), $h_\text{ref}$ (``disk\_h\_ref'') in \si{\astronomicalunit} and $\log_{10} \left( \varrho_{0,\text{env}} \right)$ (``modular\_envelope\_disk\_rho\_0\_env''). The location of the best fit per disk midplane density exponent is marked by a blue cross and the value of $\chi^2_\text{red}$ is given.}
  \label{fig:x2map_sed_thickDisk}
\end{figure*}

\begin{figure*}
  \centering
  \includegraphics[width=\textwidth]{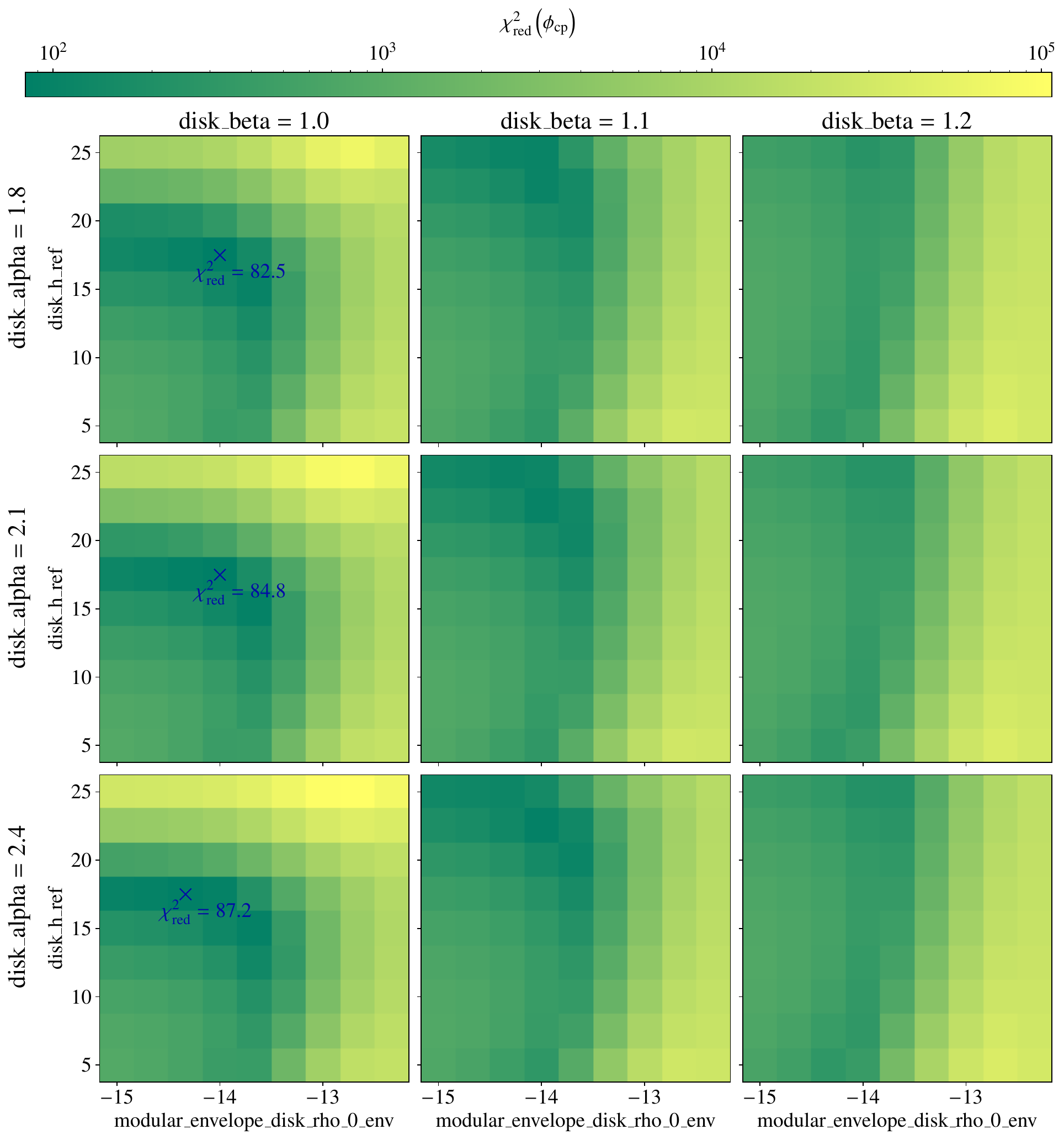}
  \caption{$\chi^2_\text{red}(\phi_\text{cp})$ map for $M_\text{disk} = \SI{1e-4}{\Msun}$ visualizing the structure of the four dimensional parameter space of $\alpha$ (``disk\_alpha''), $\beta$ (``disk\_beta''), $h_\text{ref}$ (``disk\_h\_ref'') in \si{\astronomicalunit} and $\log_{10} \left( \varrho_{0,\text{env}} \right)$ (``modular\_envelope\_disk\_rho\_0\_env''). The location of the best fit per disk midplane density exponent is marked by a blue cross and the value of $\chi^2_\text{red}$ is given.}
  \label{fig:x2map_cp_thickDisk}
\end{figure*}

\begin{figure*}
  \centering
  \begin{subfigure}[b]{0.32\textwidth}
    \includegraphics[width=\linewidth]{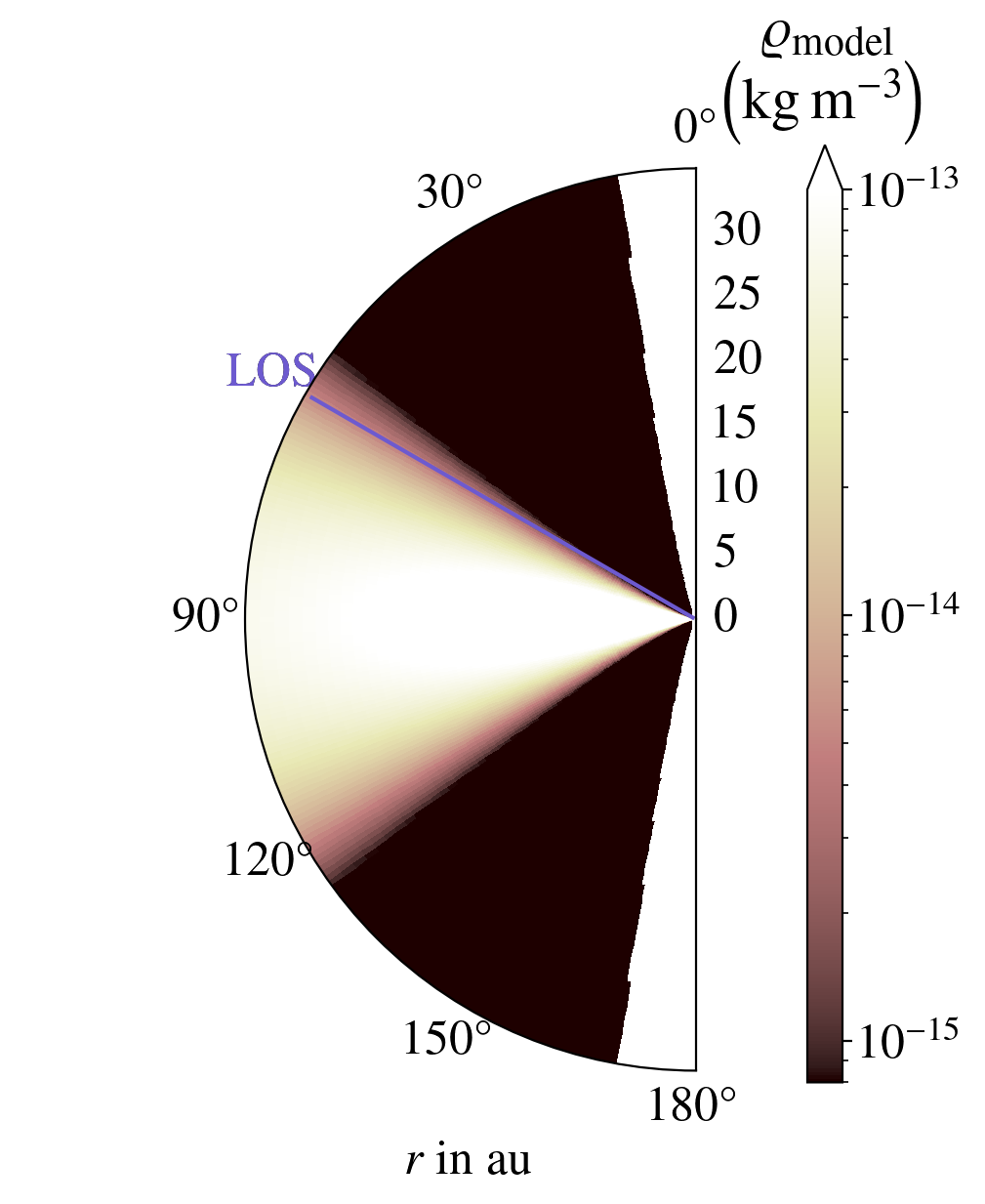}
    \caption{Disk only}
    \label{fig:dens_disk_only}
  \end{subfigure}
  \hfill
  \begin{subfigure}[b]{0.32\textwidth}
    \includegraphics[width=\linewidth]{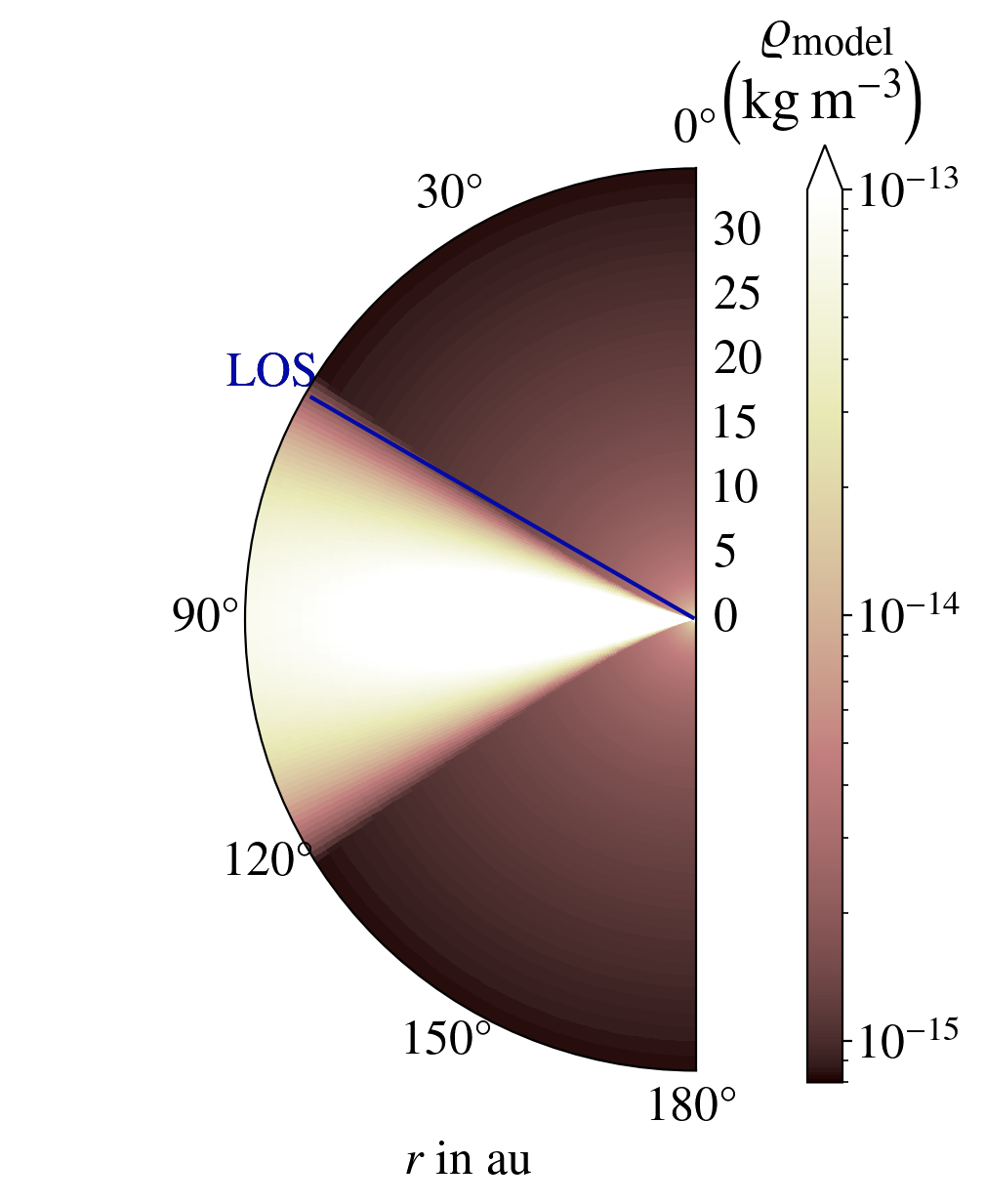}
    \caption{Disk and envelope}
    \label{fig:dens_disk_env}
  \end{subfigure}
  \hfill
  \begin{subfigure}[b]{0.32\textwidth}
    \includegraphics[width=\linewidth]{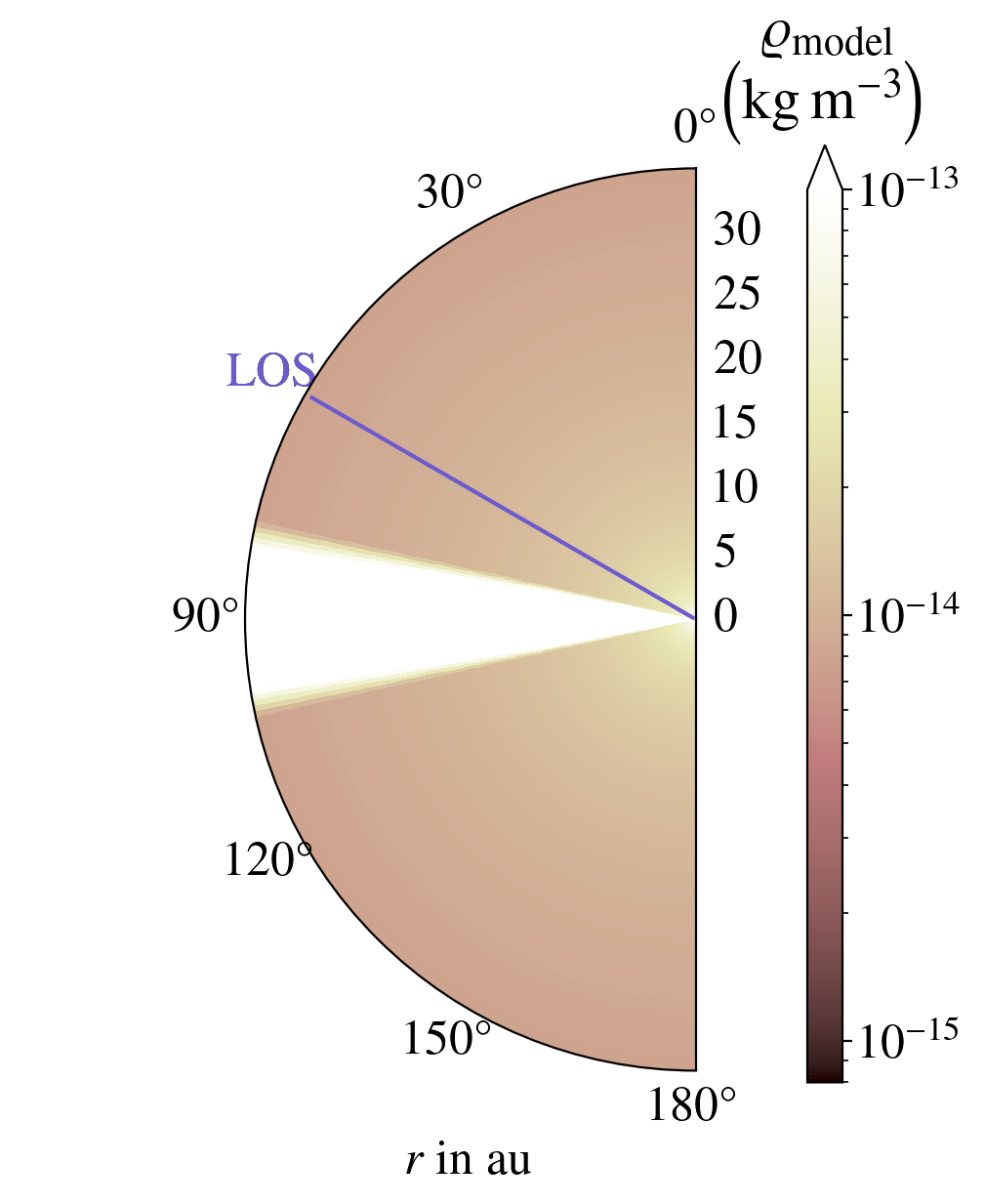}
    \caption{Dense disk and envelope}
    \label{fig:dens_dens_disk_env}
  \end{subfigure}
  \caption{Density distributions of the MCRT models listed in Table \ref{tab:mcrt_best_fit_models}.}
  \label{fig:density_distribution_disk_models}
\end{figure*}

\begin{figure*}
  \centering
  \begin{subfigure}[b]{0.32\textwidth}
    \includegraphics[width=\linewidth]{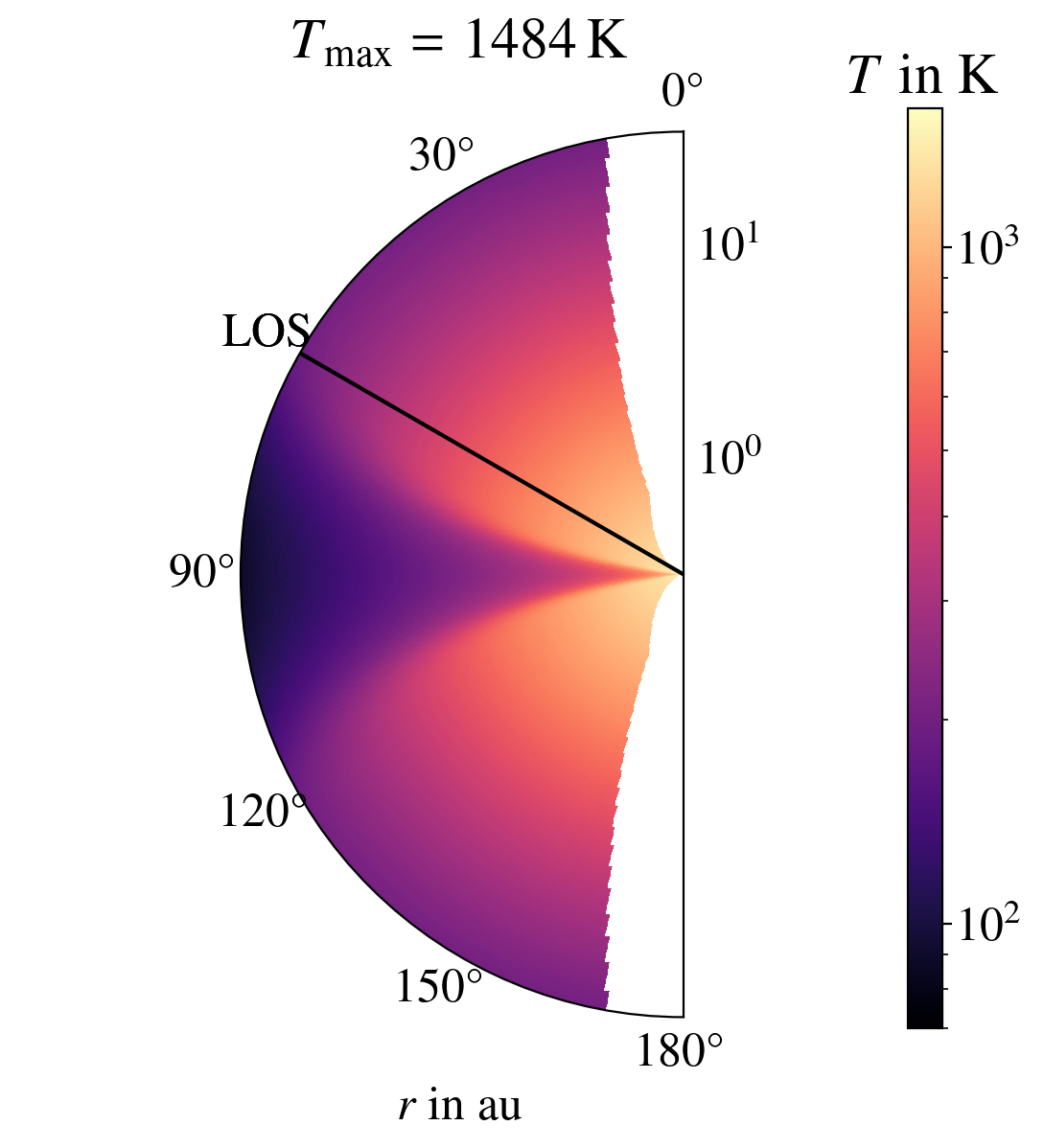}
    \caption{Disk only}
    \label{fig:temp_disk_only}
  \end{subfigure}
  \hfill
  \begin{subfigure}[b]{0.32\textwidth}
    \includegraphics[width=\linewidth]{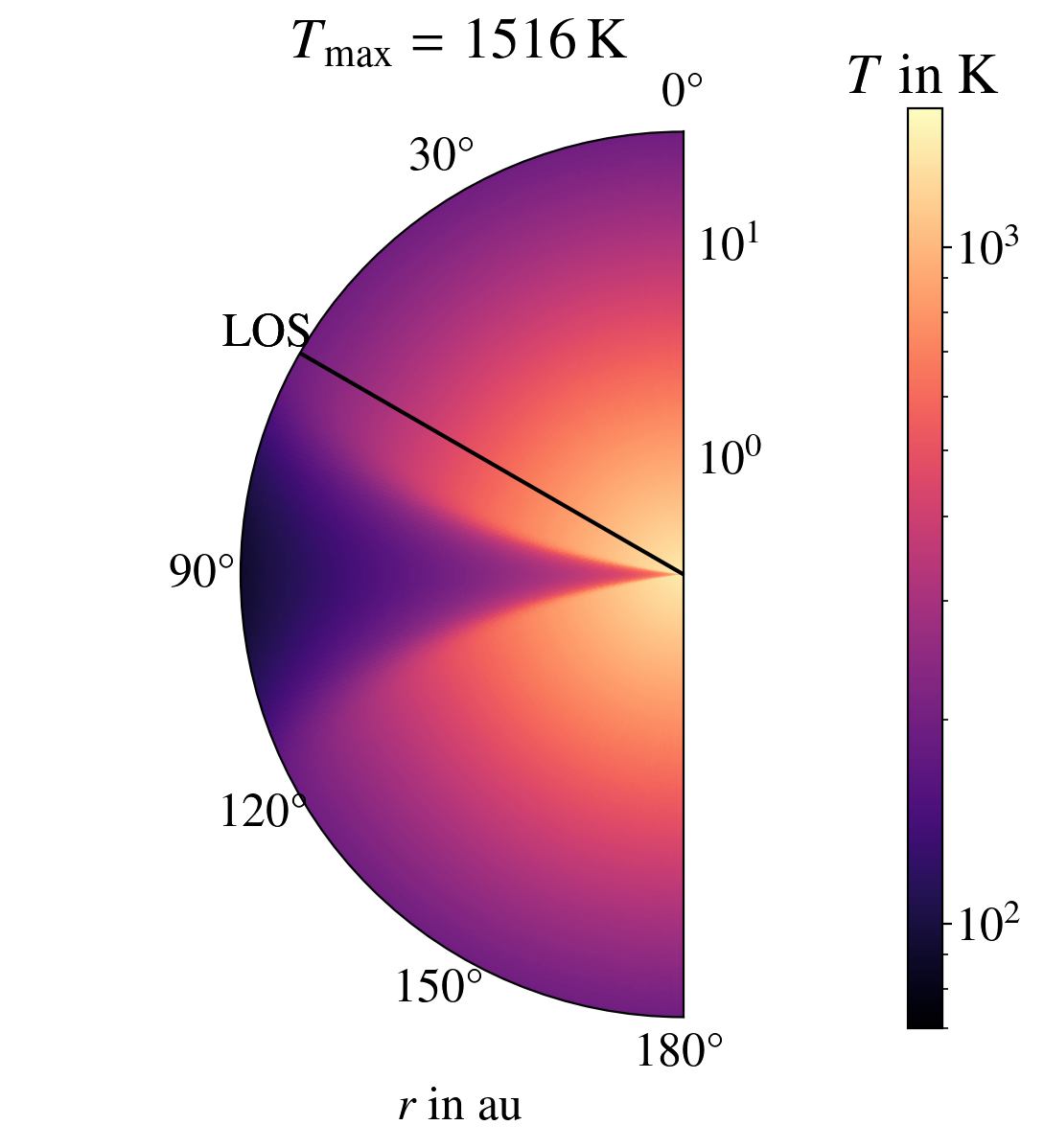}
    \caption{Disk and envelope}
    \label{fig:temp_disk_env}
  \end{subfigure}
  \hfill
  \begin{subfigure}[b]{0.32\textwidth}
    \includegraphics[width=\linewidth]{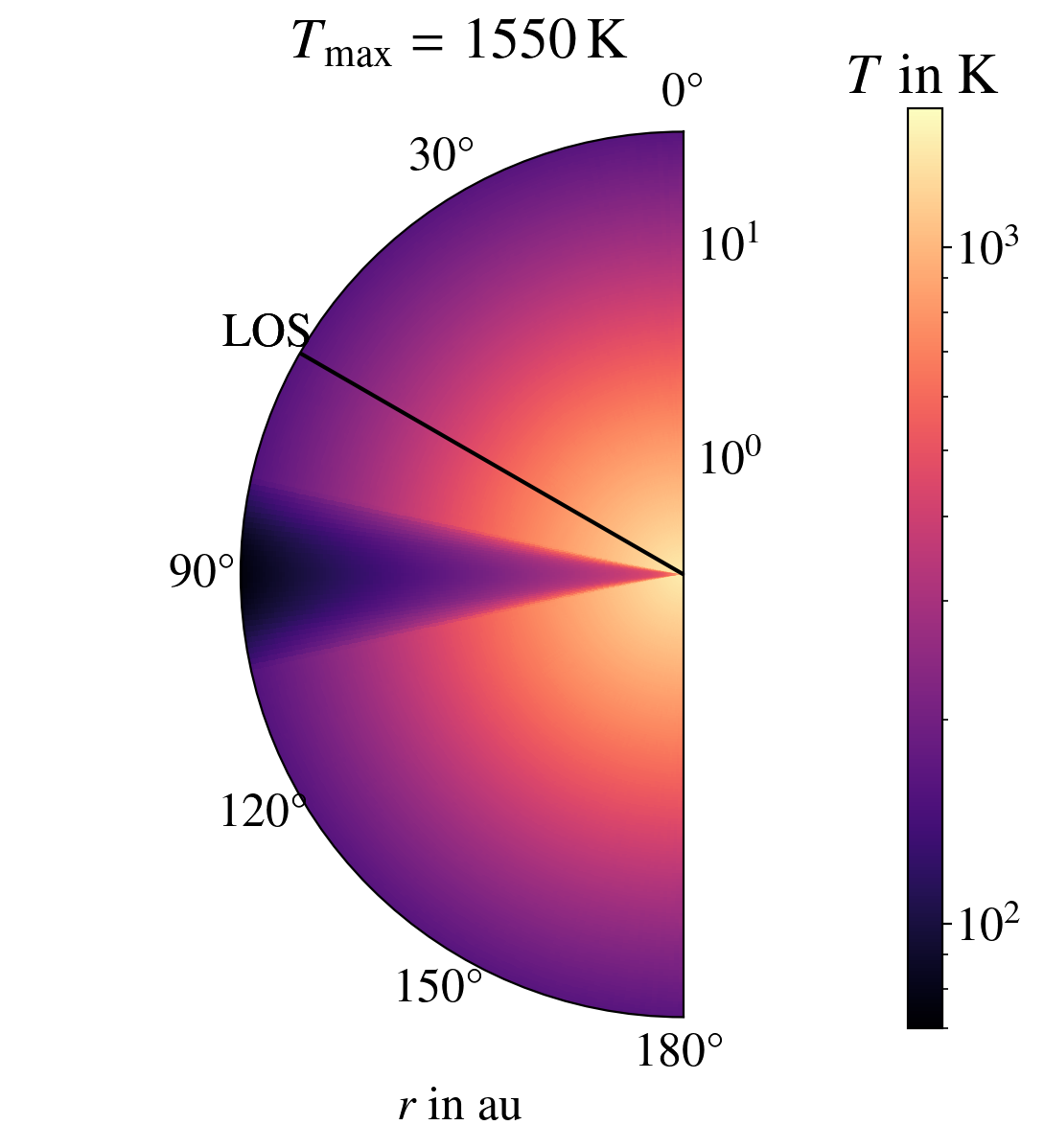}
    \caption{Dense disk and envelope}
    \label{fig:temp_dens_disk_env}
  \end{subfigure}
  \caption{Temperature distributions computed by MCRT simulations for the model density distributions shown in Fig.~\ref{fig:density_distribution_disk_models} of the MCRT models listed in Table \ref{tab:mcrt_best_fit_models}. We note that the radius $r$ is scaled logarithmically.}
  \label{fig:temperature_distribution_disk_models}
\end{figure*}

    \end{appendix}
\end{document}